\documentclass[acmsmall,screen,nonacm]{acmart}
\usepackage{amsmath}
\usepackage{amsfonts}
\usepackage{relsize}
\usepackage{enumitem}
\usepackage{stmaryrd}
\usepackage{mathpartir}
\usepackage{algorithmic}
\usepackage{quiver}
\usepackage{graphicx}
\usepackage{textcomp}
\usepackage{tikz,tikz-cd}
\usetikzlibrary{arrows.meta}
\usepackage{xcolor}
\usepackage[nameinlink,capitalise]{cleveref}
\usepackage{pedrohaa}
\newcommand{\laxslice}[2]{\cat{#1}\mathbin{/_{\ell}}#2}

\newcommand\emp{\mathrm{e}}
\newcommand\comp{\mathbin{\bullet}}
\newcommand{\topos}{\mathcal{S}}

\newcommand{\appref}[1]{Appendix~\ref{#1}}

\AddToHook{env/conjecture/begin}{\crefalias{theorem}{conjecture}}
\AddToHook{env/proposition/begin}{\crefalias{theorem}{proposition}}
\AddToHook{env/lemma/begin}{\crefalias{theorem}{lemma}}
\AddToHook{env/corollary/begin}{\crefalias{theorem}{corollary}}
\AddToHook{env/example/begin}{\crefalias{theorem}{example}}
\AddToHook{env/definition/begin}{\crefalias{theorem}{definition}}

\crefname{conjecture}{conjectures}{conjectures}
\crefformat{conjecture}{#2conjectures~#1#3} 
\Crefformat{conjecture}{#2Conjecture~#1#3} 

\crefname{proposition}{proposition}{propositions}
\crefformat{proposition}{#2proposition~#1#3} 
\Crefformat{proposition}{#2Proposition~#1#3} 

\crefname{lemma}{lemma}{lemmas}
\crefformat{lemma}{#2lemma~#1#3} 
\Crefformat{lemma}{#2Lemma~#1#3}

\crefname{corollary}{corollary}{corollarys}
\crefformat{corollary}{#2corollary~#1#3} 
\Crefformat{corollary}{#2Corollary~#1#3}

\crefname{example}{example}{examples}
\crefformat{example}{#2example~#1#3} 
\Crefformat{example}{#2Example~#1#3}

\crefname{definition}{definition}{definitions}
\crefformat{definition}{#2definition~#1#3} 
\Crefformat{definition}{#2Definition~#1#3}

\begin{document}

\theoremstyle{acmdefinition}
\newtheorem{remark}[theorem]{Remark}

\title{Fibered Separation Logic:
  Weakest Preconditions,
  Universal Constructions,
  and
  Local Independence}

\author{John M. Li}
\email{li.john@northeastern.edu}
\affiliation{
  \institution{Northeastern University}
  \city{Boston}
  \country{US}
}

\author{Ryan Doenges}
\email{doengesr@bc.edu}
\affiliation{
  \institution{Boston College}
  \city{Chestnut Hill}
  \country{US}
}

\author{Pedro H. Azevedo de Amorim}
\email{phaa20@bath.ac.uk}
\affiliation{
  \institution{University of Bath}
  \city{Bath}
  \country{UK}
}

\begin{abstract}
  This paper introduces a categorical framework for models of higher-order
  separation logic including the weakest precondition modality. Weakest
  preconditions are interpreted by strong monads on a category fibered over
  a model of Moggi's computational lambda calculus. The fibered categories
  in question are termed bunched fibrations and shown to be equivalent to BI
  hyperdoctrines.

  Well-known heuristic techniques for obtaining weakest preconditions are shown
  to be expressible in categorical terms and are proved sound. The ``baking in
  the frame rule'' trick is shown to satisfy a universal property. Constructions
  are given for extending a program logic with cost credits.
  Our largest case study is parametrized by a category having local independent
  product (LIP) structure, a notion due to Simpson that axiomatizes
  name generation, randomness, and local state.
  From an LIP category, we construct an allocation monad,
  a bunched fibration,
  and a monad lifting.
  We instantiate our construction to obtain a novel probabilistic separation
  logic for measure-theoretic probability that both simplifies and clarifies
  prior work.
\end{abstract}

\maketitle

\section{Introduction}

\label{sec:introduction}
This paper introduces a systematic approach to the semantics of separation
logic in terms of \emph{strong monads} on models of bunched implications (BI).
While BI, the logic of resources at rest, has been treated systematically by
O'Hearn~and~Pym~\cite{o1999logic}, Biering~\cite{biering2004logic}
and many others, 
we find the dynamic aspects of separation logic have not yet received
a comparable treatment.
This is not for lack of material,
as the literature on separation logic is extensive,
well understood by experts and workers in verification,
and organized, to some extent, into heuristics and rules.

Separation logic extends ordinary Hoare logic 
with substructural logical connectives that permit compositional reasoning about
resource-manipulating programs.
The separating conjunction \(P \sep Q\) holds on resources that can be separated into 
one part that satisfies \(P\) and another part that satisfies \(Q\).
The frame rule says that a verified program \(c\) respects this notion of resource
separation, in the sense that whatever \(c\) does to \(P\) is stable in the
presence of additional resources described by a separate \(F\).
\[
  \{P\}\,c\,\{Q\}
  \implies
  \{P\sep F\}\,c\,\{Q \sep F\}
\]
In the original incarnation of separation logic~\cite{reynolds2002separation},
the resources in question were
fragments of the heap. Subsequent developments vastly generalized the notion of
resource, leading to logics for higher-order 
mutable state~\cite{svendsen2014impredicative,jung2015iris},
concurrency~\cite{brookes2007semantics},
probability~\cite{barthe2019probabilistic,li2023lilac,bao2025bluebell},
entropy~\cite{haselwarter2024tachis},
error~\cite{aguirre2024error},
space~\cite{madiot2022separation,chargueraud2019verifying},
and time~\cite{atkey2011amortised,haselwarter2024tachis}.
As a result, the semantics of separation logic have become fragmented,
which presents problems both for scientific communication and the design of
new logics.

The nascent theory of separation logic has improved the situation 
by identifying and classifying models of BI, the assertion language of
separation logic~\cite{o1999logic,pym2002semantics,pym1999bunched}.
Resources have been organized into resource algebras such as partial
commutative monoids~\cite{calcagno2007local}. Cao, Cuellar, and
Appel~\cite{jungle} showed the soundness of common rules like dereliction 
depends entirely on the algebraic properties of the underlying resource
algebra. Well-behaved classes of duplicable propositions \(P \vdash P \sep P\)
have been identified with persistence modalities \(\pers\), which in turn are
related to the lattice of idempotent elements of the underlying resource
algebra~\cite{bizjak2018models,gratzer2025idempotent}. For higher-order BI, it
has been shown that introducing rules for subset types \(\{x: X \mid P(x)\}\)
leads to the disappointing principle \(P \sep Q \dashv\vdash P \land
Q\)~\cite[\S 2.2]{biering2007bi}, establishing that desirable categorical models of higher-order
BI are are not toposes but rather \emph{BI hyperdoctrines}.

Despite these insights into the logic of resources at rest, separation logics
are still designed and verified by alternately following heuristics and proving soundness by
exhaustive case analysis. This happens because the dynamic aspects of separation
logic have so far eluded systematic analysis.  In a modern logic, dynamic
assertions, i.e., assertions about program
behavior, are defined in terms of a \emph{weakest precondition
modality}: \(
  \{P\}\,c\,\{Q\}
  \iff
  P \vdash \WP\,c\,Q.
  \)
  
The weakest precondition is meant to describe the resources and
knowledge necessary for \(c\) to produce a result satisfying
postcondition \(Q\).
It is often defined by a logical formula
involving the semantics of computations, the effects
of \(c\) on resources, and the invariants the logic is meant to
enforce.
Presentations of these formulas
in the literature
either simply record the definition of weakest precondition
without comment or show how to build it up by proof and
counterexample~\cite{lakatos1976proofs} until all desired rules
are true, as in Jung~et~al.~\cite[\S\S6.3, 7.3]{jung2018iris}. This
reflects the design process, in which a designer writes down a list of
desired rules and then gradually adjusts the defining formula for \(\WP\)
as they rotate between proposing candidate definitions, proving
soundness of rules, and consulting the literature for ideas.
In the end, one obtains a \(\WP\) that validates
at the very least those rules shown in \Cref{fig:wp-rules}, which permit
decomposing monadic computations (\textsc{WP-Return} and
\textsc{WP-Bind}), framing out unused resources (\textsc{WP-Frame}),
and incorporating static reasoning from the assertion language
(\textsc{WP-Mono}).

\begin{figure}
  \begin{align*}
    Q\, v &\vdash \WP\, (\mathsf{ret}\, v)\, Q
    && \text{\textsc{WP-Return}} \\
    \WP\, c\, (\lambda x.\, \WP\, (f\ x) \, Q) &\vdash \WP\, (x \leftarrow c; f\ x)\, Q
    && \text{\textsc{WP-Bind}} \\
    F \sep \WP\, c\, Q &\vdash \WP\, c\, (F \sep Q)
    && \text{\textsc{WP-Frame}} \\
    P \vdash Q\hspace{0.7em} &\hspace{-0.7em}\implies \WP\, c\, P \vdash \WP\, c\, Q
    && \text{\textsc{WP-Mono}}
  \end{align*}
  \caption{Selected proof rules for weakest preconditions.}
  \label{fig:wp-rules}
\end{figure}

This paper presents a systematic, categorical foundation for
weakest preconditions satisfying the rules in \Cref{fig:wp-rules}.
First, we replace the BI hyperdoctrines of Biering~\cite{biering2004logic}
with an equivalent category of \emph{bunched fibrations}.
Then, following the analysis of weakest preconditions
for Hoare logic by Aguirre et al.~\cite{aguirre2022weakest},
we interpret \(\WP\) as a \emph{monad on a bunched fibration}.
We show that any such monad validates the laws \textsc{WP-Mono}, \textsc{WP-Bind},
and \textsc{WP-Ret}, and that
\textsc{WP-Frame} amounts to the existence of a \emph{tensorial strength} for
the monad (\Cref{sec:fibered-wp}).

Our bunched fibrations provide a fibered account of first- and higher-order
BI~(\Cref{sec:posetal-fibrations-for-separation-logic}). We prove that bunched
fibrations correspond to existing models of BI, based on BI
hyperdoctrines~(\Cref{thm:bi-hyp-fib}), and give conditions under which bunched
fibrations are stable under
pullback~(\Cref{prop:fo-bunched-fib-pb,prop:ho-bunched-fib-pb}).

We provide two demonstrations of the utility of our fibered framework
for defining separation logics.
First, \Cref{sec:constructions} shows how it
provides the language to precisely express
the kinds of design patterns used to build complex separation logics.
It considers three examples:
a common pattern of ``baking in'' the frame rule,
a pattern of adding credit-based cost reasoning to a program logic
by upgrading the underlying resource algebra,
and a pattern of defining a persistence modality in terms of universal quantification
over all possible resources \`a la modal necessity.
In two of these three cases we are able to find
an accompanying universal property for the construction,
giving it a canonical status with respect to
other constructions for the same purpose.

Our second demonstration
showcases how our fibered framework
can be used to build a new kind of separation logic.
First, we leverage Simpson's newly-introduced axiomatization of local
independence structure~\cite{simpson2024equivalence}
to define a new monad that captures
an abstract notion of allocation
effect arising in the context of name generation,
local store, probability, and nondeterminism.
Then, we build a bunched fibration and a bunched lifting that
generically construct a higher-order separation logic
for this abstract notion of allocation.
Finally, by using this abstract construction
in combination with our pullback-stability theorems,
we obtain a novel probabilistic separation logic for measure-theoretic probability
that simplifies and clarifies the probabilistic separation logics
put forward by prior work~\cite{li2023lilac,bao2025bluebell,ho2026bayesian}.

\subsubsection*{A Recipe for Making Separation Logics} \label{sec:recipe}
Our hope with this fibered framework is to lay the foundations of
the following three-step recipe for making new separation logics
in a systematic manner.

First, one fixes a monad $T$
  on a category $\cat{C}$ that interprets the effectful operations of interest.
  This could be plucked from the functional programming literature,
  like the state monad or the reader monad,
  or
  constructed from an algebraic signature.
  It can be a sophisticated model, like the model of full ground
  store~\cite{kammar2017monad}, or an elementary one, like an ordinary state monad
  on \(\cat{Set}\).

Next, one builds a bunched fibration $\cat{E} \to \cat{C}$,
  giving a separation logic for reasoning about
  the pure computations embodied by the objects and morphisms of \(\cat{C}\).
  This bunched fibration can be built in a number of ways.
  It could be one of the canonical kinds of bunched fibration, such as the
  versatile \emph{lax slice} fibration described in \Cref{ex:lax-slice};
  it could come from a BI hyperdoctrine, via the correspondence
  to bunched fibrations;
  or it could come from the use of a change-of-base theorem.
  The resulting separation logic can be first- or higher-order depending on
  the structure of the fibration and of the category \(\cat{C}\).

  Finally, one lifts the monad \(T\) and its effectful operations
  to a monad and operations on the total category \(\cat{E}\) 
  of the bunched fibration.
  This too can be done in a number of ways,
  either by using a stock
  lifting  --- e.g., the frame-preserving state example ---
  by the application of general constructions such as free frame, credits, or the independence
  lifting,
  or by change of base along a suitable functor.

\section{Related Work} \label{sec:related-work}

\paragraph{Resource algebras}
Models of BI are typically
based on a \emph{resource algebra}.
This umbrella term refers to algebraic
structures including Kripke resource monoids~\cite{galmiche2005semantics},
separation algebras~\cite{dockins2009fresh},
partial commutative monoids~\cite{jung2015iris},
CMRAs~\cite{jung2018iris}, and
maximal-idempotent resource algebras~\cite{gratzer2025idempotent}.
Resource algebras provide an important class of models of separation logic, as any of them
can be used to quickly build a model of higher-order BI.
However, they are less helpful when it comes to designing \(\WP\).
Our bunched fibrations,
via their equivalence to BI hyperdoctrines,
are general enough to
accommodate all models of BI that come from resource algebras,
as well as models of BI that do not~\cite{polzer2020local};
our fibered approach additionally permits
a mathematical account of \(\WP\)s as bunched monad liftings.

\paragraph{Iris}
The higher-order separation logic Iris has developed a
generic \(\WP\) that can be applied to a wide range of
operational semantics;
this combined with the ability to specify custom resource algebras
makes Iris a powerful framework for building new program logics.
In our view what distinguishes the present work from Iris
is that Iris is focused on specific kinds of languages and logics;
in particular, it is most suitable for step-indexed
operational reasoning that makes use of
features such as
higher-order impredicative invariants and L{\"o}b induction.
By contrast, our fibered framework
aims to describe the general principles shared by all separation logics,
including those that may be first-order, denotational in nature,
and not defined directly in terms of a resource
algebra---indeed, the probabilistic logic presented in \Cref{sec:lip}
provides an example.
We similarly aim to describe constructions for building new separation logics
and \(\WP\)s at this level of generality.

\paragraph{Models of effectful program logics}
The connection between Hoare logic and monads goes all the way back to
Moggi, who noted a correspondence between algebras \(T\Omega \to
\Omega\) and ``Hoare's triples''~\cite{moggi1991notions}.
Proximal to the present work are Dijkstra
monads~\cite{DBLP:conf/pldi/SwamyWSCL13,ahman2017dijkstra}, which
Maillard has formulated as liftings of relative monads~\cite{maillard2019principles},
and the fibered monad liftings of Aguirre et al.~\cite{aguirre2022weakest}.
We take up the latter approach, extending it to separation logic,
and further develop this research program by formalizing useful design
patterns and intuitions from program logics as constructions
on fibrations equipped with monad liftings.

The BI hyperdoctrines of
Biering~et~al.~\cite{biering2004logic,biering2007bi} isolate the
essential features of higher-order BI in a category of
models.  They are the theoretical foundation
of the Iris program logic and find both operational and
denotational applications in 
program logic~\cite{biering2007bi,polzer2020local,aagaard2023denotationally}.
Our 2-category of bunched fibrations is equivalent to the 2-category
of BI hyperdoctrines \(\cat{BIHy}\) by design.
The switch to fibrations significantly simplifies the presentation of
monad liftings, which are achievable in \(\cat{BIHy}\) in principle
but difficult to manipulate in practice.

\paragraph{Probabilistic separation logic}
Separation logic has been adapted to
support probabilistic reasoning, where
separating conjunction enforces
probabilistic independence.
Since PSL~\cite{barthe2019probabilistic},
there have been varied approaches to these probabilistic separation logics.
For instance, the interpretation of separation
is sometimes syntactic, based on separation of variable
names~\cite{barthe2019probabilistic,zilberstein2026probabilistic},
and sometimes semantic, based on independence of \(\sigma\)-algebras
\cite{li2023lilac,bao2025bluebell,ho2026bayesian}.
The semantic interpretation of separation has recently been reformulated
categorically in terms of monoidal categories of sample
spaces~\cite{a-nominal-approach-to-psl,simpson2024equivalence}.
The model we present in \Cref{sec:lip} recovers these models
while extending them with a new interpretation of \(\WP\).
This new \(\WP\) improves on prior work:
whereas definitions of existing \(\WP\)s for PSLs
introduce constraints that force the \(\WP\) to
validate the frame rule
and respect substitution,
these properties hold for our \(\WP\) by construction.

\section{Background}
\label{sec:background}

This section serves as a self-contained presentation of the standard heap
semantics of separation logic and its categorical formulation. First we look at
the assertion language of separation logic (\cref{fig:syntax}).

\begin{figure}
\[
\begin{array}{rcl}
  P,Q & ::= & P \land Q \mid P \lor Q \mid P \To
  Q \mid P \wand Q \mid P \sep Q \mid \top \mid \bot \mid \emp \mid \WP\,c\,Q
  \mid \forall x : A.P \mid \exists x:A.P
\end{array}
\]
  \begin{mathpar}
    \inferrule*{ }{\emp\ \prop}

    \inferrule*{
      \Gamma \vdash P\ \prop \\ 
      \Gamma \vdash Q\ \prop
    }{\Gamma \vdash P \sep Q \ \prop}

    \inferrule*{
      \Gamma \vdash c : TA \\
      \Gamma, x:A \vdash Q\ \prop
    }{
      \Gamma \vdash \WP\,c\,Q\ \prop
    }
  \end{mathpar}
\caption{Syntax and selected formation rules for first-order separation logic.
  Programs $c$ are typed by a judgment $\Gamma \vdash c : A$ as in Moggi's
  monadic metalanguage and use the type constructor $T$ for the
  monad.}\label{fig:syntax}
\end{figure}

Heaps are finite integer-valued partial maps \(H = \Loc\pfin\Z\). Two heaps are
disjoint if they have disjoint domains. The disjoint union operation \(h \uplus
h'\) is the union of the two maps if they are disjoint and undefined otherwise.

Assertions with no free variables are
interpreted in \(\hProp = H \to 2\),
the set of heap propositions, which
is partially ordered by entailment: \[
P \entails Q \iff \forall\,h.\,P(h) \To Q(h)
\]
This partially ordered set supports a collection of algebraic operations that
soundly interpret the propositional connectives in \cref{fig:syntax}.
\begin{align*}
(\top)(h) &\iff \top \\
(\bot)(h) &\iff \bot \\
\hemp(h) &\iff h = \emptyset \\
(P \land Q)(h) &\iff P(h) \land Q(h) \\
(P \lor Q)(h) &\iff P(h) \lor Q(h) \\
(P \To Q)(h) &\iff P(h) \Rightarrow Q(h) \\
(P \sep Q)(h) &\iff \exists\,h_1\,h_2.\,h=h_1\uplus h_2 \land P(h_1) \land Q(h_2) \\
(P \wand Q)(h) &\iff \forall\, h_1\,h_2. h_1 = h\uplus h_2 \land P(h_2) \Rightarrow Q(h_1)
\end{align*}
Assertions with free variables drawn from
\(\Gamma = x_1 : X_1, \dots, x_n : X_n\),
are interpreted as functions \(X_1 \times \dots \times X_n \to \hProp\).
In ordinary separation logic, the syntax also includes
all the points-to predicates \(\ell \mapsto v\),
which are interpreted by 
\begin{align*}
  (\ell \mapsto v)(h) \iff \dom(h) = \{\ell\} \land h(\ell) = v.
\end{align*}

In order to turn this assertion logic into a program logic,
we will define a weakest precondition for
computations in the state monad
\[
  \St_H A = H \To A \times H.
\]
Then we can define a weakest precondition $\WP\,c\,Q$ for every 
computation \(c \in \St_H A\) and predicate \(Q : A \to \hProp\)
as follows.
\begin{align*}
  (\WP\,c\,Q)(h)
  \iff
    \begin{aligned}
      &\text{for all \(h_f\) disjoint from \(h\)}\\
      &\text{there exists \(h'\) disjoint from \(h_f\)}\\
      &\text{with \(c(h\uplus h_f) = (a, h' \uplus h_f)\) and \(Q(a)(h')\)}
    \end{aligned}
\end{align*}
Note the use of a framing heap \(h_f\) in the definition.
Without this additional heap, the frame rule would not hold.

\paragraph{Modeling Separation Logic Categorically}

The logical operations on \(\hProp\) make it a \emph{BI algebra},
which is both a Heyting algebra and a symmetric monoidal closed poset.
This makes it a model of the intuitionistic connectives \((\top, \bot,
\land, \lor, \To)\) and the substructural connectives \((\emp, \sep, \wand)\),
respectively.
Additionally, \(\hProp\) is a \emph{complete BI algebra}: it has all set-indexed
meets and joins.
The set \(\hPred(X) = X \to \hProp\)
inherits the complete BI algebra structure from \(\hProp\)
by applying operations pointwise.
Furthermore, for each function \(s : Y \to X\),
one has a ``substitution'' map \(\hPred(X) \to \hPred(Y)\)
defined by precomposition. This forms a homomorphism of BI algebras
since the structure on \(\hPred\) is defined pointwise.
All together this means \(\hPred\) defines a contravariant functor \(\Set\op
\to \cat{BIAlg}\), where \(\cat{BIAlg}\) is the category of BI algebras and
their homomorphisms.

Categorical models of BI codify this situation
with the notion of
\emph{BI hyperdoctrine},
which encompasses all functors of the form \(\cat{C}\op \to \cat{BIAlg}\)
that additionally support quantifiers and equality.
As is typical in categorical logic, these are given by adjoints:
quantifiers are left and right adjoint to substitution along
the weakening map \(\pi_2 : A \times B \to A\),
and equality is left adjoint to substitution along
the duplication map \(\delta : A \times B \to A \times B \times B\).
A more detailed treatment of BI hyperdoctrines may be found in
Biering~et~al.~{\cite[Section 2.2]{biering2007bi}}.

A class of examples of BI hyperdoctrines that will be important to us later on
comes from the notion of an \emph{ordered pcm}---a resource algebra
of the kind described in \Cref{sec:related-work}.

\begin{definition}[ordered pcm] \label{def:ordered-pcm}
  An ordered pcm is a partial commutative monoid (pcm) $(M, \comp, \emp)$
  additionally equipped with a partial order \(\le\)
  such that the following monotonicity condition holds:
  if $x'\comp y'$ is defined
  and $x \leq x'$ and $y \leq y'$,
  then $x \comp y $ is defined
  as well
  and \(x\comp y \le x' \comp y'\).
\end{definition}

\begin{proposition}
  Given an ordered pcm $(M, \comp, \emp)$,
  the collection \(\wp^\uparrow(M)\)
  of upwards-closed subsets of \(M\)
  forms a complete BI algebra.
\end{proposition}
The Heyting algebra structure is given by the usual Kripke semantics
for intuitionistic propositional logic,
and the monoidal closed structure is given by
  \begin{align*}
    (P \sep Q)(m) &= \exists\,m_1\,m_2.\,
      (m_1 \comp m_2 \le m) \land P\,  m_1 \land Q\, m_2\\
    (P \sepimp Q)(m) &= \forall\,m'.\, (m \comp m' \downarrow) \land P\, m' \Rightarrow Q \,(m \comp m')\\
    \hemp (m) &= \emp \le m
  \end{align*}

  \begin{proposition}
    \label{prop:pcm-hyperdoctrine}
  Every ordered pcm gives rise to
  a BI hyperdoctrine \(\hPred : \Set\op \to \cat{BIAlg}\),
  defined by \(\hPred(X) = X \to \wp^\uparrow(M)\).
\end{proposition}

This recovers the standard linear heap model:
the set $L \rightharpoonup \Z$ of heaps forms an ordered pcm
under the discrete ordering,
the unit is the empty heap,
and the partial multiplication is given by
disjoint union of heaps.
Ordering heaps by inclusion works too and yields an affine separation logic.
A generalization appears in \Cref{sec:lip}, where it is necessary to define a \emph{pcm object}
in an arbitrary topos $\topos$ in place of \(\cat{Set}\). This too gives rise to
a higher-order BI hyperdoctrine, by taking the object \(\hProp\)
to be the collection of internally upwards-closed subsets of \(M\).

\section{Fibered Semantics of BI}
\label{sec:general}

In this section we pass from indexed models to fibrations,
replacing hyperdoctrines \(H: \cat{B}\op \to \cat{BIAlg}\) with a
category \(\cat{E}\) containing the entire model equipped with a
projection functor \(p: \cat{E} \to \cat{B}\).  We define a
category of bunched fibrations \(\cat{BIFib}\) and prove it is
equivalent to the category of BI hyperdoctrines \(\cat{BIHy}\).

We choose fibrations over hyperdoctrines for reasons of expediency.
BI hyperdoctrines \(F: \cat{B}\op \to \cat{BIAlg}\) are a convenient
format for models, but they achieve this by organizing predicates and
entailments into individual algebras \(F(A)\) that are functorial in
\(A\).  This makes it easy to extend an operations on algebras to an
operation on the entire hyperdoctrine.  For operations that
work across indices, such as quantification, this becomes awkward; for
constructions that involve additional data it becomes unmanageable.

Fibrations collate the algebras \(F(A)\) of a hyperdoctrine into a
single category \(\cat{E}\) that puts substitutions \(f: A \to B\) on
equal footing with entailments \(P \leq Q\).  There is a standard
construction for passing from any functor valued in posets to a
fibration.

\begin{definition}\label{def:total-cat}
  For a functor \(H: \cat{B}\op \to \cat{Poset}\) we define the \emph{total category} \(\int H\).
  \begin{itemize}
  \item The objects are pairs \((A, P)\) with \(A \in \ob{\cat{B}}\) and \(P \in H(A)\).
  \item The morphisms \((A, P) \to (B, Q)\) are morphisms \(f \in
    \cat{B}(A, B)\) such that \(P \leq_{A} H(f)(Q)\).
  \end{itemize}
\end{definition}

\begin{example}\label{ex:std-heap-fibration}
  Recall the BI hyperdoctrine \(\hPred\) of heap predicates discussed in
  \Cref{sec:background}. Its total category \(\int \hPred\) has objects \((A,
  P)\) where \(A\) is a set and \(P : A \to H \to 2\) is a heap predicate.
  Morphisms \((A, P) \to (B, Q)\) are functions \(u : A \to B\) such that
  \(P(a) \vDash Q(u(a))\).
\end{example}

\begin{proposition}\label{prop:groth-faithful}
  The projection \(\pi : \int H \to \cat{B}\) out of the total
  category is a faithful functor.
\end{proposition}

Constructions on this category are in general one level of abstraction
down from the corresponding construction on \(F\), which reduces the
bureaucratic overhead involved in proving anything about them. For
instance, morphisms of hyperdoctrines on the same base are natural
transformations (a 2-cell) \(\alpha: F \To G\), while morphisms of
fibrations on the same base are just a functor (a 1-cell).
Despite this, passing to the total category does not lose
information. To see why, we need to properly introduce fibrations.

\paragraph{Notational conventions}
We think of the total category of a fibration as being ``upstairs'' of
the base category, which is ``downstairs,'' and draw diagrams
accordingly. Variables like \(X\) and \(Y\) represent objects of the
total category and will lie over objects of the base category denoted
by variables like \(A\) and \(B\).
Our treatment of fibrations mostly follows
Jacobs~\cite{jacobs1999categorical}, but we caution the reader that
the theory of posetal fibrations is much simpler than the general
picture developed in that book.

\begin{definition}[{\cite[Definition 1.1.3]{jacobs1999categorical}}]
  Let \(p: \cat{E} \to \cat{B}\) be a functor.
  \begin{enumerate}
    \item A morphism \(f : X \to Y\) in \(\cat{E}\) is \emph{cartesian}
  if for any \(h : Z \to Y\) and \(u : pZ \to pX\) with \(pf \circ u =
ph\) there exists a unique \(g: Z\to X\) with \(pg = u\) making \(fg =
h\).
\[\begin{tikzcd}
	Z & X & Y \\
	pZ & pX & pY
	\arrow["{\exists ! g}"', dashed, from=1-1, to=1-2]
	\arrow["h", curve={height=-12pt}, from=1-1, to=1-3]
	\arrow["f"', from=1-2, to=1-3]
	\arrow["u"', from=2-1, to=2-2]
	\arrow["ph", curve={height=-12pt}, from=2-1, to=2-3]
	\arrow["pf"', from=2-2, to=2-3]
\end{tikzcd}\]
\item A morphism \(f\) of \(\cat{E}\) is \emph{vertical} if \(pf = \id{}\).
\item The functor \(p\) is a \emph{fibration} if for any morphism
  $u : A \to pY $ in $\cat{B}$, there exists a cartesian morphism \(f : X \to Y\)
  such that \(pX = A\) and \(pf = u\).
  This is called a \emph{cartesian lift} of \(f\).
  \end{enumerate}
\end{definition}
Cartesian lifts are unique up to unique isomorphism, but are not
strictly unique in general.  A fibration is \emph{cloven} if it is
equipped with a \emph{choice of lifts}, which assigns to every object
\(Y\) and morphism \(u: A \to pY\) a cartesian morphism \(f: X \to Y\)
with \(pX = A\) with \(pf = u\).
For each object \(A\) of \(\cat{B}\) there is a category \(\cat{E}_A =
p^{-1}(A)\) called the \emph{fiber over \(A\)}.  Its objects are all
the \(X \in \ob \cat{E}\) such that \(pX = A\), and the morphisms \(u:
X \to Y\) are all the morphisms \(u \in \cat{E}(X, Y)\) such that \(pu
= \id{u}\). A fibration is said to be \emph{posetal} when all of its fiber
categories are posets. The following characterization makes working with posetal
fibrations convenient.
\begin{proposition}[{\cite[Theorem 9.1(a)]{streicher2018fibered}}]
  \(p\) is posetal if and only if \(p\) is faithful.
\end{proposition}
This applies in particular to \(\int H\) (\Cref{prop:groth-faithful}).
In \(\int H\), if a morphism \(f : (A, P) \to (B, Q)\) is vertical,
then \(A = B\) and \(P \leq H(\id{A})(Q) = Q\). So the fiber over \(A\)
of \(\int H\) is isomorphic to the poset \(H(A)\). Next we
characterize the cartesian morphisms.
\begin{proposition}
  If a morphism \(f: (A, P) \to (B, Q)\) of \(\int H\) is cartesian
  over \(u: A \to B\), then \(P = H(u)(Q)\).
\end{proposition}
\begin{proof}
  We already know \(P \leq H(u)(Q)\). Consider the
  morphism \(g : (A, H(u)(Q)) \to (B, Q)\) lying over \(u\).
  There is a commuting triangle \(u = u\circ \id{A}\) in \(\cat{B}\),
  so we apply the universal property of \(f\) to get a vertical
  morphism \((A, H(u)(Q)) \to (A, P)\), i.e., a proof that \(H(u)(Q)
  \leq P\).
\end{proof}

\begin{definition}
  For each \(u : A \to B\) in the base of a cloven fibration \(p\),
  there is a \emph{reindexing functor} \(u\pb : \cat{E}_{B} \to \cat{E}_{A}\)
  defined on objects \(X\) as the chosen lift of \(u\) against \(X\).
  The action on morphisms is given by the universal property of the lift.
\end{definition}

\begin{proposition}\label{prop:posetal-facts}
  Suppose \(p: \cat{E}\to\cat{B}\) is posetal.
  \begin{enumerate}
  \item \(p\) has a unique choice of lifts.
  \item \(p\) is \emph{split}: we have \(\id{A}\pb  = \id{\cat{E}_A}\) and \(v\pb \circ u \pb = (uv) \pb\) for all \(A\) and all composable \(u, v\) in \(\cat{B}\).
  \item Every vertical morphism \(f\) is both epi and mono in \(\cat{E}\).
  \end{enumerate}
\end{proposition}

We now have enough to recover the action of \(H\) on morphisms: it is
the contravariant reindexing functor \(\blank \pb\) sending morphisms
of \(\cat{B}\) to monotone functions between the fibers. Altogether we
have proved the following, which is a special case of the more general
theory of the Grothendieck construction. See Jacobs~\cite[Theorem
1.10.7]{jacobs1999categorical} or Johnstone~\cite[\S B1.3]{elephant1}.

\begin{proposition}\label{prop:groth-facts}
  For all functors \(H : \cat{B}\op \to \cat{Poset}\),
  the total category \(\pi : \int H \to \cat{B}\) is a
  posetal fibration with the following properties.
  \begin{enumerate}
  \item \((\int H)_A \iso H(A)\)
  \item \(u: (A, P) \to (A, Q)\) is vertical if and only if \(P \leq_A Q\)
  \item \(u: (A, P) \to (B, Q)\) is cartesian if and only if \(P = H(u)(Q)\)
  \item \(u\pb (B, Q) = (A, H(u)(Q))\)
  \end{enumerate}
\end{proposition}
Fibrations generalize categories and as such come with their own
notions of functor and natural transformation. These assemble into a
2-category \(\cat{Fib}\) with a subcategory
\(\cat{Fib}_{\mathsf{pos}}\) of posetal fibrations.
\begin{definition}
  Given fibrations \(p: \cat{E} \to \cat{B}\) and \(q: \cat{D} \to \cat{A}\),
  a \emph{fibered functor} from \(p\) to \(q\) is a commutative square
  \[\begin{tikzcd}
                {\cat{E}} & {\cat{D}} \\
                {\cat{B}} & {\cat{A}}
                \arrow["G", from=1-1, to=1-2]
                \arrow["p", from=1-1, to=2-1]
                \arrow["q", from=1-2, to=2-2]
                \arrow["F", from=2-1, to=2-2]
  \end{tikzcd}\]
  in which \(G\) preserves cartesian morphisms. Inline, we write \((F,
  G) : p \to q\) or just \(G\), \emph{pars pro toto}.
\end{definition}

\begin{definition}
  Let \((F, G)\) and \((H, K)\) be fibered functors
  from \(p\) to \(q\). A fibered natural transformation from \((F,
  G)\) to \((H, K)\) is a pair of natural transformations
  \(\alpha: F \To H\) and \(\gamma: G \To K\)
  such that \(p \gamma = \alpha p\). It is \emph{vertical} if its
  components are vertical.
\end{definition}

\subsection{Modeling BI}
\label{sec:posetal-fibrations-for-separation-logic}

In this section we introduce 2-categories of \emph{bunched
fibrations} that are equivalent to the 2-category of BI hyperdoctrines
via the total category construction. We briefly review posetal
fibrations for ordinary first-order and higher-order logic, which are
known to be equivalent to ordinary first-order and higher-order
hyperdoctrines.
\begin{definition} \label{def:fo-bunched-fib}
  A posetal fibration \(p: \cat{E} \to \cat{B}\) is \emph{first-order} if
  \begin{enumerate}
  \item The base category \(\cat{B}\) has finite products.
  \item Each fiber is a Heyting algebra and each
    \(u\pb\) is a homomorphism of Heyting algebras.
  \item (Simple products) Reindexing \(\pi\pb\) along a projection \(\pi: A \times B \to A\)
    has a right adjoint \(\prod_{A,B}\) such that for all \(u\),
          \[u\pb \prod_{B,C} = \prod_{A,C} (u \times \id{C})\pb\]
  \item (Simple coproducts) Reindexing along a projection \(\pi: A \times B \to A\)
    has a left adjoint \(\coprod_{A,B}\) such that for all \(u\),
          \[u\pb \coprod_{B,C} = \coprod_{A,C} (u \times \id{C})\pb\]
  \item (Equality) Reindexing \(\delta\pb\) along a duplication \(\delta : A \times B \to A
        \times B \times B\) has a left adjoint \(\Eq_{A,B}\) such that for all
        \(u\),
        \[((u \times \id{C}) \times \id{C}) \pb \Eq_{B,C} = \Eq_{A, C} (u \times \id{C})\pb\]
  \end{enumerate}
\end{definition}
A first-order fibration is \emph{higher-order} if it has a
\emph{generic object}: an object \(\gentrue\) lying over over an
object \(\Omega\) such that for any \(X \in \cat{E}\) there is a
unique characteristic morphism \(u: pX \to \Omega\) with \(u\pb
\gentrue = X\).

We now define bunched fibrations and their morphisms, which are
certain fibered functors.  Taking 2-cells to be fibered natural
transformations between morphisms, they assemble into the 2-categories
\(\cat{BIFib}\) and \(\cat{HOBIFib}\).

\begin{definition}
  A \emph{bunched fibration} is a first-order posetal fibration
  \(p: \cat{E} \to \cat{B}\)
  equipped with an additional symmetric monoidal closed structure
  on each fiber that is preserved by reindexing.
  A \emph{morphism of bunched fibrations} \(p\) and \(q\) is a fibered
  functor \((F, G) : p \to q\) such that
  \(F\) preserves finite products,
  \(G\) strictly preserves bicartesian closed and symmetric monoidal closed structure on fibers,
  and 
  \(F\) preserves \(\prod\), \(\coprod\), and
\(\Eq\)~\cite[Defs. 1.9.13, 3.4.7]{jacobs1999categorical}.
\end{definition}

\begin{definition}\label{ho-fib}
A first-order bunched fibration \(p: \cat{E} \to \cat{B}\) is
\emph{higher-order} if it has a generic object and \(\cat{B}\) is
cartesian closed.
A first-order bunched fibration morphism \((F, G): p \to q\) is a
\emph{morphism of higher-order bunched fibrations} if the
canonical map \(G\gentrue_p \to \gentrue_q\) is an isomorphism.
\end{definition}

The equivalence with BI hyperdoctrines is a straightforward
consequence of \Cref{prop:groth-facts} and the more general theory of
the Grothendieck construction.
\begin{theorem} \label{thm:bi-hyp-fib}
  The total category construction defines equivalences
  of 2-categories \(\cat{BIHy} \simeq \cat{BIFib}\) and \(\cat{HOBIHy}
  \simeq \cat{HOBIFib}\).
\end{theorem}

This gives rise to an interpretation of BI (i.e., the syntax in
\cref{fig:syntax} aside from the weakest precondition) in any bunched fibration. Types and contexts are
interpreted by objects of $\cat{B}$ and propositions $\Gamma \vdash\ \prop$ are
interpreted by objects of $\cat{E}$ over $\cat{B}$. We give a few representative
clauses for propositions under a context $\Gamma$.
\begin{align*}
  \llbracket P \sep Q \rrbracket &= \llbracket P \rrbracket
  \otimes_{\llbracket \Gamma \rrbracket} \llbracket
  Q \rrbracket \\
  \llbracket \emp \rrbracket &= I_{\llbracket \Gamma \rrbracket} \\
  \llbracket \forall x: A. P \rrbracket &= \prod_{\llbracket \Gamma \rrbracket,
  \llbracket A \rrbracket} \llbracket P \rrbracket
\end{align*}

The result above (\Cref{thm:bi-hyp-fib}) follows easily because the definition of bunched
structure on a fibration is done fiberwise. However, in the sequel
we will frequently work with monoidal and monoidal closed structures
on all of \(\cat{E}\).
\begin{definition}
  A posetal fibration \(p: \cat{E} \to \cat{B}\) is \emph{(externally)
  symmetric monoidal} if \(\cat{E}\) is symmetric monoidal, \(\cat{B}\) is
  cartesian, and \(p\) is a strict symmetric monoidal
  functor. Likewise for
  \emph{(externally) symmetric monoidal closed}.
\end{definition}

External and fiberwise monoidal structure are tightly related provided
one has a cartesian or cartesian closed structure on the base
category.  For instance, the following is a consequence of Theorem
12.8 in Shulman's magisterial study of monoidal
fibrations~\cite[Theorem 12.8]{shulman2008framed}.
\begin{proposition}\label{prop:external-tensor}
 Every first-order bunched fibration is externally symmetric monoidal with tensor product \[
   X \otimes Y = \pi_1\pb X \otimes_{A\times B} \pi_2\pb Y
 \]
\end{proposition}
The same correspondence does not go through for the monoidal closed
structure for the simple reason that first-order BI hyperdoctrines do
not assume exponentials on their base and our definition aligns with
that one.  However, assuming \(\cat{B}\) closed suffices. This of
course applies to higher-order bunched fibrations.
\begin{proposition}
  If \(p: \cat{E}\to \cat{B}\) is a first-order bunched fibration and
  \(\cat{B}\) is cartesian closed then \(p\) is externally monoidal closed.
\end{proposition}
\begin{corollary}
  Every higher-order bunched fibration is externally monoidal closed.
\end{corollary}

\subsubsection{Base change for bunched fibrations}

\begin{definition}
  \label{def:fib-pb}
  Given a fibration \(p: \cat{E} \to \cat{B}\)
  and a functor \(F: \cat{A} \to \cat{B}\), the change-of-base fibration
  is the pullback in \(\cat{Cat}\) of \(p\) against \(F\). Explicitly, this is
  \begin{enumerate}
  \item A category \(F\pb \cat{E}\) with objects \((A, X \in \cat{E}_{FA})\)
    and morphisms \((u, f)\) with \(pf = Fu\).
  \item A fibration \(F\pb p : F\pb \cat{E} \to \cat{B}\) sending
    \((A, X)\) to \(A\) and with cartesian lifts of \(f: A \to
    B\) given by the cartesian lifts of \(Ff\) in \(p\).
  \item A projection functor \(F\pb\cat{E} \to \cat{E}\) sending \((A, X)\) to \((FA, X)\).
  \end{enumerate}
  The projection is a fibered functor \(F\pb p \to p\).
\end{definition}
We note in passing that change of base along \(F\) for a total
category fibration \(\int H\) is equivalently the total category of
the precomposition \(\int H \circ F\).

\begin{proposition}\label{prop:fo-bunched-fib-pb}
  If \(p: \cat{E} \to \cat{B}\) is a first-order bunched fibration and
  \(F: \cat{A} \to \cat{B}\) preserves products,
  then \(F\pb p\) is a first-order bunched fibration and the fibered
  functor \(F\pb p \to p\) is a morphism of first-order bunched fibrations.
\end{proposition}
\begin{proof}
  Fiberwise structure exists and is preserved by the projection \(F\pb
  p \to p\) because \(F\pb\cat{E}_A \iso \cat{E}_{FA}\).
  For quantifiers and equality, we need only observe that \(F\delta_A =
  \delta_{FA}\) and \(F\pi_{A,B} = \pi_{FA,FB}\).
\end{proof}

\begin{proposition}\label{prop:ho-bunched-fib-pb}
  If \(p: \cat{E} \to \cat{B}\) is a higher-order bunched fibration,
  \(\cat{A}\) is cartesian closed,
  and \(F : \cat{A} \to \cat{B}\) is a cartesian closed functor with a right adjoint,
  then \(F\pb p\) is a higher-order bunched fibration.
  If \(\epsilon\pb \gentrue = \gentrue\), then \(F\pb
  p \to p\) is a morphism of higher-order bunched fibrations.
\end{proposition}
\begin{proof}
  The first-order structure is handled by \Cref{prop:fo-bunched-fib-pb}.
  Let \(G : \cat{B} \to \cat{A}\) be the right adjoint to \(F\).
  Set \(\Omega' = (G\Omega, \epsilon_\Omega\pb \gentrue)\).
  Given \(X\) over \(FA\), we already have \(u : FA \to \Omega\) with \(u\pb \gentrue = X\).
  Let \(v : A \to G\Omega\) be the transpose of \(u\). We calculate \(
    v\pb \Omega' = (A, (Fv)\pb\epsilon_\Omega\pb \gentrue) = (A, u\pb\gentrue) = (A, X)\)
\end{proof}

\begin{remark}[Models of Iris]
  In a paper on BI hyperdoctrines, Bizjak and Birkedal temporarily
  pass to fibrations in order to obtain
  the model of Iris as a pullback along the inclusion
  of total presheaves (COFEs) into \(\cat{Psh}(\omega)\)~\cite[Theorem
  6.8]{bizjak2018models}.  An analysis of the adjunction between
  COFEs~\cite{di2004unifying} and the topos of trees
  \(\cat{PSh}(\omega)\)~\cite{DBLP:conf/lics/BirkedalMSS11}
  shows the hypotheses of \Cref{prop:ho-bunched-fib-pb} are satisfied,
  establishing that their pullback square is a morphism of
  higher-order bunched fibrations.
\end{remark}

We now present the lax slice, a generalization of
\Cref{prop:pcm-hyperdoctrine} to internal BI algebras.
Lax slices over resource algebras \(R\) shall be taken
over the canonical BI algebra \(\upclosed(R)\).
A similar construction appears in
Aguirre~et~al.~\cite{aguirre2022weakest}.
\begin{definition}[lax slice]
  \label{ex:lax-slice}
  Let \(U\) be a BI algebra internal to a topos \(\topos\).
  Set \(f \leq g\) if \(\langle f, g\rangle\) factors
  through the order relation \(({\leq}) \in \mathrm{Sub}(U \times U)\).
  The lax slice \(\laxslice S U\) is the category with objects \((A, p : A \to U)\)
  and morphisms laxly commuting triangles
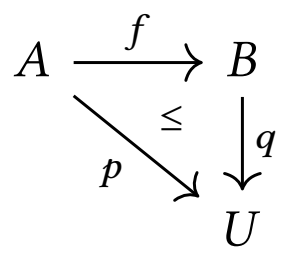
\[\begin{tikzcd}
	A & B \\
	& U
	\arrow["f", from=1-1, to=1-2]
	\arrow[""{name=0, anchor=center, inner sep=0}, "p"', from=1-1, to=2-2]
	\arrow["q", from=1-2, to=2-2]
	\arrow["\leq"{description}, draw=none, from=0, to=1-2]
\end{tikzcd}\]
\end{definition}

\section{Fibered Weakest Precondition} \label{sec:fibered-wp}

This section develops the necessary technical machinery
for monad liftings and their relationship to weakest precondition
modalities.

\subsection{Monads and Adjunctions}
We make use of the formal theory of monads in a 2-category. This
generalization of monads and adjunctions is due to
Street~\cite{street1972formal} but for an accessible tutorial consult
Hinze and Marsden~\cite{hinze2025graphical}.
While this section contains no new results, it will
be heavily used throughout this paper.

\begin{definition}
  A \emph{monad internal to a 2-category} \(\cat{K}\) consists of an
  object \(C\) and a 1-cell \(T : C \to C\) equipped with a unit \(\eta : T \To
  \id{C}\), and multiplication \(\mu: T \circ T \to T\) satisfying the
  usual monad laws.

\end{definition}
\begin{definition}\label{def:monad-morphism}
  Given monads \((S : C \to C, \eta^S, \mu^S)\) and \((T : D \to D, \eta^T, \mu^T)\),
  a \emph{monad morphism} from \(S\) to \(T\)
  is a 1-cell \(F: C \to D\) and 2-cell \(\lambda: FS \To TF\)
  satisfying the following unit and multiplication laws.
\[\begin{tikzcd}
	F & TF & FSS & TFS & FTT \\
	FS & & FS & & TF
	\arrow["{\eta^TF}", rightarrow, from=1-1, to=1-2]
	\arrow["{F\eta^S}"', rightarrow, from=1-1, to=2-1]
	\arrow["{\lambda S}", rightarrow, from=1-3, to=1-4]
	\arrow["{F\mu^S}"', rightarrow, from=1-3, to=2-3]
	\arrow["{T\lambda}", rightarrow, from=1-4, to=1-5]
	\arrow["{F\mu^T}", rightarrow, from=1-5, to=2-5]
	\arrow["\lambda", rightarrow, from=2-3, to=2-5]
	\arrow["\lambda", rightarrow, from=2-1, to=1-2]
\end{tikzcd}\]
\end{definition}

\begin{definition}
  A \emph{adjunction in a 2-category} \(\cat{C}\) consists of two objects \(C\) and \(D\)
  with a left adjoint \(L: C \to D\), a right adjoint \(R: D \to C\), a unit \(\eta: \id{C} \To R \circ L\),
  and a counit \(\epsilon: L \circ R \to \id{D}\) such that \(\epsilon L \circ L \eta = \id{L}\) and
  \(R\epsilon \circ \eta R = \id{R}\). We abbreviate this as \(L
  \dashv R : C \to D\) following the direction of the left adjoint.
\end{definition}
\begin{definition}
  A \emph{morphism of adjunctions} 
  from \(L \dashv R : C \to D\) to \(L' \dashv R' : C' \to D'\)
  is a pair \((F: C \to C', G: D \to D')\) such that \(GL = L'F\), \(GR = R'F\),
  \(G\epsilon_X = \epsilon'_{GX}\), and \(F\eta_X' = \eta_{GX}\).
\end{definition}

For \(\cat{K} = \cat{Cat}\), the internal definitions recover the
usual definitions of monad, monad morphism, and adjunction.

We briefly record some facts about the Huber envelope, which is the
construction that ``sandwiches'' a monad in an adjunction.

\begin{definition}\label{def:huber-env}
  The Huber envelope of a monad \(T : D \to D\) within an adjunction
  \(L \dashv R: C \to D\) is the monad \(RTL : C \to C\) with unit \(\eta = R
  \eta^T_L \circ \eta^{RL}\) and multiplication \(\mu = R \mu^T_L \circ
  RT\epsilon^{LR}TL\). This assignment is functorial in \(T\) and \(L
  \dashv R\).
\end{definition}

\begin{proposition}\label{prop:huber-morphism}
  Let \(L \dashv R : C \to D\) be an adjunction with counit \(\epsilon: LR \To \id{}\)
  and let \(T: D \to D\) be a monad.
  There is a monad morphism \(RTL \to T\) given by \((L, \epsilon TL)\).
\end{proposition}

We briefly recall the definition of monoidal strength due to
Kock~\cite{kock1972strong} and introduced to the semantics of effects
by Moggi~\cite{moggi1991notions}.
\begin{definition}\label{def:strong-monad}
  A \emph{monoidal strength} for a monad \(T\) on a symmetric monoidal
  category \(\cat{C}\) is a natural transformation \(\monadstr_{X,Y} :
  TX \mtimes Y \to T(X\mtimes Y)\) satisfying well-known coherence
  conditions. A \emph{strong monad} is a monad equipped with a strength,
  and a \emph{morphism of strong monads} is a monad morphism
  \((F, \lambda) : S \to T\) such that \(F\) is strong monoidal with
  natural isomorphism $\gamma_{A,B} : F(A\mtimes B) \to FA \mtimes FB$
  and the following diagram commutes.
\[\begin{tikzcd}
	{F(SA \otimes B)} & {FSA \otimes FB} & {TFA\otimes FB} \\
	{FS(A \otimes B)} & {TF(A\otimes B)} & {T(FA\otimes FB)}
	\arrow["\gamma", from=1-1, to=1-2]
	\arrow["{F(\sigma^S\otimes B)}"', from=1-1, to=2-1]
	\arrow["{\lambda \otimes FB}", from=1-2, to=1-3]
	\arrow["{\sigma^T}", from=1-3, to=2-3]
	\arrow["\lambda", from=2-1, to=2-2]
	\arrow["{T\gamma}", from=2-2, to=2-3]
\end{tikzcd}\]
\end{definition}

\subsection{Strong Monad Liftings for Weakest Preconditions}
\begin{definition}
  A \emph{monad lifting} on a fibration \(p: \cat{E} \to \cat{B}\) is
  a fibered functor \((T, W)\) where \(T\) is a monad on \(\cat{B}\),
  \(W\) is a monad on \(\cat{E}\), and \(p\) is a strict monad morphism.
\end{definition}

Monad liftings are the monads in the 2-category \(\cat{Fib}\) of
fibrations, fibered functors, and fibered natural transformations.
With this in mind, we take a morphism of monad liftings to be a monad
morphism in \(\cat{Fib}\).  These definitions accord with the theory
of monad liftings developed in
Aguirre~et~al.~\cite{aguirre2022weakest}.

The monad liftings of \emph{loc. cit.} validate a conjunctive frame rule
if they are equipped with a fibered cartesian strength. To model
the tensorial frame rule of separation logic, we shall require a
fibered \emph{monoidal} strength. Recall (\Cref{prop:external-tensor})
that although bunched fibrations are defined in terms of fiberwise
tensors, they have a canonical external monoidal structure
\(X \mtimes Y = \pi_1 \pb X \mtimes_{pX\times pY} \pi_2 \pb Y\).
\begin{definition}
  Let $p : \cat{E} \to \cat{B}$ be a bunched fibration.
  A \emph{bunched monad lifting} is a monad lifting \((T, W)\) on \(p\)
  such that \(T\) is strong, \(W\) is strong with respect to the
  external tensor on \(\cat{E}\), and \(p\) is a strict morphism of
  strong monads.  A \emph{morphism of bunched liftings} is a morphism
  of monad liftings whose upper and lower components are each
  morphisms of strong monads.
\end{definition}

We now have enough semantic scaffolding to justify useful proof rules. The
strong monad lifting implies the soundness of the rules \textsc{WP-Return},
\textsc{WP-Bind}, \textsc{WP-Frame} and \textsc{WP-Mono} from
\Cref{fig:wp-rules}.

The next few propositions follow from posetal fibrations being
faithful functors.
\begin{proposition}[{\cite[Lemma 12]{aguirre2022weakest}}]
  \label{prop:strength-leq}
  Let \((T, W)\) be a monad lifting on a bunched fibration.
  Strengths for \((T, W)\) correspond bijectively to
  strengths \(\monadstr\) for \(T\) such that
  \(
    W X \otimes F \leq_{TA\times B} \monadstr \pb W(X \otimes F)
  \)
  for all \(X\) over \(A\) and \(F\) over \(B\).
\end{proposition}
\begin{proposition}\label{prop:posetal-monad-morphism}
  Let \((S, V)\) and \((T, W)\) be monad liftings on \(p: \cat{E} \to
  \cat{B}\) with a monad (resp. strong monad) morphism \((F,
  \lambda)\) from \(S\) to \(T\).  If \(G\) is fibered over \(F\) and
  \(\gamma\) is fibered over \(\lambda\), then it is automatically a
  monad (resp. strong monad) morphism and \(((F, \lambda), (G,
  \gamma))\) is a morphism of monad liftings (resp. bunched monad liftings).
\end{proposition}
\begin{corollary}\label{prop:vertical-monad-morphism}
  If \(V\) and \(W\) are monad liftings (resp. bunched monad
  liftings) of the same monad \(T\), any vertical natural
  transformation \(\lambda : V \To W\) defines a monomorphism of monad
  liftings (resp. bunched monad liftings) \((T, V) \to (T, W)\).
\end{corollary}

\begin{theorem}
  \label{thm:bunched-monad-lifting-pb}
  Bunched monad liftings on first-order bunched fibrations are stable under pullback
  along cartesian functors.
\end{theorem}
\begin{proof}
  Apply \citet[Theorem VIII.1]{azevedodeamorim2025logical}.
  To do this, we need a 2-category whose internal monads
  are strong monads.
  Such a 2-category was defined in Example VIII.3 of
  \emph{op. cit.}
  for the cartesian case, which readily generalizes to the monoidal case.
  Furthermore, we need this 2-category to have enough pullbacks;
  this follows from
  \Cref{prop:fo-bunched-fib-pb}.
\end{proof}
Katsumata~\cite[Theorem 4.7]{katsumata2005semantic} offers a more concrete presentation of a slightly
simpler version of this construction.
We note also
that a similar theorem can be proved for higher-order fibrations,
for pullback along functors satisfying
the hypotheses of
\Cref{prop:ho-bunched-fib-pb}.

\begin{example}\label{ex:reynolds-wp}
  A monad on the base category of a bunched fibration may have several
  distinct liftings, as we now show for the standard heap model
  (\Cref{sec:background}) and its corresponding
  total category fibration \(p : \int \hPred \to \cat{Set}\)
  (\Cref{ex:std-heap-fibration}).
  The monad \(\St_H A = H \To A \times H\) on the base is an ordinary state
  monad in which all computations are total.
  A lifting has to be a monad on \(\int \hPred\), and in particular
  must be a functor sending \((A, Q) \mapsto (\St_H A, \WP Q)\) where \(\WP
  Q : \St_H A \to H \to 2\) is a heap predicate on computations that is monotone
  in \(Q\).
  \begin{enumerate}
    \item The \emph{evaluation lifting} is
  determined by setting \[
    (\WP\ Q)(c)(h) = \forall\, a\, h'.\ c(h) = (a, h') \To Q(a)(h').
  \]
\item The \emph{frame-preserving lifting}, which is equivalent to the WP given
  in \Cref{sec:background}, is determined by setting
  \begin{align*}
    (\WP\ Q)(c)(h_c) &= \forall h\, h_f.\ h = h_c\uplus h_f  \Rightarrow \\
    &\qquad\forall a\,h'.c(h) = (a, h') \Rightarrow \\
    &\qquad\quad\exists h_c'.\ h' = h_c'\uplus h_f \land Q(a, h_c').
  \end{align*}
  \end{enumerate}
\end{example}

The frame rule is unsound for the evaluation lifting. The computation
\(\mathsf{wipe} = \lambda h. (\unitpt, \emptyheap)\)
that unconditionally deallocates the entire heap satisfies
\(\vdash \WP\, \mathsf{wipe}\, \hemp\) but not
\(\ell \mapsto 1 \vdash \WP\, \mathsf{wipe}\, \{\hemp \sep \ell \mapsto 1\}\),
which ought to follow from a use of the frame rule with \(F = \ell \mapsto 1\).

The frame rule is sound for the frame-preserving lifting. While this is an
improvement on the evaluation lifting, the frame-preserving lifting permits
verifying strictly fewer programs. For example, no specification \(\WP\,
\mathsf{wipe}\, Q\) is valid. To see why, suppose such a specification holds and
select a nonempty framing heap \(h_f\). Certainly \(\mathsf{wipe}(\emptyheap
\uplus h_f) = (\unitpt, \emptyheap)\), so the spec says that there is some
\(h_c'\) making \(h_c' \uplus h_f = \emptyset\). But this is impossible.

\subsection{Interpreting Separation Logic}
We now extend the semantics of BI discussed in
\S\ref{sec:posetal-fibrations-for-separation-logic} to support the weakest
precondition modality.
Recall that types and contexts denote objects of the
base category, terms \(\Gamma \vdash e : A\) denote morphisms \(\llbracket
\Gamma \rrbracket \to \llbracket A\rrbracket\), and predicates on \(\Gamma\)
denote elements of the fiber over \(\llbracket \Gamma \rrbracket\). Entailment
\(P \vdash Q\) between predicates defined in the same context \(\Gamma\) is
modeled by \(\llbracket P \rrbracket \leq_{\llbracket\Gamma\rrbracket}
\llbracket Q \rrbracket\), the partial ordering of the fiber over \(\llbracket
\Gamma \rrbracket\).

The subexpressions of \(\WP\,c\,Q\)
are a term \(\Gamma \vdash c : TA\) and a predicate \(\Gamma,
x : A \vdash Q\ \prop\). Assume
a fibration equipped with a monad lifting \((T, W)\) where \(T\) has a strength
\(\sigma_{A, B} : A \times TB \to T(A \times B)\). 
Then define \[
  \llbracket \WP\,c\,Q \rrbracket =
  \langle \id{\llbracket\Gamma\rrbracket}, c \rangle \pb \monadstr_{\llbracket
  \Gamma\rrbracket, \llbracket A \rrbracket} \pb WQ
\]
This defines an element of the fiber over \(\llbracket\Gamma\rrbracket\) by
reindexing, as depicted in the following diagram.
\[\begin{tikzcd}
	{\llbracket \WP\,c\,Q\rrbracket} & {\sigma_{\llbracket   \Gamma\rrbracket, \llbracket A \rrbracket} ^* WQ} & WQ \\
	{\llbracket\Gamma\rrbracket} & {\llbracket\Gamma\rrbracket \times T\llbracket A\rrbracket} & {T(\llbracket \Gamma\rrbracket \times \llbracket A \rrbracket)}
	\arrow[from=1-1, to=1-2]
	\arrow[maps to, from=1-1, to=2-1]
	\arrow["\lrcorner"{anchor=center, pos=0.125}, draw=none, from=1-1, to=2-2]
	\arrow[from=1-2, to=1-3]
	\arrow[maps to, from=1-2, to=2-2]
	\arrow["\lrcorner"{anchor=center, pos=0.125}, draw=none, from=1-2, to=2-3]
	\arrow[maps to, from=1-3, to=2-3]
	\arrow["{\langle \mathbf{id}_{\llbracket\Gamma\rrbracket}, \llbracket c\rrbracket \rangle}", from=2-1, to=2-2]
	\arrow["{\sigma_{\llbracket \Gamma \rrbracket, \llbracket A \rrbracket}}", from=2-2, to=2-3]
\end{tikzcd}\]

Aside from the lifting $(T, W)$ this interpretation only requires a cartesian
base category and a strength for $T$, which are both already required in order
to model Moggi's monadic metalanguage in the base.

\section{General Constructions} \label{sec:constructions}

This section considers three generic constructions
that take unique advantage of
our fibered formulation of BI
and of bunched weakest precondition. One of the benefits
of our approach is that we can transform certain informal design
patterns from separation logic into crisp formal constructions
that can be reasoned about and reused. This contrasts with
existing approaches where even though there are guiding design
principles, they have to be proven sound for every new application.

While there are other useful constructions from separation logic
that are not presented in this work, the examples below suggest that they
might also be within reach of our formalism. We leave a more thorough investigation
of such constructions to future work.

\subsection{Imposing a Strength on a Monad Lifting}

We saw in \Cref{ex:reynolds-wp} that the evaluation lifting for the
global state allows deriving specifications like \(\vdash \WP\,
\mathsf{wipe}\, \hemp\) that cannot be framed. In that example, a
frame rule is recovered by replacing the definition of \(\WP\) with
one that quantifies over framing heaps \(h_f\). In theory,
this amendment to the defining formula of the weakest precondition
incurs a tedious re-proving of everything that had been proved about
the old definition,
but in practice those proofs tend to go through without a hiccup.

The practice of ``baking in the frame rule'' obtains a frame-preserving WP from
a non-frame-preserving one by quantifying over framing
propositions~\cite{birkedal2008simple}.
A new weakest precondition \(\Str{\WP}\) is defined as follows:
\begin{equation}\label{eqn:wpfr}
  \Str{\WP}\, c\, Q = \forall F: \Prop.\ F \sepimp \WP\, c\, (Q \sep F).
\end{equation} 
This validates \textsc{WP-Frame} and refines \(\WP\) in the sense
that \(P \vdash \Str{\WP}\, c\, Q\) implies \(P \vdash \WP\, c\, Q\).
This is the universal such refinement of \(\WP\), as it
imposes no new constraints besides the frame rule.

We will now reconstruct \(\Str{\WP}\) as an operation on monad
liftings \((T, W)\) we term the \emph{frame-preserving core}.  The
frame-preserving core is always a bunched lifting
(\Cref{prop:core-strong}), so it satisfies the frame rule, and it
comes with a morphism of monad liftings into \((T, W)\)
(\Cref{prop:core-refinement}).  It is optimal in a sense made precise by
\Cref{thm:core-universal}: it satisfies a universal property among
bunched liftings with ordinary monad lifting morphisms into \((T,
W)\).

\subsubsection{Constructing the core}
Fix a higher-order bunched fibration \(p: \cat{E} \to \cat{B}\) and a
monad lifting \((T, W)\) with a strength for \(T\).  This gives us
enough structure to interpret \Cref{eqn:wpfr} as an assignment of
objects over \(TA\) to objects \(X\) over \(A\). To wit, for \(X \in \cat{E}_A\) let
\begin{equation}\label{eqn:core-naive}
  \Str{W} X = \Pi_{TA,\Omega}\, \pi_2\pb \gentrue \mto \monadstr_{A,\Omega}\pb W(\pi_1\pb X \mtimes \pi_2\pb \gentrue)
\end{equation}
This suggests a Huber envelope, with \(W\) tucked into the indexed
monoidal closed adjunction and the simple product adjunction, but an
indexed adjunction is not a fibered one. Instead, we recover
\(\Str{W}\) as the fibered Huber envelope of the fibered adjunctions
and monad lifting shown in \Cref{fig:core-envelope}, which we now explain
from left to right.

\begin{figure}
\[\begin{tikzcd}
	{\mathcal{E}} & {\invcat{\cat{E}}} & {\invcat{\cat{E}}} \\
	{\mathcal{B}} & {\mathcal{B}} & {\mathcal{B}}
	\arrow[""{name=0, anchor=center, inner sep=0}, "{\mathsf{wk}_\Omega}", shift left=2, from=1-1, to=1-2]
	\arrow[from=1-1, to=2-1]
	\arrow[""{name=1, anchor=center, inner sep=0}, "{\Pi_\Omega}", shift left=2, from=1-2, to=1-1]
	\arrow[""{name=2, anchor=center, inner sep=0}, "{{-} \otimes \invariant}", shift left=2, from=1-2, to=1-3]
	\arrow[from=1-2, to=2-2]
	\arrow[""{name=3, anchor=center, inner sep=0}, "{\invariant \multimap {-}}", shift left=2, from=1-3, to=1-2]
	\arrow["{\obliv W}", shift left, from=1-3, to=1-3, loop, in=325, out=35, distance=10mm]
	\arrow[from=1-3, to=2-3]
	\arrow[equals, from=2-1, to=2-2]
	\arrow[equals, from=2-2, to=2-3]
	\arrow["T", shift left, from=2-3, to=2-3, loop, in=325, out=35, distance=10mm]
	\arrow["\dashv"{anchor=center, rotate=-90}, draw=none, from=0, to=1]
	\arrow["\dashv"{anchor=center, rotate=-90}, draw=none, from=2, to=3]
\end{tikzcd}\]
\vspace{-2em}
\caption{The adjunctions and monads that define the frame-preserving core.}
\label{fig:core-envelope}
\end{figure}

The first adjunction in \Cref{fig:core-envelope} is the fibered
analogue of the indexed simple product adjunction.  It relies on a
construction that adds a fixed proposition representing the frame to a
higher-order fibration as follows.
\[\begin{tikzcd}
	{\invcat{\cat{E}}} & {\cat{E}} \\
	{\cat{B}} & {\cat{B}}
	\arrow[from=1-1, to=1-2]
	\arrow["{\invcat{p}}"', from=1-1, to=2-1]
	\arrow["\lrcorner"{anchor=center, pos=0.125}, draw=none, from=1-1, to=2-2]
	\arrow["p"', from=1-2, to=2-2]
	\arrow["{{-} \times \Omega}", from=2-1, to=2-2]
      \end{tikzcd}\]

As noted by Hermida~\cite{hermida1999some}, the simple product
adjunctions \(\pi^{A,\Omega}_1 \dashv \Pi_{A,\Omega}\) assemble into a
fibered weakening functor \(\wk{\Omega} : \cat{E} \to
\invcat{\cat{E}}\) that sends \(X \in \invcat{\cat{E}}_{A}\) to \((A,
\pi_1\pb X)\). Its action on morphisms is obtained from the universal
property of cartesian morphisms and the naturality of \(\pi_{\blank,
\Omega}\). By the same token, the right adjoints assemble into a
fibered right adjoint \(\Pi_\Omega\).

A few words about \(\invcat{\cat{E}}\) are in order. The fiber over
\(A\) is isomorphic to \(\cat{E}_{A\times\Omega}\), so each fiber
inherits monoidal closed and cartesian closed structure from \(p\).
Reindexing functors are also inherited: for \(u: B \to A\), we have
\(u\pb(A, X) = (B, (u \times \id{\Omega})\pb X)\). As a consequence,
the fiberwise structure is preserved by reindexing.  It also contains
an object \(\invariant_A = (A, \pi_2\pb \gentrue)\) we call the
\emph{invariant}.  All reindexing functors of \(\invcat{\cat{E}}\)
preserve the invariant.

The second adjunction in \Cref{fig:core-envelope} is the
\emph{invariant adjunction}. We already have an indexed
family of vertical adjunctions \(\blank \mtimes_A \invariant_A \dashv
\invariant_A \mto_A \blank\). Since the indexed adjoints are preserved
by reindexing, they assemble into the adjunction \(\blank \mtimes
\invariant \dashv \invariant \mto \blank\) fibered over the identity.

The last ingredient is the monad lifting. We cannot take the Huber
envelope of \((T, W)\) within the fibered adjunction we just
constructed because \(W\) is a monad on \(\cat{E}\). In any case, we
do not want monadic computations downstairs to alter the invariant
\(\Omega\).  We address this by reindexing along the strength of \(T\)
to ignore the invariant as follows.
\begin{definition}
  The \emph{oblivious monad} \(\obliv\, W\) on \(\invcat{\cat{E}}\)
  is the functor \(
    \obliv(W) (A, X) = (TA, \monadstr_{A\Omega}\pb W X)
  \).
\end{definition}
The action of \(\obliv\, W\) on morphisms into \(X\) is obtained from
the naturality of the strength and the universal property of
\(\monadstr_{A,\Omega}\pb W X\). The unit and multiplication at \(X\)
are determined by the unit and multiplication laws for \(\monadstr\)
and the universal property of \(\monadstr_{A,\Omega}\pb W X\).  The
monad laws hold automatically because the fibration is faithful.

We can now construct a lifting for \(T\) that
recovers the indexed definition from \Cref{eqn:core-naive}.
Applying \Cref{def:huber-env} and \Cref{prop:huber-morphism} to
the adjunction and monad in \Cref{fig:core-envelope} gives:
\begin{lemma}\label{lem:core-lifting}
  \((T, \Str{W})\) is a monad lifting.
\end{lemma}
Our fibrations are posetal, so making this into a bunched lifting
amounts to proving an inequality \(\Str{W} X \otimes Y \leq
\monadstr_{A,B}\pb(\Str{W}(X \otimes Y))\)
(\Cref{prop:strength-leq}). We leave these proofs to
\appref{sec:core-strong-appendix}, as they are an algebraic version
of the usual syntactic proof that a weakest precondition defined by
\Cref{eqn:wpfr} validates the frame rule.
\begin{proposition}\label{prop:core-strong}
  \((T, \Str{W})\) is a bunched monad lifting.
\end{proposition}

\subsubsection{The frame-preserving core is universal} The
frame-preserving core is in some sense the smallest change one could make
to a monad lifting in order to get a strong monad lifting. We make this precise
with a universal property that says any (non-strong) monad lifting morphism
from a strong monad lifting into \((T, W)\) factors uniquely through
\((T, \Str{W})\). The argument hinges on classifying \(\Str{\blank}\)
as the right adjoint to an inclusion.
\begin{definition}\label{def:strong-lift-cat}
  The category $\prestrongc$ of \emph{pre-strong bunched liftings} is all the
  higher-order bunched fibrations \(p : \cat{E} \to \cat{B}\)
  equipped with monad liftings \((T, W)\) where \(T\) is strong. Morphisms
  \((q, S, V) \to (p, T, W)\) are monad lifting morphisms \(((F,
  \lambda), (G, \gamma))\) such that
    \((G, F): p \to q\) is a morphism of higher-order bunched fibrations
    and
    \((F, \lambda) : S \to T\) is a morphism of strong monads.
    The category of \emph{strong bunched liftings} $\strongc$ 
    restricts $\prestrongc$ to bunched liftings and their morphisms.
\end{definition}
The strong bunched liftings form a full subcategory of the pre-strong
bunched liftings because any morphism that preserves strengths
downstairs will preserve them upstairs if they exist.
\begin{proposition}\label{prop:invcat-functor}
  The category with an invariant construction is a functor
  \(\invcat{\blank} : \cat{HOBIFib} \to \cat{HOBIFib}\). Moreover,
  if \(F : p \to q\) in \(\cat{HOBIFib}\) and \(L_p \dashv R_p\) and
  \(L_q \dashv R_q\) are instances of \Cref{fig:core-envelope}
  then \((F, \invcat{F})\) is a morphism of adjunctions.
\end{proposition}
\begin{proof}
  First, \((\blank \times \Omega)\pb p\) is a higher-order bunched
  fibration. Quantifiers are straightforward to construct even though
  \(\blank \times \Omega\) is not product preserving.  The generic
  object is \(\Omega \To \Omega\).

  Let \((F, G) : \cat{D} \to \cat{E}\) be a morphism of higher-order bunched fibrations.
  Since \(F(A\times \Omega_{\cat{D}}) = FA \times \Omega_{\cat{E}}\), we define
  \[
    \invcat{G}(A, X \in \cat{D}_{A\times\Omega_{\cat{D}}}) = (FA, GX \in \cat{E}_{FA \times \Omega_{\cat{E}}}).
  \]
  Composition and identity laws are immediate.

  The morphism of adjunctions comes from the fact that the adjunctions
  in question are defined using only higher-order bunched structure
  and \(F\) preserves all of that structure.
\end{proof}

\begin{proposition}\label{prop:obliv-functor}
  The oblivious monad construction defines a functor \(\obliv: \prestrongc \to \prestrongc\).
\end{proposition}

\begin{proposition}\label{prop:core-functor}
The frame-preserving core defines a functor \(
  \LiftS : \prestrongc \to \strongc.
\)
\end{proposition}
We will now construct a monad morphism into \(W\) showing that the
frame-preserving core refines the original monad.
\begin{corollary}\label{prop:core-refinement}
  For any \(X\), there is a vertical morphism \(\xi_X : \Str{W}{X} \leq WX\).
  This defines a morphism \(\xi: \Str{W} \to W\) in \(\prestrongc\).
  If \(W\) is strong, \(\xi\) is in \(\strongc\).
\end{corollary}

\begin{lemma}\label{prop:core-fp}
  If \((T, W)\) is a bunched monad lifting then \(\Str{W} = W\).
\end{lemma}
Together, \Cref{prop:core-fp} and \Cref{prop:core-strong} give:
\begin{lemma}\label{lem:core-idempotent}
  If \((T, W)\) is a monad lifting then \(\Str{(\Str{W})} = \Str{W}\).
\end{lemma}
Details can be found in \appref{app:sec:core-universal}.

\begin{theorem}\label{thm:core-adjunction}
  The functor \(\LiftS : \prestrongc \to \strongc\) that constructs
  the frame-preserving core of a monad lifting is right adjoint to the
  functor \(U: \strongc \to \prestrongc\) that forgets the tensorial
  strength of a bunched lifting.
\end{theorem}
\begin{proof}
  The unit \(\eta : X \to \LiftS U X\) for \(X\) strong is an identity
  (\Cref{prop:core-fp}) and the counit \(\epsilon : U \LiftS Y \to Y\)
  for \(Y\) pre-strong is the refinement vertical natural transformation
  (\Cref{prop:core-refinement}). Now for the laws.
\[\begin{tikzcd}
	{\LiftS\, X} & {\LiftS\, U \LiftS X} & UY & {U \LiftS U Y} \\
	             & {\LiftS\, X}          &    & {U Y}
	\arrow["{\eta}", from=1-1, to=1-2]
	\arrow["{\id{}}"', from=1-1, to=2-2]
	\arrow["{\LiftS\,\epsilon_X}", from=1-2, to=2-2]
	\arrow["{U\eta}", from=1-3, to=1-4]
	\arrow["{\id{}}"', from=1-3, to=2-4]
	\arrow["{\epsilon}", from=1-4, to=2-4]
\end{tikzcd}\]
The unit \(\eta\) is always an identity.
In the first diagram, the interesting morphism for \(\LiftS\,
\epsilon_X\) is a vertical natural transformation with components \(
  \Str{(\Str{W})} Z \leq \Str{W} Z,
\)
where \(Z\) is an actual object of some total category and \(W\) is
the lifted monad associated with \(X\).  The inequality is actually
an identity (\Cref{lem:core-idempotent}). In the second
diagram, the morphism \(\epsilon_Y\) is an identity because \(Y\) is
already strong (\Cref{prop:core-fp}).\end{proof} Since \(U\) is full
and faithful, \Cref{thm:core-adjunction} exhibits \(\strongc\) as a
coreflective subcategory of \(\prestrongc\).  As a right adjoint, the
frame-preserving core enjoys a co-free universal property with respect
to morphisms of bunched liftings.
\begin{corollary}\label{thm:core-universal}
  For any higher-order bunched fibration \(q: \cat{D} \to \cat{A}\)
  with a strong monad lifting \((S, U)\), every morphism of monad
  liftings \(((F, \lambda), (G, \gamma)): (S, U) \to (T, W)\) factors
  uniquely through the refinement \(\xi_W: (T, \Str{W}) \to (T, W)\).
\end{corollary}

\subsection{Credit-Based Cost Reasoning}

This section shows how to extend monad liftings to reason about cost. This is inspired by work
on credits for separation logic, pioneered by Atkey \cite{atkey2011amortised}
and widely applied since then \cite{zhan2018verifying,chargueraud2019verifying,mevel2019time,pottier2024thunks,aguirre2024error,haselwarter2024tachis,li2025modular}.
We follow the methodology
explained in \Cref{sec:recipe}.

\paragraph*{Semantic monad.} Assuming the existence of a monad $T$,
we want a monad $T_c$ that extends $T$ with a cost model.
There has been much work done in finding such extensions~\cite{azevedodeamorim2025cert,grodin2024decalf,kaminski2018weakest,kura2023higher}.
We base our semantics on
the writer monad transformer used in the $\mathbf{decalf}$ type theory~\cite{grodin2024decalf}. Therefore, assuming a commutative partially ordered monoid $M$ internal to
a topos $\topos$, we use the monad $T(M \times -)$.

\paragraph*{Bunched fibration}
For the sake of making our construction more concrete, we assume that our base
bunched fibration is of the form $\laxslice S {\upclosed(R)}$, where $R$ is a resource algebra
internal to the topos $\topos$. In order to model the cost information, we want
to also grade our predicates by elements of $M$. Therefore, we take as our new
bunched fibration $\laxslice S {\upclosed(M \times R)}$ --- i.e. predicates are graded by
the base resource algebra $R$ and by the cost monoid $M$.

\paragraph*{Monad lifting}
We now define a lifting of $T(M \times -)$. Assuming that
we have a monad lifting $(T, \hat{T}) : \laxslice S {\upclosed(R)} \to \laxslice
S {\upclosed(R)}$,
we define our lifting in three steps. First, we apply the following change of base
transformation that threads the cost information through $\hat{T}$.

\begin{theorem}
  \label{th:error-credit}
  The lifting $\hat{T}$ transports to a lifting on $\laxslice S {\upclosed(M \times R)}$.
\end{theorem}
\begin{proof}
  We define the following monad lifting
  \[\hat{T}'(A, p : M \times R \times A \to \Omega) = (TA, T' p : M \times R \times T A \to \Omega),\] where \((T' p)(m, r, c) = T (\lambda r'.\, p \,m\, r')\, r\, c\).
  We have to show that this is indeed a monad and that it lies over $T$. Because our bunched fibrations are assumed faithful, these properties hold by construction. It remains to show that the underlying monad structure does indeed lift. The proof can be found in \appref{app:error-credit}. \qedhere
\end{proof}

Then, we define a lifting of the object $M$ as $\lambda m\, m'\, r.\, m = m'$,
where we are using the equality internal to $\topos$. We conclude this construction
by combining these two liftings as $\hat{T}'(\hat{M} \otimes -)$. Concretely,
applying it to $(A : \topos, p: M \times R \to \mathcal{P}(A))$, results
in the following $M \times R$-graded predicate over $T(M \times A)$:
\[
T(\lambda r'.\, \exists m_1, m_2, (m = (m_1 \cdot m_2)) \land p(m_1, r'))\, r
\]
We now illustrate what this weakest precondition modality gives us in the context
of costful global state where the predicates are indexed by sets of heaps.
\begin{example}
  We assume the bunched fibration defined in \Cref{ex:reynolds-wp}. Given a
  postcondition $q$ and a computation $c : A \to TB$, the weakest precondition
  modality given above unfolds to
  \begin{align*}
  (\WP\,c\,Q)(h, m)
  \iff
    \begin{aligned}
      &\text{for all \(h_f\) disjoint from \(h\)}\\
      &\text{there exists \(h'\) disjoint from \(h_f\)}\\
      &\text{there exists \(m_1, m_2\) such that  \(m_1 \cdot m_2 = m\)}\\
      &\text{with \(c(h\uplus h_f, m_2) = (a, h' \uplus h_f)\) and \(Q(a)(h', m_1)\)}
    \end{aligned}
\end{align*}
\end{example}

\subsection{Persistence via Fibered Limits}

In this section we show how to obtain
persistence-like modalities . These modalities are an important component of
modern higher-order separation logics, as they are used for handling
persistent knowledge, including Hoare triples and invariants. We propose
two generic models for persistence-like modalities within our formalism.
The first one uses familiar constructions from fibered category theory
and the second one uses the idempotent core of affine quantales. We
conclude by showing our persistence modality interacts with monad liftings.

\begin{example}[{cf. \cite[Definition 3.3.4]{DBLP:books/daglib/0072949}}]
  If \(p : \cat E \to \cat B\) is a bunched fibration
  and \(R\) a monoidal category then there is
  a bunched fibration \(q = [R,p]: [R, \cat E] \to [R, \cat B]\). The intuitionistic
  connectives are given pointwise;
  the substructural ones are given
  by Day's convolution.
  Pulling this back along \(\Delta\), the constant presheaf functor \(\cat{B} \to
  [R, \cat{B}]\), produces a fibration \(\Delta^* q : \cat{E}' \to \cat{B}\).
\end{example}
\begin{figure}
\[\begin{tikzcd}
	{\cat{E}} \\
	& {\cat{E}'} & {[R,\cat{E}]} \\
	& {\cat{B}} & {[R, \cat{B}]}
	\arrow["{\Delta^*}"{description}, dashed, from=1-1, to=2-2]
	\arrow["\Delta", curve={height=-6pt}, from=1-1, to=2-3]
	\arrow["p", curve={height=6pt}, from=1-1, to=3-2]
	\arrow[from=2-2, to=2-3]
	\arrow[from=2-2, to=3-2]
	\arrow["\lrcorner"{anchor=center, pos=0.125}, draw=none, from=2-2, to=3-3]
	\arrow["{[R,p]}", from=2-3, to=3-3]
	\arrow["\Delta"', from=3-2, to=3-3]
\end{tikzcd}\]
\caption{Pullback definition of fibered inclusion functor}
\label{fig:gradedpb}
\end{figure}

The universal property of pullbacks then gives a fibered
inclusion $\Delta^* : \cat{E} \to \cat{E}'$, shown in \Cref{fig:gradedpb}.
To define adjoints to $\Delta^*$, we have to remember that, in
ordinary category theory, (co)limits of a given shape $\cat{J}$
in a category $\cat{C}$
are equivalently
adjoints to the constant-functor
functor $\Delta : \cat{C} \to [\cat{J}, \cat{C}]$. In the case of fibered
category theory, this generalizes as follows:
fibered (co)limits are equivalently defined as
fibered adjoints to functors defined just like $\Delta^*$.
these fibered adjoints exist whenever $\cat{E}$ and $\cat{B}$ have $M$-shaped
(co)limits that are strictly preserved by $p$~\cite[Exercise 9.2.4]{jacobs1999categorical}.
These assumptions are
satisfied by the kinds of fibrations we have encountered so far.
Taking the right adjoint performs
universal quantification
over the base resource, as in ``for every resource $m \in M$, this proposition
holds''.
The following example illustrates.

\begin{example}
  Consider the heap model from \Cref{sec:background}. In this case, the starting
  fibration is the predicate fibration over $\Set$. The total category
  $\cat{E}'$ then becomes the category whose objects are pairs of a set $A$
  and a heap-graded predicate $\mathbb{H} \to \pow(A)$, and morphisms are
  predicate-preserving functions. The inclusion $\cat{E} \to \cat{E}'$ maps a
  predicate to the constant graded predicate. Its right adjoint maps a graded
  predicate $\varphi$ to the predicate ``for every heap $h$, $\varphi(h)$ holds''.
  Resolving this adjunction yields the comonad that maps a graded predicate
  to its resource-invariant fragment. Since the empty heap is the least heap,
  this is equal to mapping a predicate $\varphi$ to $\varphi(\emptyset)$, the usual
  definition of the persistent modality in the heap model.
\end{example}

Therefore, it is sensible to define our persistence modality as an adjunction
between an intuitionistic fibration and a bunched one. We call them
resourceful-non-resourceful (RNR) adjunctions as a nod to linear-non-linear
adjunctions from the linear logic literature~\cite{benton1994mixed}.
We begin by defining monoidal adjunctions.

\begin{definition}[{\cite[Proposition 13]{mellies2009categorical}}]
  If $(\cat{C}, \otimes_{\cat{C}}, I_{\cat{C}})$ and $(\cat{D}, \otimes_{\cat{D}}, I_{\cat{D}})$ are monoidal categories, a monoidal adjunction is a pair of adjoint functors $L \dashv R : \cat{C} \to \cat{D}$ such that $L$ is strong monoidal.
\end{definition}

We now present the definition of RNR adjunctions. For the sake of simplicity,
we have defined first-order RNR adjunctions, but this readily generalizes to
the higher-order case as well.

\begin{definition}
  An RNR adjunction consists of a first-order fibration $p_I : \cat{E}_I \to \cat{B}$,
  a first-order bunched fibration $p_B : \cat{E}_B \to \cat{B}$ and a fibered
  monoidal coreflective adjunction:
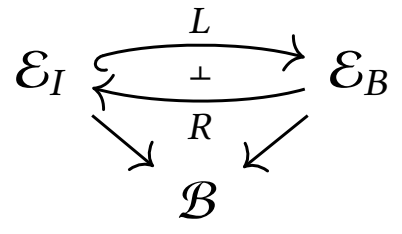
\[\begin{tikzcd}[sep=small]
	{\cat{E}_I} && {\cat{E}_B} \\
	& {\cat{B}}
	\arrow[""{name=0, anchor=center, inner sep=0}, "L", curve={height=-6pt}, hook', from=1-1, to=1-3]
	\arrow[from=1-1, to=2-2]
	\arrow[""{name=1, anchor=center, inner sep=0}, "R", curve={height=-6pt}, from=1-3, to=1-1]
	\arrow[from=1-3, to=2-2]
	\arrow["\dashv"{anchor=center, rotate=-90}, draw=none, from=0, to=1]
\end{tikzcd}\]
\end{definition}
\paragraph{Idempotent core of a quantale}

There is a nice proportionality analogy ``Heyting algebras are to
locales like BI algebras are to quantales''.  Explicitly, quantales
are complete BI algebras. This structure has received much attention
as a way of capturing non-commutative notions of space and algebra.
We now show how known constructions from the quantale literature
can be used to construct RNR adjunctions. We start with some preliminary
and familiar definitions.
\begin{definition}
  A quantale is affine if its monoidal unit is the top element.
\end{definition}

\begin{definition}
  A quantale morphism is a function that preserves the quantalic structure.
\end{definition}

\begin{definition}
  An element $x$ of a quantale is \emph{idempotent} if $x * x = x$.
\end{definition}

While the definitions above are standard, we now show one of the first
surprising basic facts about quantales: every element can be maximally
approximated by an idempotent element.
\begin{definition}
  Given an element $x$ of an affine quantale $Q$, its idempotent part is
  $x^- = \bigvee \{q : \mathsf{Idem}(Q) \mid q \leq x\}$.
\end{definition}
Maximality of this construction is justified by the lemma below.
\begin{lemma}
  For every idempotent element $q$, $q \leq x$ if, and only if, $q \leq x^-$.
  More abstractly, the $(-)^-$ operation is right adjoint to the
  inclusion $\iota$.
\end{lemma}
A direct calculation shows that every quantale morphism maps idempotents to idempotents: $f(x) * f(x) = f(x * x) = f(x)$. Therefore, when we restrict a
quantalic bunched fibration---i.e. a bunched fibration that is fiberwise a
quantale---we recover a first-order fibration
\begin{lemma}
  Given an affine quantalic bunched fibration $p : \cat{E} \to \cat{C}$,
  its idempotent restriction $p_{idem} : \cat{E}_{idem} \to \cat{C}$
  is a first-order fibration.
\end{lemma}
Furthermore, this restriction is universal and we can get an RNR adjunction
from it.
\begin{theorem}
  Given an affine quantalic bunched fibration $p : \cat{E} \to \cat{C}$,
  there is an RNR adjunction between $p$ and its idempotent restriction
  $p_{idem} : \cat{E}_{idem} \to \cat{C}$.
\end{theorem}
\begin{proof}
  There is a fibered inclusion $p_{idem} \hookrightarrow p$.
  We now use the $(-)^-$ construction to define the right
  adjoint $p \hookrightarrow p_{idem}$. On objects, it maps
  $(X : \cat{C}, p : \cat{E}_X)$ to $(X, p^{-})$ and it is
  identity on morphisms. We now have to show that this is
  a fibered adjunction. We start by showing that it is an
  adjunction. The unit and counit can be readily defined
  using the $\iota \dashv (-)^-$ adjunction.

  We now show that the functors are fibered. While the inclusion is
  fibered by construction, the right adjoint needs a bit more work,
  which we now show. In this case, fibered-ness reduces to showing
  that for every morphism $f : X \to Y$ $f\pb(x^-) = f\pb(x)^-$. For
  the direction $f\pb(x)^- \leq f\pb(x^{-})$, consider an idempotent
  element $q' \leq f\pb(x)$.  It suffices to show that $q' \leq f^*(x^-)$.
  We do this by noting that by the adjoint functor theorem,
  the quantale morphism $f^*$ has a left adjoint $f^!$ and, therefore,
  $q' \leq f^*(f^!(q'))\leq f^*(x^-)$. The first inequality is
  the unit of the $f^! \dashv f^*$ adjunction. As for the second one,
  we use monotonicity of $f^*$ and $q' \leq f^*(x)$ which implies
  $f^!(q') \leq x$ and, therefore, $f^!(q') \leq x^-$.
  
  As for the direction $f\pb(x^{-}) \leq f\pb(x)^-$, we begin by
  unfolding the definition of $x^-$ and using the fact that $f\pb$
  preserves suprema, meaning that we have to show
  \[
  \bigvee f\pb(q) \leq f\pb(x)^-,
  \]
  For every idempotent $q$ less than or equal to $x$. By
  the definition of $(-)^-$ and because $f(q)$ is an
  idempotent, it suffices to show that $f^{\#}(q) \leq f\pb(x)$,
  which follows from monotonicity of $f\pb$ and $q \leq x$.
\end{proof}

\paragraph{From RNR adjunctions to weakest preconditions}
Though RNR adjunctions capture the interaction between
structural and substructural logics, it does not account for the
weakest precondition modality, nor does it capture the interaction
between traditional Hoare logic and separation logic. To remedy this,
we incorporate monad liftings into our definition by requiring two
monad liftings $\hat{T}$ and $\hat{T_g}$, respectively, with a
mediating monad morphism $\hat{T} \to \hat{T_g}$.

We conclude this section by showing that given an RNR adjunction and
a bunched lifting, there is a universal non-bunched
lifting we can construct for the intuitionistic fibration
$\cat{E}_I \to \cat{B}$, given by the Huber envelope. We now
formalize this universal property.
\begin{definition}
  The category $\cat{RNRLift}(p_I, p_B)$ is defined as
  \begin{itemize}[leftmargin=*]
  \item The objects are RNR adjunctions over $p_I$ and $p_B$
    with two liftings $\overline{T}_I$ and $\overline{T}_B$,
    connected by a monad morphism $\gamma : L \overline{T}_I \to \overline{T}_B L$.
  \item The morphisms are monad morphisms between the liftings.
  \end{itemize}
\end{definition}

In order to prove our universal property, we define a
variant of $\cat{RNRLift}(p_I, p_B)$ where the objects only require existence
of the bunched lifting $\overline{T}_B$. We call such category
$\cat{PreRNRLift}(p_I, p_B)$. We can now state the main theorem of this section.

\begin{theorem}[cf. \Cref{app:huber}]
  \label{th:huber}
  The Huber envelope is the left adjoint to the forgetful functor $\cat{RNRLift}(p_I, p_B) \to \cat{PreRNRLift}(p_I, p_B)$.
\end{theorem}

\section{Bunched Liftings for Local Independence} \label{sec:lip}

\newcommand\Psh[1]{[{#1}^{\mathrm{op}};\Set]}
\newcommand\Sh{\operatorname{\mathrm{Sh}}}
\newcommand\Res{{\mathrm{Res}}}
\newcommand\RV{\operatorname{\mathrm{RV}}}
\newcommand\Alloc{\mathop{\mathrm{Alloc}}}
\newcommand\Liftalloc{\overline{\mathrm{Alloc}}}

This section introduces a generic \emph{allocation monad}
for any situation involving
local independence as axiomatized by Simpson~\cite{simpson2024equivalence},
alongside a higher-order separation logic for it.
Then, it combines this with
the pullback-stability theorems presented in the preceding sections
to give a novel separation logic for the Giry monad.

\subsection{A Generic Allocation Monad}

In brief, local independence structure
equips a category \(\cat{C}\) with what is called
a \emph{system of independent pullbacks}.
This consists of a collection of commuting squares in $\cat{C}$,
called \emph{independence squares},
with some squares
called \emph{independent pullbacks}
if they are universal among independent squares with the same bottom cospan.
Independent squares and independent pullbacks must satisfy five
axioms resembling properties of ordinary pullbacks (e.g., the two-pullbacks lemma).
Details can be found in Simpson~\cite[Definition~6.1]{simpson2024equivalence}.

In this section we work with an arbitrary category \(\cat{C}\)
that has a system of independent pullbacks as well as a terminal object.

\begin{definition} \label{def:Alloc}
  The \emph{allocation monad}
  \(\Alloc\)
  on \(\Psh{\cat{C}}\)
  has action on objects
  defined by a colimit (\(\pi\) denotes the canonical projection
  out of a slice category):
  \[\Alloc(P) = c\mapsto
    \colim \left(
      ({\cat{C}}/c)\op
      \xrightarrow{\pi\op} {\cat{C}}\op
      \xrightarrow{P} \Set \right)
    \]
  Explicitly, elements of \(\Alloc(P)(c)\)
  are equivalence classes \([d,f,x]\)
  where \(f\) is a morphism \(d \to c\)
  and \(x \in P(d)\),
  with \([d,f,x]=[d',f',x']\)
  if and only if the cospan \((f,f')\)
  completes to a square \(fg = f'g'\)
  such that \(P(g)(x) = P(g')(x')\).

  Notably, the colimit just given
  does not come with an evident contravariant action,
  as it uses \(c\) covariantly to form \(\cat{C}/c\).
  To make \(\Alloc(P)\) a functor,
  its action on morphisms \(f : c'\to c\) is defined by
  independent pullback:
  \(\Alloc(P)(f)[d, g, x] = [d', gf', P(f')(x)]\)
  where \(f' : d' \to d\) is the independent pullback of \(f\) along \(g\).
\end{definition}
Thinking of objects of \(\cat{C}\) as possible worlds,
each \([d,f,x] \in \Alloc(P)(c)\)
represents an
expansion \(f\) of the current world \(c\) into a new world
\(d\) where new resources can be allocated before producing \(x\).
The strong monad structure on \(\Alloc\) reflects this:
\begin{align*}
  \eta_{P,c}(x) &= [c,\id{}, x]\\
  \mu_{P,c}[e,f,[d,g,x]] &= [e,fg,x]\\
  \sigma_{P,Q,c}(x,[d,f,y]) &= [d,f,(P(p)(x),y)]
\end{align*}
The unit yields
computations that perform no expansion,
the multiplication combines expansions using
composition,
and the strength uses the action of \(P\) to transport
\(x\) to the expanded world \(d\).

Suitable choices of \(\cat{C}\) recover
monads used in the semantics of name generation,
local store, and synthetic probability.
\begin{itemize}
\item
  When \(\cat{C} = \cat{FinInj}\op\),
  the opposite of the category of finite sets and injective functions,
  \(\Alloc\) becomes the \emph{name generation monad}
  used by Stark~\cite{stark1994names} to model the \(\nu\)-calculus.
\item
  When \(\cat{C} = \mathbb{E}\),
  the category of heap layouts and initializations from Kammar~et.~al.~\cite[\S IV.B]{kammar2017monad},
  \(\Alloc\) becomes the \emph{hiding monad} that models locality
  of full ground reference cells.
\item
  And when \(\cat{C} = \cat{Prob}_{\rm{Pol}}\),
  the category of Polish probability spaces and measure-preserving functions,
  \(\Alloc\) becomes the \emph{probability monad} defined by
  Simpson~\cite{simpson2017probability} that provides an analog of the Giry monad
  on his topos of probability sheaves.
\end{itemize}

This final example
  has recently been generalized to an arbitrary Markov category with conditionals,
  providing a notion of abstract sample space for discrete and Gaussian probability
  as well as nondeterminism~\cite{stein2025random}.
  Each such choice of Markov category makes \(\Alloc\) an accompanying notion
  of probabilistic or nondeterministic computation.
We note in passing that \(\Alloc\) always produces atomic
sheaves---a proof can be found in \appref{app:prop:alloc-atomic}.

\subsection{Local Independence as Bunched Fibration}

\newcommand\RProp{{\mathrm{RProp}}}
This section presents a higher-order bunched fibration over \(\Psh{\cat{C}}\).
The starting point for this construction is the following observation,
which provides a monoidal analog of the standard fact that
a category with pullbacks and a terminal object has products:
\begin{proposition} \label{prop:lip-semicart}
  \(\cat{C}\) is symmetric semicartesian.
\end{proposition}
As in the cartesian case,
the tensor product \(c_1 \otimes c_2\)
is defined by taking the independent pullback
of the unique morphisms \(c_1 \to 1\) and \(c_2 \to 1\)
into the terminal object of \(\cat{C}\).
Though we will not use it,
we note a stronger form of Proposition~\ref{prop:lip-semicart}
showing that independent pullbacks create
locally symmetric semicartesian structure.
\begin{proposition} \label{prop:lip-locally-semicart}
  Let \(\cat{C}\) be an arbitrary category with a system of independent pullbacks. Then
  \begin{enumerate}
  \item
    each slice \(\cat{C}/c\) is symmetric semicartesian;
  \item
    independent pullback along \(f : d\to c\)
    defines a strong monoidal functor \(f^* : \cat{C}/c \to \cat{C}/d\);
  \item
    \((\id{})^* \cong \id{}\) and \((fg)^* \cong g^* f^*\),
    making the slice construction a contravariant pseudofunctor
    from \(\cat{C}\) to the category of symmetric semicartesian categories.
  \end{enumerate}
\end{proposition}
The proof is analogous to the case of ordinary pullbacks,
except one must take care to check that mediating maps
into a pullback are given by independent squares.
Details can be found in \appref{app:sec:lip}.

Proposition~\ref{prop:lip-semicart}
makes
\(\Psh{\cat{C}}\)
a symmetric semicartesian category
as well,
by Day convolution.
This structure can be reflected into a
pcm object \(\Res\) of resources:
the idea is that a morphism \(f : d \to c\)
models local knowledge \(c\) of the current world \(d\).
For instance, when \(\cat{C} = \cat{FinInj}\op\),
such a morphism
is an injective map \(c \rightarrowtail d\) picking out
a subset of \(d\) visible to a particular observer.

\begin{definition}
  The \emph{presheaf of resources} is
  defined by the colimit
  \(\Res = \colim_{c:\cat{C}_{\cong}} \yo(c)\), where \(\yo\) is the Yoneda embedding
  and \(\cat{C}_{\cong}\) is the groupoid of isomorphisms in \(\cat{C}\).
\end{definition}
Explicitly, elements of \(\Res(c)\)
are equivalence classes
\([r,f:c\to r]\)
with \([r,f]=[r',f']\) if and only if
\(if = f'\) for some
isomorphism \(i : r \to r'\).
The quotient makes
the specific \(r\) appearing in a
given resource irrelevant,
in much the same way that a subobject is an equivalence class of monomorphisms.

\begin{proposition}
  \(\Res\) forms a commutative monoid object
  in the monoidal category \((\Psh{\cat{C}},\otimes,1)\).
\end{proposition}
\begin{proof}
Via the coend formula for \(\Res\otimes\Res\),
the product operation \(\comp : \Res \otimes \Res \to \Res\)
of this monoid object is given
by the cowedge
\(\left(\comp_{c_1,c_2} : \cat{C}(c,c_1\otimes c_2)\times\Res(c_1) \times \Res(c_2) \to \Res(c)\right)_{c_1,c_2:\cat{C}}\)
that sends \((f,[r,g],[s,h])\)
to \([r\otimes s,(g\otimes h)f]\).
The unit \(\emp : 1 \to \Res\)
is \(\emp_c(\star) = [1,c\xrightarrow{!}1]\).
The commutative monoid laws follow from
the symmetric semicartesian structure of \(\cat{C}\).
\end{proof}

The following characterization
of the Day convolution \(\Res\otimes\Res\)
shows that
this commutative monoid
can be reexpressed
as a \emph{partial} commutative monoid
in the cartesian monoidal category \(\Psh{\cat{C}}\).
\begin{proposition}
  \((\Res\otimes\Res)(c)\)
  is isomorphic to the subset of \((\Res\times\Res)(c)\)
  consisting of
  pairs \(([r,f],[s,g])\)
  such that \({!} \circ f = {!} \circ g\)
  is an independence square
  (\(!\) denotes the unique morphism into \(1\)),
  and this isomorphism is natural in \(c\).
\end{proposition}

\begin{proposition}
  \((\Res,\comp,\emp)\) forms a pcm in \(\Psh{\cat{C}}\).
\end{proposition}

The multiplication \((\comp)\) is the partial function
\(\Res\times\Res\rightharpoonup\Res\)
that is defined if and only if
the resources that it is given
are independent.
Finally, elements of \(\Res\) are ordered
by the amount of knowledge that they capture
to give an ordered pcm object.
\begin{definition}
  Let \((\sqsubseteq)\)
  be the subpresheaf of \(\Res \times \Res\)
  consisting of those elements \(([r,f],[r',f']) \in (\Res\times\Res)(c)\)
  such that \(gf' = f\) for some \(g : r' \to r\).
\end{definition}
\begin{proposition} \label{prop:ordered-pcm}
  \((\Res,\sqsubseteq,\comp,\emp)\)
  is an ordered pcm in \(\Psh{\cat{C}}\).
\end{proposition}
A proof that \(\comp\) satisfies the requisite monotonicity condition
can be found in \appref{app:sec:ordered-pcm}.
With \(\Res\) in hand, one can define
a presheaf \(\RProp\) of resourceful propositions
as the presheaf
of internally upwards-closed subsets of \(\Res\).
\(\RProp\) is a complete BI algebra
object in \(\Psh{\cat{C}}\) by construction, so
the lax slice over \(\RProp\) is a higher-order bunched fibration.

In what follows, it will be convenient to have the following
explicit definition of this bunched fibration.
Say that a \emph{predicate} on a presheaf \(P\) is a subfunctor \(\varphi\)
of \(P \times \coprod_{c:\cat{C}}\yo(c)\) such that
if \((x,(r,f)) \in \varphi\) for some
\(x \in Pc\) and \(r:\cat{C}\) and \(f:c\to r\),
and \(p\) factors as \(f=pf'\) for some \(f' : c \to r'\) and \(p : r' \to r\),
then \((x,(r',f')) \in \varphi\) as well.
A calculation using the Kripke-Joyal semantics for \(\Psh{\cat{C}}\) shows:
\begin{proposition}
  The lax slice of \(\Psh{\cat{C}}\) over \(\RProp\) is equivalent to
  the fibration \(p : \cat{E} \to \Psh{\cat{C}}\) where
  \begin{itemize}[leftmargin=*]
  \item Objects of \(\cat{E}\) are pairs \((P,\varphi)\) with \(\varphi\) a predicate on \(P\);
  \item Morphisms of \(\cat{E}\) from \((P,\varphi)\) to \((Q,\psi)\)
    are maps of presheaves \(\alpha : P \to Q\)
    such that \((x,(r,f)) \in \varphi\) implies \((\alpha_c(x),(r,f)) \in \psi\)
    for all objects \(c\) of \(\cat{C}\), \(x \in Pc\), and \((r,f) \in \Res(c)\);
  \item The functor \(p\) is the projection \((P,\varphi) \mapsto P\).
  \end{itemize}
\end{proposition}

\subsection{Lifting the Allocation Monad}

\begin{definition} \label{def:Liftalloc}
  The lifting of \(\Alloc\)
  is defined by
  \(\Liftalloc(P,\varphi) = (\Alloc(P),\psi)\),
  where the predicate \(\psi\)
  contains a pair \((m,(r,f))\)
  if and only if
  there exists a representative \((c',p,x)\)
  of \(m\) and a new resource \((r',f')\in\Res(c')\)
  such that \((x,(r',f')) \in \varphi\)
  and there is a morphism \(r'\to r\) such that
    \(\begin{tikzcd}[sep=small]
      c' \arrow[r, "f'"] \arrow[d, "p"'] & r' \arrow[d] \\
      c \arrow[r, "f"']                  & r
    \end{tikzcd}\)
    forms an independence\vspace{-1.5em}
    square.
\end{definition}
Intuitively, \(\psi\)
says that a piece of local knowledge \((r,f)\)
about the current world \(c\)
is enough to establish \(\varphi\) as a postcondition for
a computation \(m\)
if the result \(x\) produced by \(m\)
satisfies \(\varphi\)
given upgraded knowledge \((r',f')\),
formed by amalgamating
\((r,f)\) with any new resources
used to produce \(x\).
That this amalgamation forms an independence square
captures the idea that one's
state of knowledge before running \(m\) is independent
of the resources \(m\) will allocate.
This independence is crucial for establishing
that \(\Liftalloc\) validates the frame rule.

\begin{theorem} \label{thm:Liftalloc}
  \(\Liftalloc\) forms a bunched monad lifting
  of \(\Alloc\) along the higher-order bunched fibration \(p : \cat{E} \to \Psh{\cat{C}}\).
\end{theorem}
\begin{proof}
  The axioms for a system of independent pullbacks
  ensure that the strong monad structure
  for \(\Alloc\) lifts to a strong monad structure for \(\Liftalloc\).
  For instance, the unit of \(\Alloc\) lifts to a map
  \(\eta : (P,\varphi) \to \Liftalloc(P,\varphi)\)
  by axiom IP1.
  Similarly, the multiplication of \(\Alloc\) lifts
  because independence squares can be pasted together by axiom IP3,
  and the strength lifts because independence squares satisfy a
  two-pullbacks-lemma-type property by axiom IP4.
  The monad laws for these lifted unit, multiplication, and strength maps follow from the monad laws for \(\Alloc\) because the lax slice fibration is posetal.
  Finally, an explicit calculation verifies that \(\Liftalloc\) is a fibered functor.
  Details can be found in \appref{app:sec:Liftalloc}.
\end{proof}

\newcommand\Gen{\operatorname{\mathrm{Gen}}}
We conclude this section by showing how this generic construction
yields logics for
name generation, used in the semantics of dynamic allocation
and abstract garbage collection,
and for continuous probability.

\begin{example}[name generation]
When \(\cat{C} = \cat{FinInj}\op\),
objects of \(\cat{C}\) are finite sets \(L\)
that can be thought of as sets of names
that are in use at a given point in time.
Morphisms \(L' \to L\)
are injective functions \(L\rightarrowtail L'\).
These functions have two readings: as an expansion of
\(L\) into \(L'\),
or as local knowledge of the part of \(L'\) hit by \(L\).
A resource in \(\Res(L)\) is precisely a subobject of \(L\),
hence equivalent to a subset \(U\) of \(L\).

The bunched fibration \(p : \cat{E}\to\Psh{\cat{C}}\)
models a separation logic for disjointness of names.
Objects of \(\cat{E}\)
are pairs \((P,\varphi)\)
where \(P:\cat{FinInj}\to\cat{Set}\)
and \(\varphi\) is a predicate
on pairs \((x\in P(L),U\subseteq L)\)
satisfying the property that if \((x,U)\in \varphi\)
and \(U\subseteq U'\) then \((x,U') \in \varphi\).
The separating conjunction
of two predicates \(\varphi,\psi\) on a presheaf \(P\)
is defined by
\((x,U)\in(\varphi\sep\psi)\)
if and only if there exist disjoint subsets \(V,W\) of \(U\)
such that \((x,V)\in\varphi\) and \((x,W)\in\psi\).

The allocation monad
\(\Alloc\) is equivalent to the name generation monad \(\Gen\)
defined by Stark~\cite{stark1994names},
where
\(\Gen(P)(L)\) is
the set of pairs \((L_{\rm{new}},x\in P(L+L_{\rm{new}}))\)
quotiented by an equivalence relation
that identifies the results \(x\) up to
renaming of names in \(L_{\rm{new}}\).

The lifted monad \(\Liftalloc\)
is defined by \(\Liftalloc(P,\varphi) = (\Gen P, \psi)\)
where \(([L_{\rm{new}},x], U \subseteq L)\in\psi\)
if and only if \((x, U + L_{\rm{new}}) \in \varphi\).
Intuitively, this states that a piece of local knowledge \(U\)
suffices to establish a postcondition \(\varphi\)
if and only if
\(\varphi\) holds of the result \(x\)
when \(U\) is upgraded with knowledge of
the newly generated names.
\end{example}

\begin{example}[continuous probability] \label{example:continuous-probability}
Let \(\cat{C}\) be the category of Polish probability spaces
and measure-preserving maps.
Objects of \(\cat{C}\)
are the sample spaces of standard probability theory.
Morphisms \(\Omega'\to\Omega\)
can be interpreted either as an expansion of
\(\Omega\) into \(\Omega'\),
or as local knowledge of \(\Omega'\)
restricted to those observable
quantities on \(\Omega'\) that factor through \(\Omega\).

The bunched fibration over \(\Psh{\cat{C}}\)
models a separation logic where separating conjunction
means probabilistic independence.
Objects of \(\cat{E}\) are pairs
\((P,\varphi)\)
where \(\varphi\) is a predicate
on pairs of shape \((x\in P(\Omega),(\Omega_g,g:\Omega\to\Omega_g))\)
such that if \((x,(\Omega_g,g))\in \varphi\)
and
\(g = fg'\) for some \(g':\Omega\to\Omega_{g'}\)
then \((x,(\Omega_{g'},g')) \in \varphi\).
The separating conjunction
of two predicates \(\varphi,\psi\) on a presheaf \(P\)
is defined by
\((x,(\Omega_g,g))\in(\varphi\sep\psi)\)
if and only if there exist
random variables \(X,Y\) with domain \(\Omega_g\)
such that \(X\) and \(Y\)
are probabilistically independent and \((x,(\cod(X),Xg))\in\varphi\)
and \((x,(\cod(Y),Yg))\in\psi\).

The allocation monad
\(\Alloc\) is the probability monad \(\underline{\mathcal M}\)
on \(\Psh{\cat{C}}\) defined by Simpson~\cite{simpson2017probability}.
The lifting given by Theorem~\ref{thm:Liftalloc}
yields a bunched weakest precondition for \(\underline{\mathcal M}\).
From here, a bunched weakest precondition for the more familiar
Giry monad \(\mathcal M\), restricted to the category
\(\cat{Pol}_{\mathrm{um}}\) of Polish spaces
and universally-measurable maps,
can be obtained
by taking a series of pullbacks:
\[\begin{tikzcd}[sep=small]
	{\cat{G}} & {\cat{F}} & {\cat{E}} \\
	\\
	{\cat{Pol}_{\mathrm{um}}} & {\Sh(\cat{C},J_{\mathrm{at}})} & {\Psh{\cat{C}}}
	\arrow[from=1-1, to=1-2]
	\arrow[from=1-1, to=3-1]
	\arrow["\lrcorner"{anchor=center, pos=0.125}, draw=none, from=1-1, to=3-2]
	\arrow[from=1-2, to=1-3]
	\arrow[from=1-2, to=3-2]
	\arrow["\lrcorner"{anchor=center, pos=0.125}, draw=none, from=1-2, to=3-3]
	\arrow[from=1-3, to=3-3]
	\arrow[from=3-1, to=3-2]
	\arrow[from=3-2, to=3-3]
\end{tikzcd}\]

The square on the right
pulls back along the
inclusion of atomic sheaves on \(\cat{C}\) into presheaves.
The inclusion preserves finite products
because it has a left adjoint (sheafification);
hence the fibration in the middle is a first-order bunched fibration
by Proposition~\ref{prop:fo-bunched-fib-pb}.
Moreover, inclusion of sheaves into presheaves
forms a strong monad morphism because
\(\underline{\mathcal M}\) sends sheaves to
sheaves~\cite[Theorem 7]{simpson2017probability},
giving a bunched monad lifting
for the middle fibration
by Theorem~\ref{thm:bunched-monad-lifting-pb}.

The square on the left depicts a pullback
along the functor \(\RV\)
that sends a Polish space \(A\) to the atomic sheaf
of \(A\)-valued random variables quotiented by almost-everywhere
equality~\cite[Prop. 1]{simpson2017probability}.
This functor preserves finite products,
making the fibration on the left
a first-order bunched fibration
by Proposition~\ref{prop:fo-bunched-fib-pb}.
\(\RV\) is also
strong-monad-preserving~\cite[Theorem 8]{simpson2017probability},
so applying Theorem~\ref{thm:bunched-monad-lifting-pb} a
second time gives a bunched monad lifting for the fibration on the left.

All told, this gives a model of first-order BI
where separating conjunction models probabilistic independence,
complete with a bunched weakest precondition
for the Giry monad.

An explicit description of this weakest precondition is as follows.
Objects of \(\cat{E}\) are
pairs \((A,\varphi)\) where \(A\) is a Polish space
and \(\varphi\) is a predicate on \(\RV(A)\).
The lifting of \(\mathcal M\) then
sends \((A,\varphi)\)
to \((\mathcal M A, \psi)\)
where \(\psi\)
contains those tuples \((k,(\Omega_g,g))\)
consisting of (equivalence classes of) Markov kernels
\(k : \Omega \to \mathcal M A\),
Polish probability spaces \(\Omega_g\),
and measure-preserving maps \(g : \Omega\to \Omega_g\)
such that:
\begin{enumerate}[leftmargin=*]
\item \(k\)
  factors as the Bayesian inversion of
  a measure-preserving map \(p : \Omega'\to\Omega\)
  followed by pushforward along a random variable \(X : \Omega'\to A\);
\item
  \((X,(\Omega_{g'},g')) \in \varphi\)
  for some Polish probability space \(\Omega_{g'}\)
  and measure-preserving map \(g' : \Omega\to\Omega_{g'}\);
\item \(p\), \(g\), and \(g'\)
  complete to a commutative square
  and \(p\) and \(g'\) are conditionally independent
  given \(pg\).
\end{enumerate}
\end{example}
\noindent It is fruitful to compare this weakest precondition
with the weakest preconditions
used by existing probabilistic separation logics
for measure-theoretic probability~\cite{li2023lilac,bao2025bluebell,ho2026bayesian}.
Points (1) and (2) in the above definition recover
the usual conditions that can be found in
existing work, stating that the kernel \(k\)
is simulated by random variable \(X\)
defined on an expansion of the underlying resource,
and also that \(X\) satisfies the postcondition \(\varphi\).
Point (3) is new and follows from our generic construction
of \(\Liftalloc\).
The strength and fiberedness of \(\Liftalloc\)
ensure that our weakest precondition
has a frame rule and commutes with substitution by default---both properties
that had to be imposed via additional side conditions in prior work.

\section{Conclusion and Future Work}

We introduced and developed the general theory
of \emph{bunched fibrations} and \emph{bunched monad liftings},
a fibered framework for separation logic with weakest precondition.
We showed that this framework
captures design patterns of workers in separation logic,
and introduced a new construction based on Simpson's categories
with local independence structure that
simplifies and clarifies existing work on intricate
separation logics for measure-theoretic probability.

As future work, we hope to generalize our notion of bunched fibration
to support proof-relevance,
and to consider further applications of our fibered framework
(e.g., to modeling step-indexed logics and logics with invariants).
We also hope to relate our bunched fibrations
with the theory of Dijkstra monads~\cite{maillard2019principles}
and our liftings for local independence with
the recently-introduced epi-regular categories~\cite{di2025dagger}.

\bibliographystyle{ACM-Reference-Format}
\bibliography{mybib}

\onecolumn
\appendix

\section{The frame-preserving core of a monad lifting}

\subsection{Strength}
\label{sec:core-strong-appendix}
The following lemma is almost a strength for \(\Str{W}\), but the right hand
side uses \(W\) instead of \(\Str{W}\).
\begin{lemma}\label{lem:frame-monad-elim}
  The inequality 
  \(
  \Str{W}(X) \otimes K \leq \monadstr_{A,B}\pb W (X \mtimes K)
  \)
  holds in the fiber \(\cat{E}_{TA\times B}\)
  for any \(X\) over \(A\) and \(K\) over \(D\).
\end{lemma}
\begin{proof}
  As a preliminary remark, the counit of the \(\wk\Omega \dashv \Pi_\Omega\)
  adjunction is a vertical natural transformation \(\eta : \wk\Omega \Pi_{\Omega} (A, X) \to (A, X)\)
in \(\invcat{\cat{E}}\). for all \(A\) in the base and all \(X \in
  \cat{E}_{A\times \Omega}\). The image in \(\cat{E}\) of the counit at
  \(\invariant \mto \obliv\,W(\wk\Omega X \mtimes \invariant)\) is a vertical proof 
  \begin{equation} \label{eqn:monad-forall-elim}
    \pi_1\pb \Str{W} X \leq_{TA\times\Omega}
     \pi_2\pb \gentrue \mto_{TA\times\Omega}
     \monadstr_{A,\Omega}\pb 
     W (\pi_1\pb X \mtimes_{A\times\Omega} \pi_2\pb \gentrue).
  \end{equation}
  
  We now address the lemma itself. Since we are working in a higher-order
  fibration, there is a characteristic morphism \(\chi_K : D \to \Omega\) with
  \(K = \chi_K \pb \gentrue\). 
  We now derive a sequence of inequalities in the fiber over $TA \times D$.
  \begin{align*}
    \pi_1\pb \Str{W} X &\leq
     \pi_2\pb \gentrue \mto_{TA\times\Omega}
     \monadstr_{A,\Omega}\pb 
     W (\pi_1\pb X \mtimes_{A\times\Omega} \pi_2\pb \gentrue) \\
     &\text{by \Cref{eqn:monad-forall-elim}}
   \\
    (\id{TA}\times\chi_K)\pb \pi_1\pb \Str{W} X &\leq
    (\id{TA}\times\chi_K)\pb\left[
     \pi_2\pb \gentrue \mto_{TA\times\Omega}
     \monadstr_{A,\Omega}\pb 
     W (\pi_1\pb X \mtimes_{A\times\Omega} \pi_2\pb \gentrue)\right] \\
   &\text{by monotonicity of reindexing}
   \\
    \pi_1\pb \Str{W} X &\leq
    (\id{TA}\times\chi_K)\pb\left[
     \pi_2\pb \gentrue \mto_{TA\times\Omega}
     \monadstr_{A,\Omega}\pb 
     W (\pi_1\pb X \mtimes_{A\times\Omega} \pi_2\pb \gentrue)\right] \\
   &\text{projection law}
   \\
    \pi_1\pb \Str{W} X &\leq
    (\id{TA}\times\chi_K)\pb
     \pi_2\pb \gentrue \mto_{TA\times D}
    (\id{TA}\times\chi_K)\pb
     \monadstr_{A,\Omega}\pb 
     W (\pi_1\pb X \mtimes_{A\times\Omega} \pi_2\pb \gentrue) \\
    &\text{reindexing \(\mto\)}
   \\
    \pi_1\pb \Str{W} X &\leq
    \pi_2\pb\chi_K\pb \gentrue \mto_{TA\times D}
    (\id{TA}\times\chi_K)\pb
     \monadstr_{A,\Omega}\pb 
     W (\pi_1\pb X \mtimes_{A\times\Omega} \pi_2\pb \gentrue) \\
    &\text{by projection law}
   \\
    \pi_1\pb \Str{W} X &\leq
    \pi_2\pb K \mto_{TA\times D}
    (\id{TA}\times\chi_K)\pb
     \monadstr_{A,\Omega}\pb 
     W (\pi_1\pb X \mtimes_{A\times\Omega} \pi_2\pb \gentrue) \\
    &\text{by characteristic morphism law}
   \\
    \pi_1\pb \Str{W} X \mtimes_{TA\times D}
    \pi_2\pb K
    &\leq
    (\id{TA}\times\chi_K)\pb
     \monadstr_{A,\Omega}\pb 
     W (\pi_1\pb X \mtimes_{A\times\Omega} \pi_2\pb \gentrue) \\
    &\text{by adjoint transpose}
   \\
    \pi_1\pb \Str{W} X \mtimes_{TA\times D}
    \pi_2\pb K
    &\leq
    \monadstr_{A,D}\pb
    (T(\id{A}\times\chi_K))\pb
     W (\pi_1\pb X \mtimes_{A\times\Omega} \pi_2\pb \gentrue) \\
    &\text{by naturality of \(\monadstr\)}
   \\
    \pi_1\pb \Str{W} X \mtimes_{TA\times D}
    \pi_2\pb K
    &\leq
    \monadstr_{A,D}\pb
     W ((\id{A}\times\chi_K)\pb(
     \pi_1\pb X \mtimes_{A\times\Omega} \pi_2\pb \gentrue)) \\
    &\text{by fiberedness of \(W\) over \(T\)}
   \\
    \pi_1\pb \Str{W} X \mtimes_{TA\times D}
    \pi_2\pb K
    &\leq
    \monadstr_{A,D}\pb
     W (\pi_1\pb X \mtimes_{A\times\D} \pi_2\pb K)) \\
    &\text{by reind. \(\otimes\), proj. and char. laws}
  \end{align*}
  This concludes the proof.
\end{proof}

\begin{proof}[Proof of \Cref{prop:core-strong}]
  We have already shown \((T, \Str{W})\) is a monad lifting (\Cref{lem:core-lifting}).
  By \Cref{prop:strength-leq} it suffices to show \(\Str{W} X \otimes Y \leq
  \monadstr_{A,B}\pb(\Str{W}(X \otimes Y))\).
  We construct a chain of logical equivalences by working backwards from our
  goal.
  \begin{align*}
    \Str{W} X \otimes Y &\leq_{TA\times B} \monadstr_{AB}\pb(\Str{W} (X \otimes Y)) 
    \\
    & \text{defn. of \(\Str{W}\)} \\
    \Str{W} X \otimes Y &\leq_{TA \times B}  \monadstr_{AB}\pb(\Pi_\Omega \invariant
    \mto \obliv\,W((\wk\Omega(X \otimes Y)) \mtimes \invariant) \\
    & \text{laws for \(\Pi_\Omega\)} \\
    \Str{W} X \otimes Y &\leq_{TA \times B}  \Pi_\Omega \monadstr_{AB}\pb \left[\invariant
    \mto \obliv\,W((\wk\Omega(X \otimes Y)) \mtimes
    \invariant)\right]\\
    & \text{reind. preserves \(\invariant\)} \\
    \Str{W} X \otimes Y &\leq_{TA \times B}  \Pi_\Omega 
    \left[\invariant
    \mto \monadstr_{AB} \pb \obliv\,W(\wk\Omega(X \otimes Y) \mtimes
    \invariant)\right] \\
    & \text{\(\Pi\) right adjoint}
    \\
    \wk\Omega (\Str{W} X \otimes Y) &\leq_{TA \times B} 
    \invariant
    \mto \monadstr_{AB} \pb \obliv\,W(\wk\Omega(X \otimes Y) \mtimes
    \invariant) \\
    & \text{\(\mto\) right adjoint}
    \\
    \wk\Omega (\Str{W} X \otimes Y)
    \mtimes 
    \invariant
    &\leq_{TA \times B} 
    \monadstr_{AB} \pb \obliv\,W(\wk\Omega(X \otimes Y) \mtimes
    \invariant) \\
    & \text{using \(\invcat{\cat{E}}_{U} \iso \cat{E}_{A\times\Omega}\)}
    \\
    \pi_1 \pb (\pi_1 \pb \Str{W} X \otimes \pi_2\pb Y)
    \mtimes 
    \pi_2 \pb \gentrue
    &\leq_{(TA \times B) \times \Omega} 
    (\monadstr_{A,B} \times \id\Omega) \pb \monadstr_{A\times B, \Omega}\pb W(\pi_1\pb (\pi_1\pb
    X \otimes \pi_2 \pb Y) \mtimes
    \pi_2\pb \gentrue)
    \\
    & \text{use associator, assoc. law for \(\monadstr\)} \\
    \pi_1 \pb \Str{W} X \otimes \pi_2\pb (\pi_1\pb Y
    \mtimes 
    \pi_2 \pb \gentrue)
    &\leq_{TA \times (B \times \Omega)} 
    \monadstr_{A, B\times \Omega}\pb W(\pi_1\pb X \otimes \pi_2(\pi_1 \pb
    Y \mtimes \pi_2\pb \gentrue))
  \end{align*}
  Taking \(K = \pi_1\pb Y \mtimes \pi_2\pb \gentrue\) and \(D = B \times
  \Omega\) in \Cref{lem:frame-monad-elim} concludes the proof.
\end{proof}

\subsection{Universality}
\label{app:sec:core-universal}

\begin{proof}[Proof of \Cref{prop:obliv-functor}]
  We only need to describe the action on the upper half of a monad
  morphism drawn from \(\prestrongc\).
  Let \(q: \cat{D} \to \cat{A}\) and \(p: \cat{E} \to \cat{B}\) be
  higher-order bunched fibrations.  Let \((G, \gamma: GV \To WG)\) be
  a monad morphism from \(V\) to \(W\) fibered over \((F, \lambda: FS
  \To TF)\).  We seek a fibered natural transformation \[
    \invcat{G} (\obliv\, V) X \to (\obliv\, W) \invcat{G} X.
  \]
  We construct it by a backwards calculation.
  \begin{align*}
    \invcat{G} (\obliv\, V) X &\leq \lambda\pb (\obliv\, W) \invcat{G} X)  \\
    (FSA, G \pi_1\pb V X) &\leq \lambda\pb (TFA, \pi_1\pb W G X) & \text{by defn.} \\
    (FSA, \pi_1\pb G V X) &\leq \lambda\pb (TFA, \pi_1\pb W G X) & \text{\(G\) fib'd. over \(F\) cart.} \\
    \pi_1\pb G V X &\leq (\lambda \times \id\Omega)\pb (\pi_1\pb W G X) & \text{reindexing in \(\invcat{\cat{E}}\)} \\
    \pi_1\pb G V X &\leq \pi_1 \pb \lambda \pb W G X & \text{projection law}
  \end{align*}
  but this follows from the monotonicity of \(\pi_1\pb\).
\end{proof}

\begin{proof}[Proof of \Cref{prop:core-functor}]
  Huber envelopes are functorial in both their adjunction and monad
  arguments, so this follows from \Cref{prop:invcat-functor,prop:obliv-functor}.
\end{proof}

\begin{proof}[Proof of \Cref{prop:core-refinement}]
  By \Cref{prop:posetal-monad-morphism} it suffices to construct \(\lambda\).
  Isomorphisms in \(\cat{E}\) are always cartesian arrows. In
particular, the right unitor isomorphism is cartesian over \(\pi_1 : A
\times 1 \to A\) and its inverse is cartesian over \(\langle \id{A},
!_A \rangle : A \to A \times 1\). As a consequence, \(\pi_1\pb X = X\)
for any \(X\) and \(\rho\pb Y = Y\) for any \(Y\) over a product \(B
\times 1\). We calculate
  \begin{align*}
    \Str{W}(X) &\iso \Str{W}{X} \mtimes I && \text{right unitor} \\
               &\leq \monadstr_{A,1}\pb W(X \mtimes I) && \text{\Cref{lem:frame-monad-elim}}\\
               &= \pi_1\pb (T\langle \id{A}, !_A \rangle)\pb W(X \mtimes I) && \text{laws for strength} \\
               &= \pi_1\pb W(\langle \id{A}, !_A \rangle\pb (X \mtimes I)) && \text{\(W\) fibered} \\
               &= W(X) && \text{right unitor cartesian} && \qedhere
  \end{align*}
\end{proof}

\begin{proof}[Proof of \Cref{prop:core-fp}]
  We already have \(\Str{W} \leq W\) (\Cref{prop:core-refinement}), so we prove the other direction.
  Let \((T, W)\) be a bunched monad lifting and let \(X\) lie over \(A\).
  Using the defining adjunctions from \Cref{fig:core-envelope} and the fact that \(((\blank
\times \Omega)\pb \cat{E})_{X} \iso \cat{E}_{X\times\Omega}\),
  we calculate a chain of equivalent inequalities beginning with \(WX \leq \Str{W} X\).
  \begin{align*}
    WX &\leq \Pi_\Omega \invariant \mto \obliv\, W (\wk\Omega X \mtimes \invariant) \\
    \wk\Omega WX &\leq \invariant \mto \obliv\, W (\wk\Omega X \mtimes \invariant) \\
    \wk\Omega WX \otimes \invariant &\leq \obliv\, W (\wk\Omega X \mtimes \invariant) \\
    WX \otimes \gentrue &\leq \monadstr_{A,\Omega} \pb W (X \mtimes \gentrue) \\
  \end{align*}
  The last inequality follows from \Cref{prop:strength-leq}.
\end{proof}

\section{Proof of Proposition~\ref{prop:lip-locally-semicart}} \label{app:sec:lip}

The monoidal product in \(\cat{C}/c\) is defined by taking independent pullbacks
  of cospans into \(c\). The action on morphisms is defined by:
  
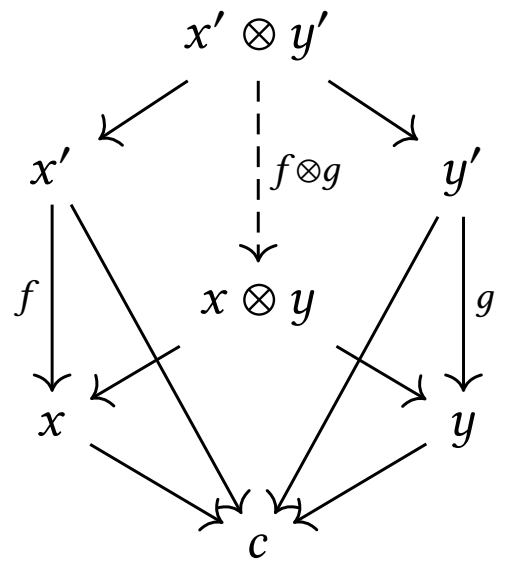
\[
  \begin{tikzcd}[sep=small]
    & x'\otimes y' \arrow[ld] \arrow[rd] \arrow[dd, "f\otimes g", dashed] &                                 \\
    x' \arrow[dd, "f"'] \arrow[rddd] &                                                                     & y' \arrow[dd, "g"] \arrow[lddd] \\
    & x\otimes y \arrow[ld] \arrow[rd]                                    &                                 \\
    x \arrow[rd]                     &                                                                     & y \arrow[ld]                    \\
    & c                                                                   &
  \end{tikzcd}\]
  Given two morphisms \(f\) and \(g\) in the slice over \(c\),
  their product \(f \otimes g\) is the dashed arrow into the independent pullback \(x \otimes y\).
  To establish the existence of \(f \otimes g\), the following diagrams show that
  the two morphisms from \(x' \otimes y'\) to \(x\) and \(y\)
  are independent over the morphisms from \(x\) and \(y\) to \(c\).
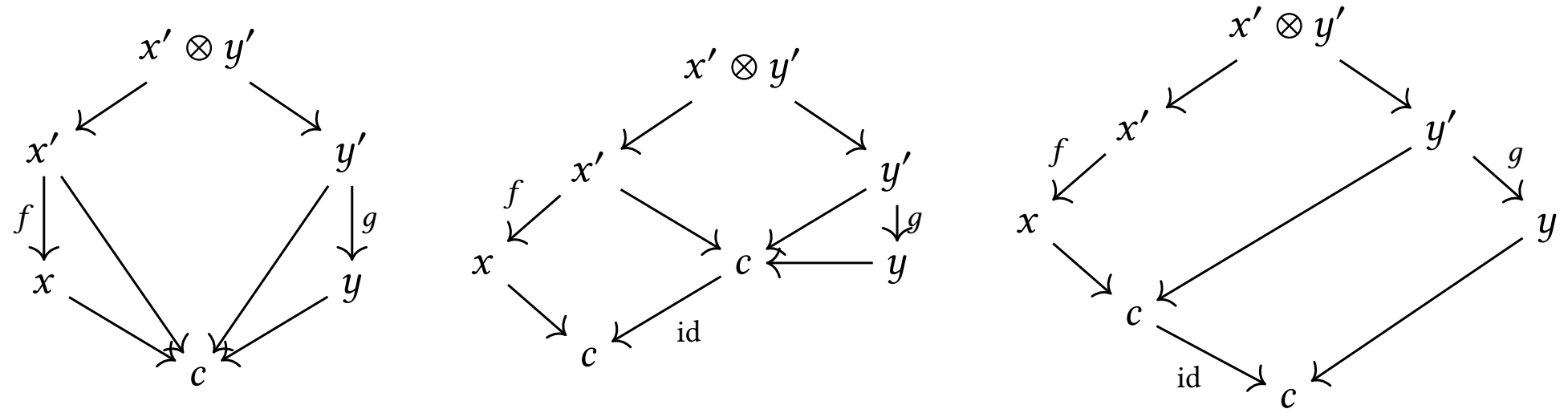
  \[
  \begin{tikzcd}[sep=small,cramped]
    & x'\otimes y' \arrow[ld] \arrow[rd] &                                 \\
    x' \arrow[dd, "f"'] \arrow[rddd] &                                    & y' \arrow[dd, "g"] \arrow[lddd] \\
    &                                    &                                 \\
    x \arrow[rd]                     &                                    & y \arrow[ld]                    \\
    & c                                  &
  \end{tikzcd}
  \qquad
  \begin{tikzcd}[sep=small,cramped]
    &                                & x'\otimes y' \arrow[ld] \arrow[rd] &                              \\
    & x' \arrow[ld, "f"'] \arrow[rd] &                                    & y' \arrow[d, "g"] \arrow[ld] \\
    x \arrow[rd] &                                & c \arrow[ld, "\mathrm{id}"]        & y \arrow[l]                  \\
    & c                              &                                    &
  \end{tikzcd}
  \qquad
  \begin{tikzcd}[sep=small,cramped]
    &                              & x'\otimes y' \arrow[ld] \arrow[rd] &                                 &                \\
    & x' \arrow[ld, "f"']          &                                    & y' \arrow[rd, "g"] \arrow[lldd] &                \\
    x \arrow[rd] &                              &                                    &                                 & y \arrow[lldd] \\
    & c \arrow[rd, "\mathrm{id}"'] &                                    &                                 &                \\
    &                              & c                                  &                                 &
  \end{tikzcd}
  \]
  The diagram on the left is copied from above. The triangles on the side commute by assumption
  and the inner quadrilateral is the independent pullback used to define \(x'\otimes y'\).
  The diagram in the middle performs a rearrangement and introduces an identity morphism;
  this makes it clear the independent pullback square for \(x'\otimes y'\) pastes with
  the square beneath it, which is independent by IP1, making an independent rectangle.
  The diagram on the right simplifies this independent rectangle,
  and then performs a similar operation on the remaining triangle,
  introducing a second identity morphism so that it becomes an independent square by IP1;
  the pasting of these two rectangles gives the independence square needed to define \(f \otimes g\).

Each functor \(f^* : \cat{C}/c \to \cat{C}/c'\)
has action on morphisms defined by:
  \(
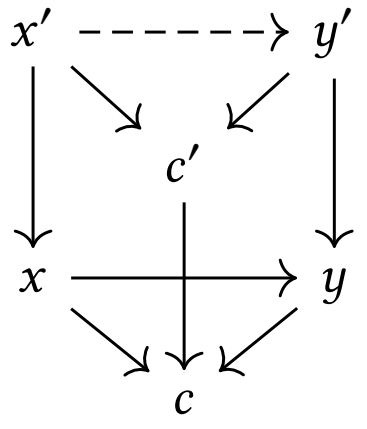
  \begin{tikzcd}[sep=small]
    x' \arrow[rd] \arrow[dd] \arrow[rr, dashed] &               & y' \arrow[ld] \arrow[dd] \\
    & c' \arrow[dd] &                          \\
    x \arrow[rd] \arrow[rr]                     &               & y \arrow[ld]             \\
    & c             &
  \end{tikzcd}\)

In this diagram
the left and right rectangles depict the independent pullback of two objects \(x\) and \(y\) of \(\cat{C}/c\)
  along \(f : c' \to c\).
  The bottom triangle depicts a morphism from \(x\) to \(y\).
  As with ordinary pullbacks,
  the action of \(f^*\) on the morphism \(x\to y\)
  is to yield the dashed morphism, defined by
  the universal property of \(y'\) as an independent pullback.
  The only extra proof obligation is to check
  the independence of the two morphisms from \(x'\) to \(c'\) and \(y\)
  with respect to the two morphisms from \(c'\) and \(y\) to \(c\).
  This independence is exhibited by the following transformation of diagrams.
  
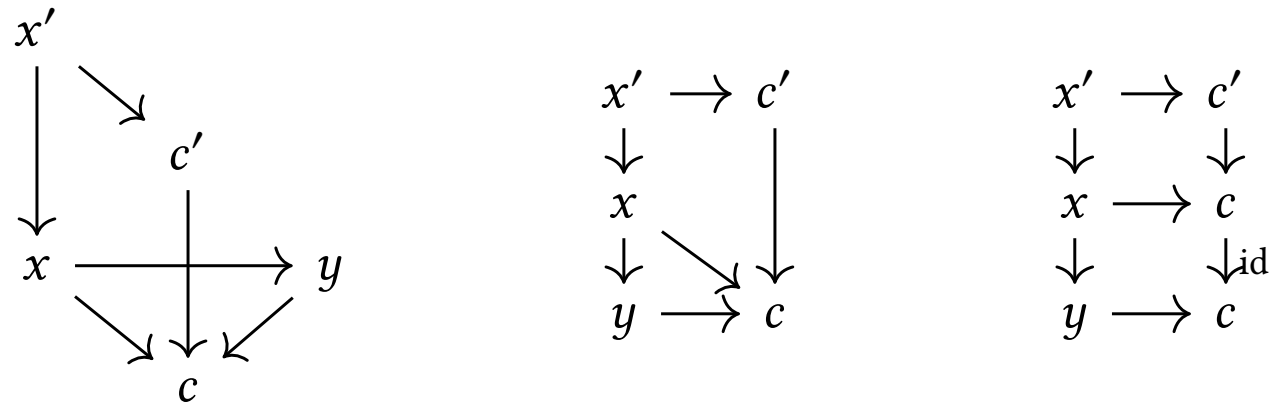
\[
  \begin{tikzcd}[sep=small]
    x' \arrow[dd] \arrow[rd] &               &              \\
    & c' \arrow[dd] &              \\
    x \arrow[rd] \arrow[rr]  &               & y \arrow[ld] \\
    & c             &
  \end{tikzcd}
  \qquad\qquad
  \begin{tikzcd}[sep=small]
    x' \arrow[d] \arrow[r] & c' \arrow[dd] \\
    x \arrow[rd] \arrow[d] &               \\
    y \arrow[r]            & c
  \end{tikzcd}
  \qquad\qquad
  \begin{tikzcd}[sep=small]
    x' \arrow[d] \arrow[r] & c' \arrow[d]               \\
    x \arrow[d] \arrow[r]  & c \arrow[d, "\mathrm{id}"] \\
    y \arrow[r]            & c
  \end{tikzcd}
  \]
  The diagram on the left is copied from above: the top quadrilateral is an independent pullback
  and the triangle commutes. The diagram in the middle pulls the bottom triangular face out front.
  The diagram on the right introduces an identity morphism.
  This shows that the original diagram is a composite of an independent pullback
  and a square that is independent by IP1;
  hence, it is independent with respect to the two morphisms from \(c'\) and \(y\) to \(c\).

From these definitions,
the standard proofs for pullbacks work to show
the axioms of a symmetric semicartesian category hold,
that each \(f^*\) satisfies the functoriality laws,
that each \(f^* : \cat{C}/c \to \cat{C}/c'\) is strong monoidal,
and that pullback forms a pseudofunctor.

\section{Proof that \(\Alloc\) produces atomic sheaves} \label{app:prop:alloc-atomic}

First note that \(\Alloc(P)\) is separated:
if
\(\Alloc(P)(p)[d_1,f_1,x_1] = \Alloc(P)(p)[d_1,f_1,x_2]\)
for some
\([d_1,f_1,x_1]\) and \([d_2,f_2,x_2]\) in \(\Alloc(P)(c)\)
and some \(p : c'\to c\),
then there is a commutative diagram of the following
shape:
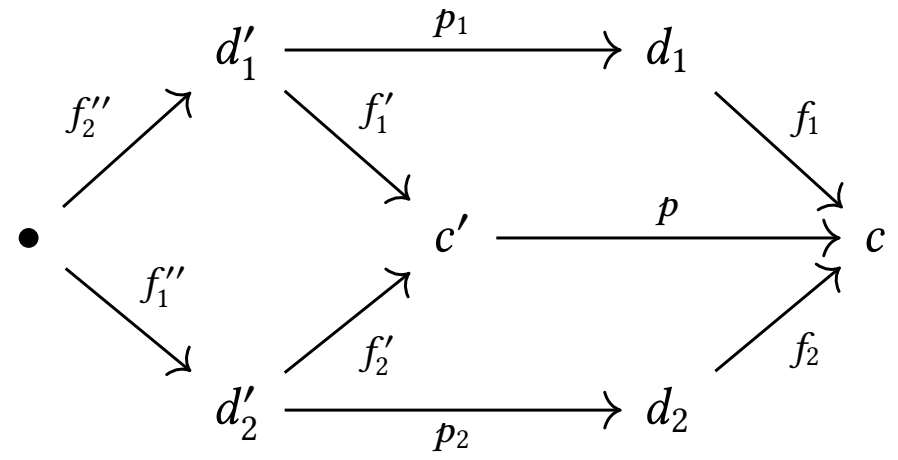
\[
\begin{tikzcd}
                                                & d_1' \arrow[rr, "p_1"] \arrow[rd, "f_1'"]   &                    & d_1 \arrow[rd, "f_1"]  &   \\
\bullet \arrow[ru, "f_2''"] \arrow[rd, "f_1''"] &                                             & c' \arrow[rr, "p"] &                        & c \\
                                                & d_2' \arrow[rr, "p_2"'] \arrow[ru, "f_2'"'] &                    & d_2 \arrow[ru, "f_2"'] &
\end{tikzcd}\]
In this diagram we have that the rhombuses are independent pullbacks,
\(\Alloc(P)(p)[d_1,f_1,x_1] = [d_1',f_1',P(p_1)(x_1)]\)
and
\(\Alloc(P)(p)[d_2,f_2,x_2] = [d_2',f_2',P(p_2)(x_2)]\),
and \(P(p_1)(x_1)\) and \(P(p_2)(x_2)\) are made equal
by applying \(f_2''\) and \(f_1''\) respectively.
From this it follows that \(p_1f_2''\) and \(p_2f_1''\)
witness \([d_1,f_1,x_1] = [d_2,f_2,x_2]\)
as needed.

For the existence of amalgamations,
suppose \(p : c' \to c\)
and
\([d',f',x']\)
is an \(p\)-invariant
element of \(\Alloc(P)(c')\)
in the sense of Simpson~\cite{simpson2017probability}.
Form the independent pullback of \(p\) against itself,
giving \(pq = pr\) for some \(q,r : c'' \to c'\).
The \(p\)-invariance of \([d',f',x']\) then implies
\(\Alloc(P)(q)(x') = \Alloc(P)(r)(x')\),
which gives the dashed arrows
in the following commutative diagram,
where all solid squares
are independent pullbacks
and
\(P(q')(x')\) and \(P(r')(x')\)
are made equal
by
\(P(f'''_1)\) and \(P(f'''_2)\).
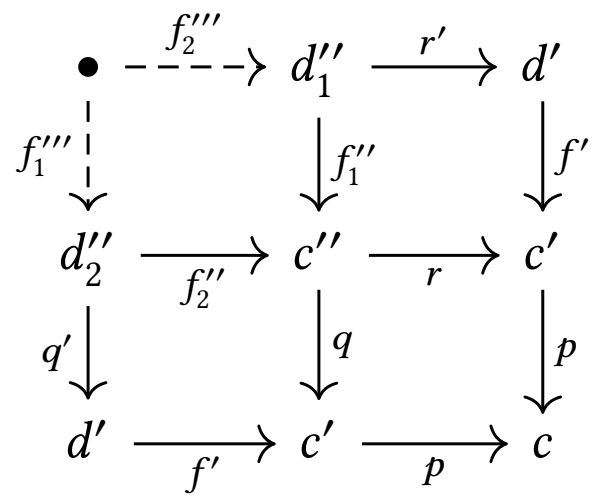
\begin{equation}\label{eqn:amalg}
\begin{tikzcd}
\bullet \arrow[d, "f'''_1"', dashed] \arrow[r, "f'''_2", dashed] & d''_1 \arrow[d, "f''_1"] \arrow[r, "r'"] & d' \arrow[d, "f'"] \\
d''_2 \arrow[d, "q'"'] \arrow[r, "f''_2"']                       & c'' \arrow[r, "r"'] \arrow[d, "q"]       & c' \arrow[d, "p"]  \\
d' \arrow[r, "f'"']                                              & c' \arrow[r, "p"']                       & c
\end{tikzcd}
\end{equation}
Let \(d = d'\), \(f = pf'\), and \(x = x'\),
so \([d,f,x]\) forms an element of \(\Alloc(P)(c)\).
The following calculation then shows that
\([d,f,x]\) is an amalgamation for \([d',f',x']\):
\begin{align*}
\Alloc(P)(p)[d,f,x]
&= [d_1'',qf_1'',P(r')(x)]
  && \text{the right half of (\ref{eqn:amalg}) is an independent pullback}
\\[.7em]
&= [d',f'q',P(q')(x')]
  && \text{the left half of (\ref{eqn:amalg}) commutes and \(P(q')(x')\)}
\\
&&&
\text{and \(P(r')(x')\)
  are identified by \(P(f'''_1)\) and \(P(f'''_2)\)}
\\[.7em]
&= [d',f',x']
&& \text{by the colimit used to define \(\Alloc(P)\)}
\end{align*}

\section{Proof of \Cref{prop:ordered-pcm}}
\label{app:sec:ordered-pcm}

\newcommand\isdefined[1]{{#1}{\downarrow}}
Monotonicity reduces to showing that
if the front face of a commutative cube
    
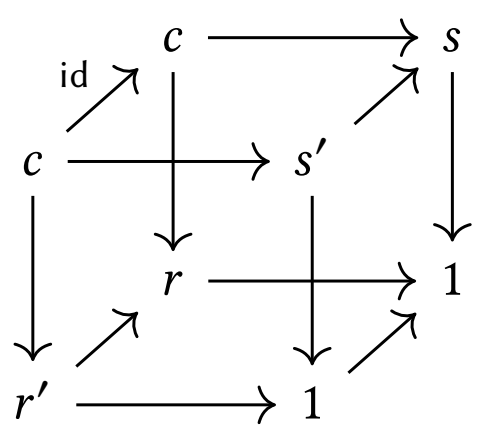
\[
    \begin{tikzcd}[sep=small]
      & c \arrow[dd] \arrow[rr] &                          & s \arrow[dd] \\
      c \arrow[dd] \arrow[rr] \arrow[ru, "\mathsf{id}"] &                         & s' \arrow[dd] \arrow[ru] &              \\
      & r \arrow[rr]            &                          & 1            \\
      r' \arrow[rr] \arrow[ru]                          &                         & 1 \arrow[ru]             &
    \end{tikzcd}
  \]
is an independence square, then the back face is an independence square too,
and that
there is an induced morphism \(r\otimes s \to r'\otimes s'\) between the independent pullbacks.

Consider the following diagrams.
  \begin{center}
    \(
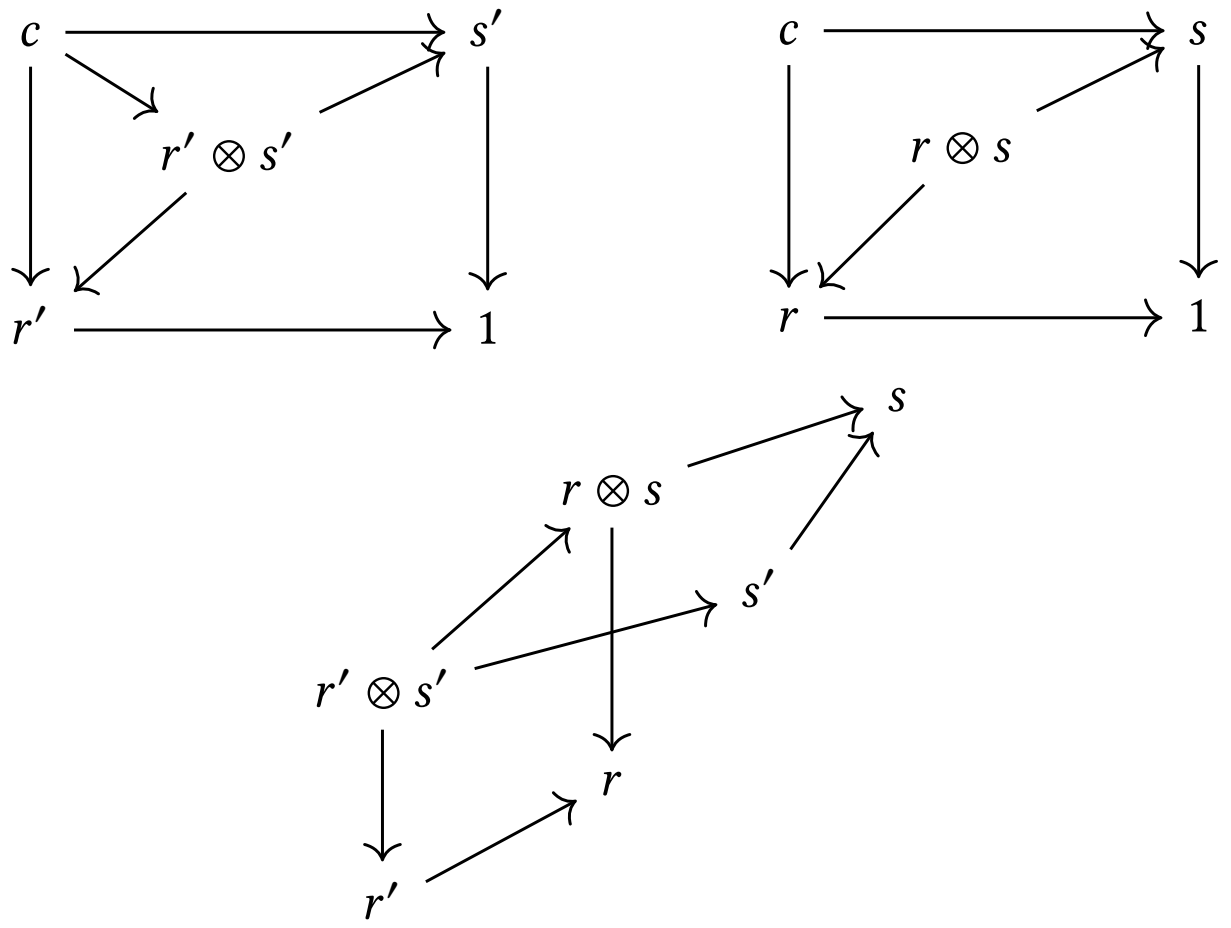
    \begin{tikzcd}[sep=small]
      c \arrow[ddd] \arrow[rrr] \arrow[rd] &                                      &  & s' \arrow[ddd] \\
      & r'\otimes s' \arrow[ldd] \arrow[rru] &  &                \\
      &                                      &  &                \\
      r' \arrow[rrr]                       &                                      &  & 1
    \end{tikzcd}
    \)
    \qquad\qquad
    \(
    \begin{tikzcd}[sep=small]
      c \arrow[ddd] \arrow[rrr] &                                    &  & s \arrow[ddd] \\
      & r\otimes s \arrow[ldd] \arrow[rru] &  &               \\
      &                                    &  &               \\
      r \arrow[rrr]             &                                    &  & 1
    \end{tikzcd}
    \)
    \qquad\qquad
    \(
    \begin{tikzcd}[sep=tiny]
      &  &                                     &                 &  & s \\
      &  & r\otimes s \arrow[ddd] \arrow[rrru] &                 &  &   \\
      &  &                                     & s' \arrow[rruu] &  &   \\
      r'\otimes s' \arrow[ddd] \arrow[rrru] \arrow[rruu] &  &                                     &                 &  &   \\
      &  & r                                   &                 &  &   \\
      &  &                                     &                 &  &   \\
      r' \arrow[rruu]                                    &  &                                     &                 &  &
    \end{tikzcd}
    \)
  \end{center}
  By the universal property of \(r'\otimes s'\), the front face decomposes into the diagram shown on the left.
  Similarly, the back space decomposes into the diagram in the middle, although we do not yet know whether the upper-left boomerang commutes.
  Finally, because \(\otimes\) is semicartesian~(\Cref{prop:lip-locally-semicart}),
  the morphisms \(r' \to r\) and \(s' \to s\) form a morphism \(r'\otimes s' \to r\otimes s\)
  making the diagram on the right, whose shape is that of two rectangles glued together.
  Placing all of these diagrams into the cube gives:
  
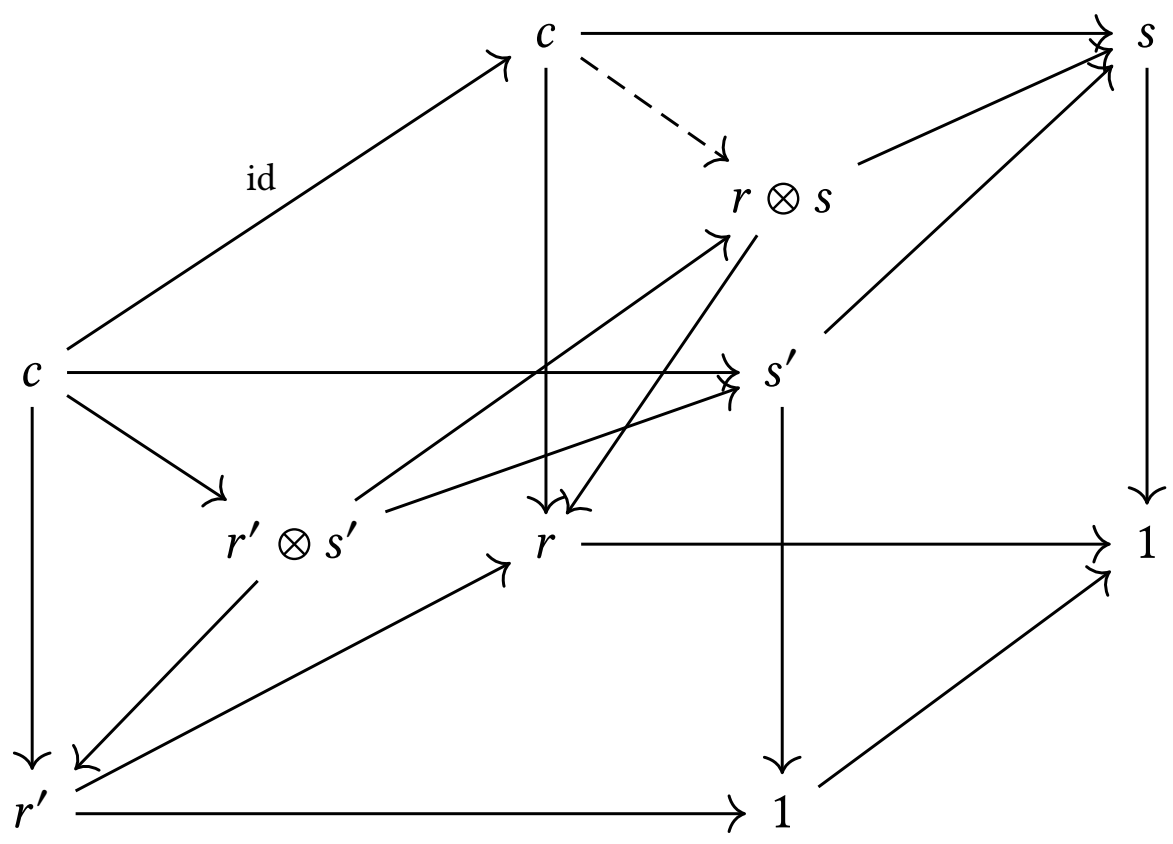
\[
  \begin{tikzcd}
    &                                                   & c \arrow[ddd] \arrow[rrr] \arrow[rd, dashed] &                                    &  & s \arrow[ddd] \\
    &                                                   &                                              & r\otimes s \arrow[ldd] \arrow[rru] &  &               \\
    c \arrow[ddd] \arrow[rrr] \arrow[rruu, "\mathrm{id}"] \arrow[rd] &                                                   &                                              & s' \arrow[ddd] \arrow[rruu]        &  &               \\
    & r'\otimes s' \arrow[ldd] \arrow[rru] \arrow[rruu] & r \arrow[rrr]                                &                                    &  & 1             \\
    &                                                   &                                              &                                    &  &               \\
    r' \arrow[rrr] \arrow[rruu]                                      &                                                   &                                              & 1 \arrow[rruu]                     &  &
  \end{tikzcd}\]
  The dashed morphism is defined to be the composite \(c \to r'\otimes s'\to r\otimes s\),
  making the rectangle that it is incident to commute by fiat.
  But now this makes the dashed morphism a morphism into an independent pullback
  that recovers the morphisms \(c \to r\) and \(c \to s\) when composed with the projections out of \(r\otimes s\).
  Since the independent pullback classifies independence squares with this property,
  it must be that \(c \to r\) and \(c \to s\) form an independence square with respect to \(r \to 1\) and \(s \to 1\).

\section{Proof of \Cref{thm:Liftalloc}} \label{app:sec:Liftalloc}

First we check naturality and monotonicity.
\begin{enumerate}
  \item
    For naturality, suppose \(([c,p,x],(r,f))\in\Liftalloc(\varphi)\)
    is witnessed by
    \((x,(r',f'))\in\varphi\) for some \(r',f'\) fitting into
    an independence square of the form just shown above,
    and let \(g : d \to c\) be an arbitrary morphism of \(\cat{C}\).
    Taking the independent pullback of \(g\) and \(p\) and pasting with the above square
    gives the independence square
    \(
    \begin{tikzcd}[sep=tiny]
      d' \arrow[r, "g'"] \arrow[d] & c' \arrow[d] \arrow[r, "f'"] & r' \arrow[d] \\
      d \arrow[r, "g"]             & c \arrow[r, "f"]             & r
    \end{tikzcd}\)
    which, combined with naturality of \(\varphi\),
    establishes \((P(g')(x),(r',f'g')) \in \varphi\)
    as needed to show \((\Alloc(P)(g)[c',p,x],\Res(g)(r,f)) \in \Liftalloc(\varphi)\).

  \item For monotonicity, consider the following diagram:
    \(
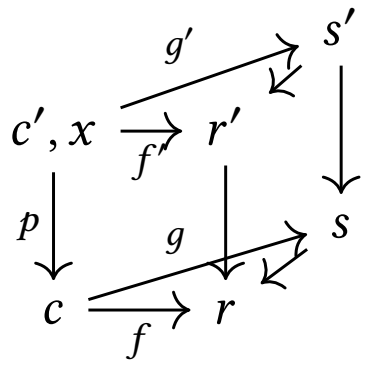
    \begin{tikzcd}[sep=tiny]
      &  &               & s' \arrow[ld] \arrow[dd] \\
      {c',x} \arrow[dd, "p"'] \arrow[rr, "f'"'] \arrow[rrru, "g'"] &  & r' \arrow[dd] &                          \\
      &  &               & s \arrow[ld]             \\
      c \arrow[rr, "f"'] \arrow[rrru, "g"]                         &  & r             &
    \end{tikzcd}\)
    The front face of the square shows an independence square witnessing
    \(((c',p,x),(r,f)) \in \psi\), and
    the bottom triangle shows a hypothetical new resource \((s,g)\)
    that is larger than \((r,f)\). To show that this new resource also satisfies \(\psi\),
    form the right square using independent pullback.
    Now \(f'\) and \(gp\) are independent given the unnamed morphisms \(r' \to r\) and \(s' \to s\),
    giving the morphism \(g'\) into the independent pullback and making the whole prism commute.
    Now the back rectangle is an independence square by
    IP4.
    Hence \(((c',p,x),(g,s)) \in \psi\).
  \end{enumerate}
  For the monad structure on \(\Liftalloc\),
  \begin{itemize}
  \item The unit is \(\eta : (P,\varphi) \to \Liftalloc(P,\varphi)\),
    well-defined because
    if \((x,(r,f)) \in \varphi\)
    then the independence square
    \(
    \begin{tikzcd}
      c \arrow[r, "f"] \arrow[d, "\mathrm{id}"'] & r \arrow[d, "\mathrm{id}"] \\
      c \arrow[r, "f"']                          & r
    \end{tikzcd}\)
    witnesses \(([\mathrm{id},x],(r,f)) \in \Liftalloc(\varphi)\).

  \item The multiplication is \(\mu : \Liftalloc(\Liftalloc(P,\varphi)) \to \Liftalloc(P,\varphi)\),
    well-defined because if \((x,(r'',f'')) \in \varphi\)
    for some \(r'',f''\) coming from a tower of independence squares of the form
    \[
    \begin{tikzcd}
      c'' \arrow[d, "p'"'] \arrow[r, "f''"] & r'' \arrow[d] \\
      c' \arrow[d, "p"'] \arrow[r, "f'"]    & r' \arrow[d]  \\
      c \arrow[r, "f"']                     & r
    \end{tikzcd}\]
    witnesses \(([c',p,[c'',p',x]],(r,f)) \in \Liftalloc(\Liftalloc(P,\varphi))\)
    then the pasting of these two squares into a rectangle
    witnesses \(([c'',pp',x],(r,f)) \in \Liftalloc(P,\varphi)\).
  \end{itemize}
  To show that the strength map \(\sigma : P \times \Alloc(Q) \to \Alloc(P\times Q)\)
  lifts to a morphism \((P,\varphi) \otimes \Liftalloc(Q,\psi) \to \Liftalloc((P,\varphi)\otimes(Q,\psi))\),
  consider the following diagram.
  
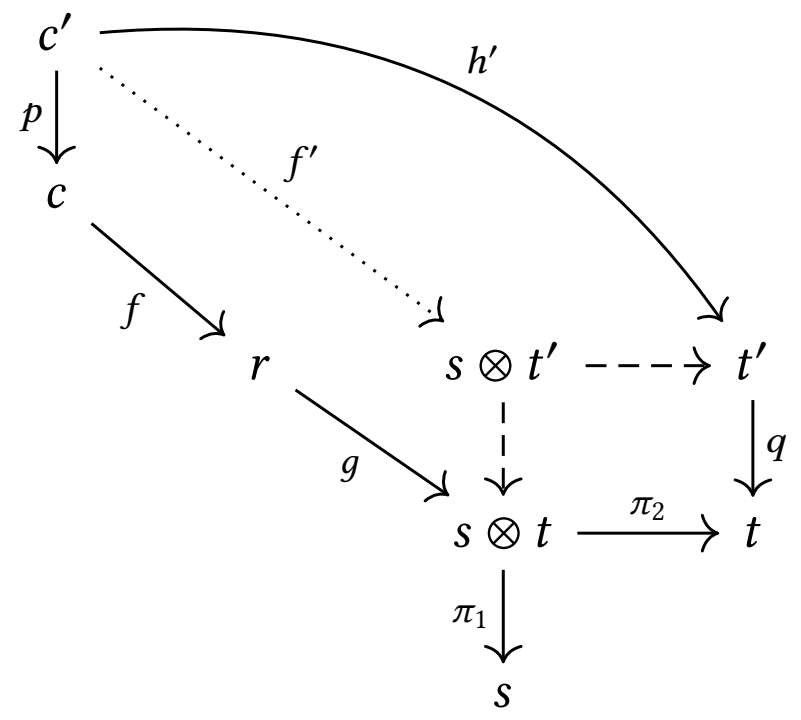
\[
  \begin{tikzcd}
    c' \arrow[d, "p"'] \arrow[rrrdd, "h'", bend left] \arrow[rrdd, "f'", dotted] &                    &                                                   &                   \\
    c \arrow[rd, "f"']                                                           &                    &                                                   &                   \\
    & r \arrow[rd, "g"'] & s\otimes t' \arrow[d, dashed] \arrow[r, dashed]   & t' \arrow[d, "q"] \\
    &                    & s\otimes t \arrow[d, "\pi_1"'] \arrow[r, "\pi_2"] & t                 \\
    &                    & s                                                 &
  \end{tikzcd}\]
  The solid lines depict
  a given witness to \(((x,[c',p,y]),(r,f)) \in (P,\varphi) \otimes \Liftalloc(Q,\psi)\),
  consisting of some \(g : r \to s \otimes t\)
  such that \((x,(s,\pi_1 gf)) \in \varphi\)
  and some \(t',h'\) fitting into an independence quadrilateral,
  with bottom cospan \((\pi_2gf,q)\),
  such that \((y,(t',h')) \in \psi\).
  This quadrilateral can be rearranged into an independent square
  with bottom cospan \((\pi_2,q)\), as follows:
  
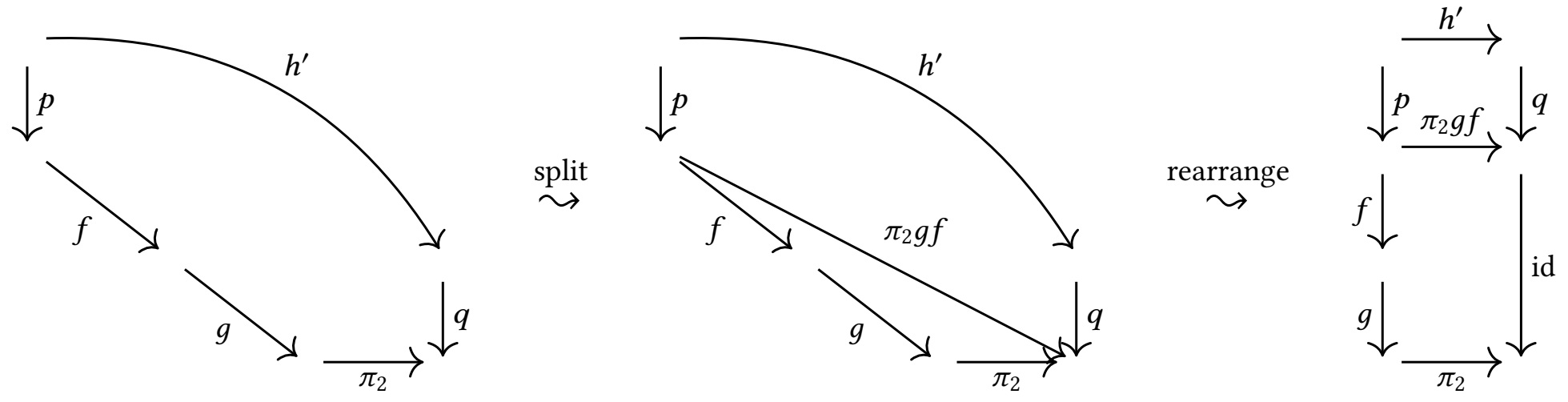
\[
  \begin{tikzcd}
    {} \arrow[d, "p"] \arrow[rrrdd, "h'", bend left] &                     &                        &                   \\
    {} \arrow[rd, "f"']                              &                     &                        &                   \\
    & {} \arrow[rd, "g"'] &                        & {} \arrow[d, "q"] \\
    &                     & {} \arrow[r, "\pi_2"'] & {}
  \end{tikzcd}
  \quad\stackrel{\text{split}}\leadsto\quad
  \begin{tikzcd}
    {} \arrow[d, "p"] \arrow[rrrdd, "h'", bend left] &                     &                        &                   \\
    {} \arrow[rd, "f"'] \arrow[rrrdd, "\pi_2gf"]     &                     &                        &                   \\
    & {} \arrow[rd, "g"'] &                        & {} \arrow[d, "q"] \\
    &                     & {} \arrow[r, "\pi_2"'] & {}
  \end{tikzcd}
  \quad\stackrel{\text{rearrange}}\leadsto\quad
  \begin{tikzcd}
    {} \arrow[d, "p"] \arrow[r, "h'"]       & {} \arrow[d, "q"]            \\
    {} \arrow[d, "f"'] \arrow[r, "\pi_2gf"] & {} \arrow[dd, "\mathrm{id}"] \\
    {} \arrow[d, "g"']                      &                              \\
    {} \arrow[r, "\pi_2"']                  & {}
  \end{tikzcd}
  \]
  The top square is just the independent quadrilateral drawn another way, hence an independent square;
  the bottom square is independent by axiom IP1. Hence the pasting of these
  two polygons forms an independent rectangle.
  This gives a mediating dotted morphism \(f'\) above,
  into the independent pullback \(s\otimes t'\) whose projections are the dashed morphisms.
  Now applying IP4 to the back quadrilateral
  and the dashed square shows that the rectangle involving \(f'\)
  is independent.
  Putting this together with \(((P(p)(x),y),(s\otimes t', f')) \in (P,\varphi)\otimes(Q,\psi)\)
  gives \(([c',p,(P(p)(x),y)],(s\otimes t,gf)) \in \Liftalloc((P,\varphi) \otimes(Q,\psi))\)
  as needed to lift the strength \(\sigma\).

  To show that the lifting is fibered,
  let \(\alpha : (P,\varphi) \to (Q,\psi)\) be a cartesian arrow above \(\alpha : P \to Q\).
  Unwinding what it means to be cartesian in a lax slice,
  this says that for any \(x \in P(c)\) and \((r,f)\in \Res(c)\),
  it holds that \((x,(r,f)) \in \varphi\) if and only if \((\alpha_c(x),(r,f)) \in \psi\).
  We now show that for any \(t \in (\Alloc P)(c)\) and \((r,f) \in \Res(c)\),
  it holds that \((t,(r,f)) \in \Liftalloc(P,\varphi)\) if and only if \((\Alloc(\alpha)(t),(r,f)) \in \Liftalloc(Q,\psi)\).
  \begin{itemize}
  \item[\(\Rightarrow\)]
    Suppose \(t = [p : c' \to c, x : P(c')]\)
    and
    \(
    \begin{tikzcd}[sep=small]
      c' \arrow[r, "f'"] \arrow[d, "p"'] & r' \arrow[d] \\
      c \arrow[r, "f"']                  & r
    \end{tikzcd}\)
    is an independence square
    and \((x,(r',f')) \in \varphi\).
    Then by assumption \((\alpha_{c'}(x),(r',f')) \in \psi\),
    which shows \(([p, \alpha(x)],(r,f)) = (\Alloc(\alpha)(t),(r,f)) \in \Liftalloc(Q,\psi)\).
  \item[\(\Leftarrow\)]
    Let \(t = [p:c_0\to c,x:P(c_0)]\).
    Suppose \(\Alloc(\alpha)(t) = [q:c'\to c,y:Q(c')]\)
    and
    \(\begin{tikzcd}[sep=small]
    c' \arrow[r, "f'"] \arrow[d, "q"'] & r' \arrow[d] \\
    c \arrow[r, "f"']                  & r
    \end{tikzcd}\)
    is an independence square
    and \((y,(r',f')) \in \psi\).
    The equality \(\Alloc(\alpha)(t) = [q,y]\)
    implies that there is some zig-zag
    \(
    \begin{tikzcd}[sep=tiny]
      & {} \arrow[ld] \arrow[rd] &       & {} \arrow[ld] \arrow[rd] &    \\
      c_0 &                          & \dots &                          & c'
    \end{tikzcd}\)
    in the slice over \(c\)
    connecting \(\alpha(x)\) to \(y\).
    By repeatedly taking independent pullbacks, this zig-zag can be straightened into
    a span
    \(
    \begin{tikzcd}[sep=tiny]
      &                             & c_* \arrow[ld] \arrow[rd] &                             &    \\
      & \dots \arrow[ld] \arrow[rd] &                           & \dots \arrow[ld] \arrow[rd] &    \\
      c_0 &                             & \dots                     &                             & c'
    \end{tikzcd}\). Let \(h\) and \(k\) be the left and right legs of this span respectively,
    so we have that \(Q(h)(\alpha(x)) = Q(k)(y)\).
    Now we have
    \begin{align*}
      &(Q(k)(y),(r',f'k)) \in \psi && \text{by \((y,(r',f'))\in\psi\) and naturality of \(\psi\)} \\
      \implies &(Q(h)(\alpha(x)),(r',f'k)) \in \psi && \text{by \(Q(h)(\alpha(x)) = Q(k)(y)\)} \\
      \implies &(\alpha(P(h)(x)),(r',f'k)) \in \psi && \text{by naturality of \(\alpha\)} \\
      \implies &(P(h)(x),(r',f'k)) \in \varphi && \text{by \(\alpha : (P,\varphi)\to(Q,\psi)\) cartesian}
    \end{align*}
    Putting this all together, we have that (1) \(t = [qk : c_* \to c, P(h)(x)]\) and (2) the outer kite of
    \(
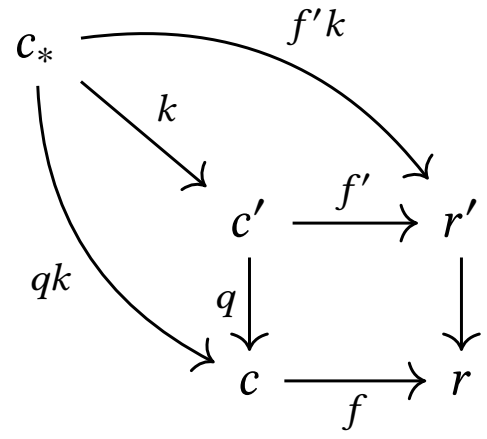
    \begin{tikzcd}
      c_* \arrow[rd, "k"] \arrow[rdd, "qk"', bend right] \arrow[rrd, "f'k", bend left] &                                    &              \\
      & c' \arrow[d, "q"'] \arrow[r, "f'"] & r' \arrow[d] \\
      & c \arrow[r, "f"']                  & r
    \end{tikzcd}\)
    is independent because the square is and (3)
    \((P(h)(x),(r',f'k)) \in \varphi\),
    which is enough to show \((t,(r,f)) \in \Liftalloc(P,\varphi)\).
    \qedhere
  \end{itemize}

\section{{Proof of Theorem~\ref{th:error-credit}}}
\label{app:error-credit}

In our proof we will implicitly use the internal higher order logic of the topos $\topos$ and the resource algebra semantics of separation logic.
\begin{description}
\item[Functorial action] Let $f : (A, p) \to (B, q)$ be a morphism. We want to show that $T f : (A, T q) \to (B, T q)$. Assuming that $a \in (T' p) (m,r)$, we want to show that $Tf (b) \in (T' q) (m, r)$. Unfolding the definitions, we want to show that if $a \in T(\lambda r'.\, p\, m, r')\, r$, then $Tf (a) \in T(\lambda r'.\, q\, m, r')\, r$. This follows by the assumption that $(T, \hat{T})$ is a monad lifting and by instantiating its functorial action to the $R$-predicate $p \, m$.
\item[Unit] Assume that $a \in p(m, r)$. We want to show that $\eta(a) \in (T'p)(m, r)$. By assumption, for every predicate $q :R \times A \to \Omega$, if $a \in q(r)$, then $\eta(a) \in T(q)(r)$. By instantiating this with the predicate $p(m, -)$, we recover what we want to prove.
\item[Multiplication] Assume that $c \in T'T'p(m, r)$. We want to show that $\mu(c) \in T'p(m,r)$. Unfolding the definitions, we have to prove that if $c \in TT(\lambda r'.\, p(m,r'))\, r$ then $\mu(c) \in T(\lambda r'.\, p(m, r'))\, r$.
\item[Strength] Assuming the usual BI semantics, when we unfold the separating conjunction, we get 
  \begin{align*}
    &\exists m_1, m_2, r_1, r_2. (m = m_1 \cdot m_2) \land (r = r_1 \cdot r_2) \land q(m_1, r_1) \land T(\lambda r'.\, p(m_1, r'))\, r_2\\
    &\leq \exists m_1, m_2. (m = m_1 \cdot m_2) \land T(\lambda r'.\, \exists r_1, r_2. (r' = r_1\cdot r_2) \land q(m_1, r_1) \land p(m_2, r_2))\, r \\
    &\leq T(\lambda r'.\, \exists m_1, m_2, r_1, r_2. (m = m_1 \cdot m_2) \land (r' = r_1 \cdot r_2) \land q(m_1, r_1) \land p(m_2, r_2))\, r
  \end{align*}
  The first inequality follows from the assumption that $T$
  satisfies the frame rule for the resource $R$.  The last
  inequality follows from using the elimination rules of
  existential quantification --- introducing free variables $m_1$
  and $m_2$ --- applying the assumed monotonicity of $T$ and
  concluding by introducing the first two RHS existential
  quantifications with $m_1$ and $m_2$, respectively.\qedhere
\end{description}

\section{Proof of Theorem~\ref{th:huber}}
\label{app:huber}

We approach this proof by proving a 2-categorical generalization that specializes
to \Cref{th:huber}. At a high-level, we want to show that the Huber envelope has
a natural universal that can be proved at the 2-categorical level.
Let $\twocat{K}$ be a 2-category and $l \dashv r : a \to b$ an adjunction internal
to $\twocat{K}$. The categories $\cat{MndAdj}_1$ and $\cat{MndAdj}_2$ are defined
as follows:
\begin{definition}
  The category of $\cat{MndAdj}_1$ is defined as
  \begin{itemize}
  \item The objects are monads over $b$
  \item The morphisms are monad morphisms. 
  \end{itemize}
  Identity and composition are given componentwise.
\end{definition}

\begin{definition}
  The category $\cat{MndAdj}_2$ is defined as.
  \begin{itemize}
  \item The objects are pairs of monads over $a$ and $b$, and a monad morphism $s \to t$.
  \item The morphisms are pairs monad liftings morphisms
    such that it commutes with the underlying monad morphism $l s \to t l$.
  \end{itemize}
\end{definition}

These categories can be seen as generalizations of $\cat{RNRLift}(p_I, p_F)$
and $\cat{PreRNRLift}(p_I, p_B)$. We now prove the following universal
property of the Huber envelope.
\begin{theorem}
  The forgetful functor $\cat{MndAdj}_2 \to \cat{MndAdj}_1$ has a left adjoint.
\end{theorem}
\begin{proof}
  A monad $t : b \to b$ is mapped to its Huber envelope $rtl$.
  It is straightforward to establish the functoriality of this
  operation, which we denote $F$. We now have to show that there
  is a natural isomorphism $\cat{MndAdj}_2(F - , =) \cong \cat{MndAdj}_1(-, U =)$,
  which follows from a simple calculation.
\end{proof}

We can see that our categories of RNR structures are almost of the shape
$\cat{MndAdj}_1$ and $\cat{MndAdj}_2$ for an appropriate underlying 2-category.
We finish the proof by proving a lemma that allows us to restrict
adjunctions up restrictions, in a sense we make precise now.

\begin{definition}
  A 1-cell $f : a \to b$ is an ff-morphism if for every object $x$,
  the induced functor $\twocat{K}(x, a) \to \twocat{K}(x, b)$ is
  full an faithful.
\end{definition}

When a 1-cell satisfies this property, we use $\hookrightarrow$. We now need the following lemma.
\begin{lemma}
  Consider the pullback below:
  
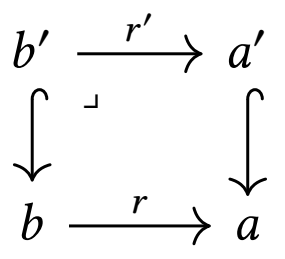
\[\begin{tikzcd}
	        {b'} & {a'} \\
	        b & a
	        \arrow["{r'}", from=1-1, to=1-2]
	        \arrow[hook, from=1-1, to=2-1]
	        \arrow["\lrcorner"{anchor=center, pos=0.125}, draw=none, from=1-1, to=2-2]
	        \arrow[hook, from=1-2, to=2-2]
	        \arrow["r", from=2-1, to=2-2]
  \end{tikzcd}\]
  If the left adjoint $l$ factors through the inclusion $a' \hookrightarrow a$
  as a morphism $l'$, then the adjunction $l \dashv r$ lifts to an adjunction
  $l' \dashv r'$.
\end{lemma}
\begin{proof}
  The unit and counit are defined using the full and faithful
  functors $\twocat{K}(a', a') \to \twocat{K}{(a', a)}$ and
  $\twocat{K}(b', b') \to \twocat{K}{(b', b)}$ given by postcomposition
  with the inclusions
  $a' \hookrightarrow a$ and $b' \hookrightarrow b$.
  Showing that they satisfy the adjunction triangle laws follows by injectivity
  and the triangle laws of the adjunction $l \dashv r$.
\end{proof}

We want to finish the proof by applying the theorem and lemma
above. Given a cartesian base category $\cat{B}$, consider the
2-category $\twocat{Cat}/\cat{B}$. We can now construct the
adjunction given by the theorem above, which almost gives us the
theorem we want. In order to bridge this gap, we restrict to the
appropriate subcategories where we ask the right hand fibration to be
a bunched fibration and the left hand fibration to be a first-order
fibration, and then we apply the lemma above. In order to
do so, we need to show that the left adjoint does indeed factor
through the restriction. Concretely, we want to show that the Huber
envelope of a strong monad is a strong monad, when the adjunction is
monoidal. This generalizes a result by \citet[Lemma~6.1]{kammar2022fully}
  \begin{figure}
    \[\begin{tikzcd}
	        {A \otimes B} & {A \otimes T B} & {A \otimes T^2B} & {T(A \otimes TB)} & {T^2(A \otimes B)} \\
	        & {T(A \otimes B)} & {A \otimes TB} && {T(A\otimes B)}
	        \arrow["{A \otimes \eta_B}", from=1-1, to=1-2]
	        \arrow["{\eta_{A \otimes B}}"', from=1-1, to=2-2]
	        \arrow["{\sigma_{A, B}}", from=1-2, to=2-2]
	        \arrow["{\sigma_{A, TB}}", from=1-3, to=1-4]
	        \arrow["{A \otimes \mu_B}"', from=1-3, to=2-3]
	        \arrow["{T\sigma_{A, B}}", from=1-4, to=1-5]
	        \arrow["{\mu_{A \otimes B}}", from=1-5, to=2-5]
	        \arrow["{\sigma_{A, B}}"', from=2-3, to=2-5]
    \end{tikzcd}\]
    \caption{Unit and multiplication strength laws}
    \label{fig:strength}
  \end{figure}

\begin{lemma}
  Assuming a pre RNR structure, the Huber envelope of the monad is
  a strong monad.
\end{lemma}
\begin{proof}
  Let $\overline{T}_I = R \circ \overline{T}_B\circ L$. The monad
  structure is defined using the Huber envelope. The strength at $A$ and $B$ is
  the dashed morphism below, which is the composition of the other morphisms.
\[\begin{tikzcd}
	{A \times R\overline{T}_B L(B)} && {R(\overline{T}_B (L(A\times B))} \\
	{R L(A) \times R \overline{T}_B L(B)} & {R(L(A)\otimes \overline{T}_B L(B))} & {R(\overline{T}_B (L(A)\otimes L(B))}
	\arrow[dashed, from=1-1, to=1-3]
	\arrow["{\eta\times id}", from=1-1, to=2-1]
	\arrow["{\mathit{lax}}", from=2-1, to=2-2]
	\arrow["\sigma", from=2-2, to=2-3]
	\arrow["\cong", from=2-3, to=1-3]
\end{tikzcd}\]

  Where $\mathit{lax}$ is the multiplicative part of the lax monoidal structure we
can define for $R$, which can be defined using our assumption that $L$
is left adjoint to $R$ and that it is strong monoidal,
i.e. it has coherent natural isomorphisms $L(A) \otimes_{\cat{D}} L(B) \to L(A \otimes_{\cat{C}} B)$ and $e : I_{\cat{D}} \to L(I_{\cat{C}})$:
\[R(X) \otimes R(Y) \xto{\eta} RL(R(X) \otimes R(Y)) \xto{R\cong} R(LR(X) \otimes LR(Y)) \xto{R(\varepsilon \otimes \varepsilon)} R(X \otimes Y)\]
\[I \xto {\eta} RL(I) \xto{\cong} R(I)\]

Strong monads must satisfy the following laws depicted in \Cref{fig:strength}.
The strength laws follow from some diagram chases which we now present. In order to
simplify notation, we refer to $\overline{T}_B$ as simply $T$. We begin by showing that our proposed strength satisfies the unit law.

  
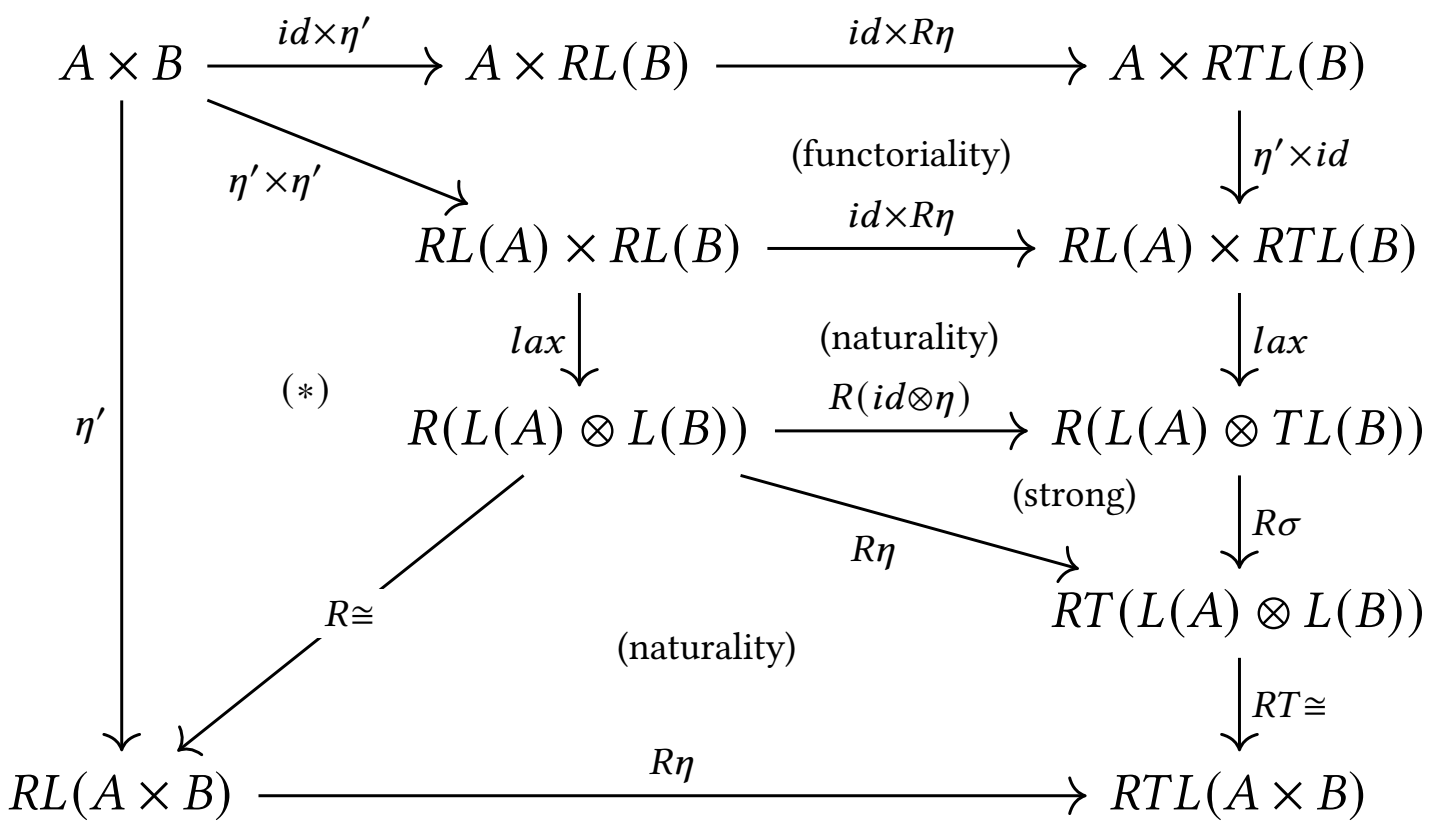
\[\begin{tikzcd}
	    {A \times B} & {A \times RL (B)} && {A \times RTL (B)} \\
	    & {RL(A) \times RL(B)} && {RL(A) \times RTL(B)} \\
	    & {R(L(A) \otimes L(B))} && {R(L(A) \otimes TL(B))} \\
	    &&& {RT(L(A) \otimes L(B))} \\
	    {RL(A\times B)} &&& {RTL(A \times B)}
	    \arrow["{id \times \eta'}", from=1-1, to=1-2]
	    \arrow["{\eta' \times \eta'}"', from=1-1, to=2-2]
	    \arrow[""{name=0, anchor=center, inner sep=0}, "{\eta'}"', from=1-1, to=5-1]
	    \arrow[""{name=1, anchor=center, inner sep=0}, "{id \times R\eta}", from=1-2, to=1-4]
	    \arrow["{\eta'\times id}", from=1-4, to=2-4]
	    \arrow[""{name=2, anchor=center, inner sep=0}, "{id \times R \eta}", from=2-2, to=2-4]
	    \arrow[""{name=3, anchor=center, inner sep=0}, "lax"', from=2-2, to=3-2]
	    \arrow[""{name=4, anchor=center, inner sep=0}, "lax", from=2-4, to=3-4]
	    \arrow["{R(id \otimes \eta)}", from=3-2, to=3-4]
	    \arrow[""{name=5, anchor=center, inner sep=0}, "{R\eta}"', from=3-2, to=4-4]
	    \arrow[""{name=6, anchor=center, inner sep=0}, "{R\cong}"{description}, from=3-2, to=5-1]
	    \arrow[""{name=7, anchor=center, inner sep=0}, "{R \sigma}", from=3-4, to=4-4]
	    \arrow[""{name=8, anchor=center, inner sep=0}, "{RT \cong}", from=4-4, to=5-4]
	    \arrow["{R\eta}", from=5-1, to=5-4]
	    \arrow["{(*)}"{description, pos=0.4}, draw=none, from=0, to=3]
	    \arrow["{\text{(functoriality)}}"{description}, draw=none, from=1, to=2]
	    \arrow["{\text{(naturality)}}"{description}, draw=none, from=3, to=4]
	    \arrow["{\text{(strong)}}"{description}, shift left=2, draw=none, from=5, to=7]
	    \arrow["{\text{(naturality)}}"{description, pos=0.4}, draw=none, from=6, to=8]
  \end{tikzcd}\]
  We now describe the commutativity of the diagram $(*)$. As a reminder, the natural transformation $lax$ is the lax monoidal structure of $R$.
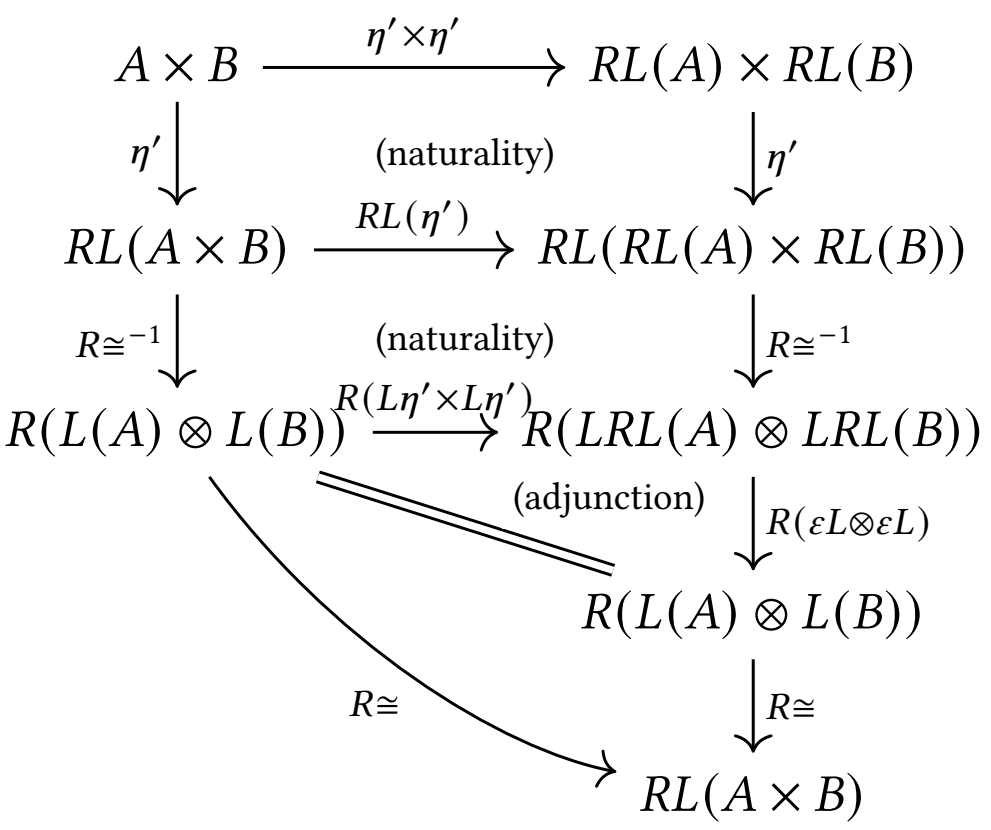
  \[\begin{tikzcd}
	    {A \times B} & {RL(A) \times RL(B)} \\
	    {RL(A \times B)} & {RL(RL(A) \times RL(B))} \\
	    {R(L(A) \otimes L(B))} & {R(LRL(A) \otimes LRL(B))} \\
	    & {R(L(A) \otimes L(B))} \\
	    & {RL(A \times B)}
	    \arrow["{\eta'\times \eta'}", from=1-1, to=1-2]
	    \arrow[""{name=0, anchor=center, inner sep=0}, "{\eta'}"', from=1-1, to=2-1]
	    \arrow[""{name=1, anchor=center, inner sep=0}, "{\eta'}", from=1-2, to=2-2]
	    \arrow["{RL(\eta')}", from=2-1, to=2-2]
	    \arrow[""{name=2, anchor=center, inner sep=0}, "{R\cong^{-1}}"', from=2-1, to=3-1]
	    \arrow[""{name=3, anchor=center, inner sep=0}, "{R \cong^{-1}}", from=2-2, to=3-2]
	    \arrow["{R(L\eta' \times L\eta')}", from=3-1, to=3-2]
	    \arrow[""{name=4, anchor=center, inner sep=0}, equals, from=3-1, to=4-2]
	    \arrow["{R\cong}"', curve={height=18pt}, from=3-1, to=5-2]
	    \arrow[""{name=5, anchor=center, inner sep=0}, "{R(\varepsilon L \otimes \varepsilon L)}", from=3-2, to=4-2]
	    \arrow["{R\cong}", from=4-2, to=5-2]
	    \arrow["{\text{(naturality)}}"{description}, draw=none, from=0, to=1]
	    \arrow["{\text{(naturality)}}"{description}, draw=none, from=2, to=3]
	    \arrow["{\text{(adjunction)}}"{description}, curve={height=-6pt}, draw=none, from=4, to=5]
  \end{tikzcd}\]
  We now present the multiplication diagram
  
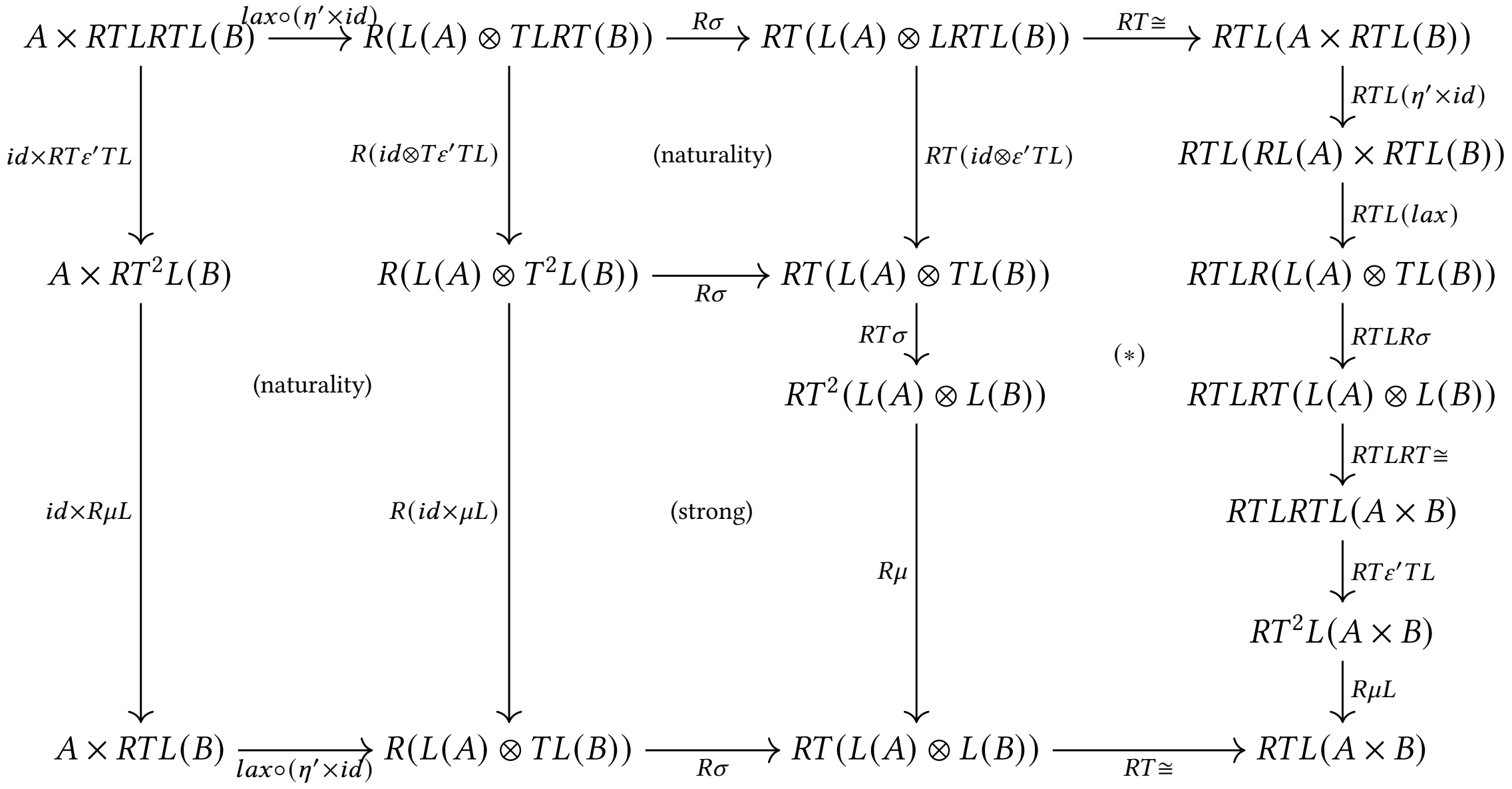
\[\begin{tikzcd}[cramped]
	    {A \times RTLRTL(B)} & {R(L(A) \otimes TLRT(B))} & {RT(L(A) \otimes LRTL(B))} & {RTL(A \times RTL(B))} \\
	    &&& {RTL(RL(A) \times RTL(B))} \\
	    {A \times RT^2L(B)} & {R(L(A) \otimes T^2L(B))} & {RT(L(A) \otimes TL(B))} & {RTLR(L(A) \otimes TL(B))} \\
	    && {RT^2(L(A) \otimes L(B))} & {RTLRT(L(A) \otimes L(B))} \\
	    &&& {RTLRTL(A\times B)} \\
	    &&& {RT^2L(A \times B)} \\
	    {A\times RTL(B)} & {R(L(A) \otimes TL(B))} & {RT(L(A) \otimes L(B))} & {RTL(A \times B)}
	    \arrow["{lax \circ(\eta'\times id)}", from=1-1, to=1-2]
	    \arrow["{id \times RT\varepsilon'TL}"', from=1-1, to=3-1]
	    \arrow["{R\sigma}", from=1-2, to=1-3]
	    \arrow[""{name=0, anchor=center, inner sep=0}, "{R(id \otimes T\varepsilon' T L)}"', from=1-2, to=3-2]
	    \arrow["{RT\cong}", from=1-3, to=1-4]
	    \arrow[""{name=1, anchor=center, inner sep=0}, "{RT(id \otimes \varepsilon' TL)}", from=1-3, to=3-3]
	    \arrow["{RTL(\eta' \times id)}", from=1-4, to=2-4]
	    \arrow["{RTL(lax)}", from=2-4, to=3-4]
	    \arrow["{id \times R\mu L}"', from=3-1, to=7-1]
	    \arrow[""{name=2, anchor=center, inner sep=0}, "{R\sigma}"', from=3-2, to=3-3]
	    \arrow[""{name=3, anchor=center, inner sep=0}, "{R(id \times \mu L)}"', from=3-2, to=7-2]
	    \arrow[""{name=4, anchor=center, inner sep=0}, "{RT\sigma}"', from=3-3, to=4-3]
	    \arrow[""{name=5, anchor=center, inner sep=0}, "{RTLR \sigma}", from=3-4, to=4-4]
	    \arrow["{R\mu}"', from=4-3, to=7-3]
	    \arrow["{RTLRT\cong}", from=4-4, to=5-4]
	    \arrow["{RT\varepsilon'TL}", from=5-4, to=6-4]
	    \arrow["{R\mu L}", from=6-4, to=7-4]
	    \arrow["{lax \circ(\eta'\times id)}"', from=7-1, to=7-2]
	    \arrow[""{name=6, anchor=center, inner sep=0}, "{R\sigma}"', from=7-2, to=7-3]
	    \arrow["{RT\cong}"', from=7-3, to=7-4]
	    \arrow["{\text{(naturality)}}"{description}, draw=none, from=0, to=1]
	    \arrow["{\text{(naturality)}}"{description, pos=0.4}, draw=none, from=3-1, to=3]
	    \arrow["{\text{(strong)}}"{description}, draw=none, from=2, to=6]
	    \arrow["{(*)}"', draw=none, from=4, to=5]
  \end{tikzcd}\]
  We now present the proof of the diagram $(*)$.
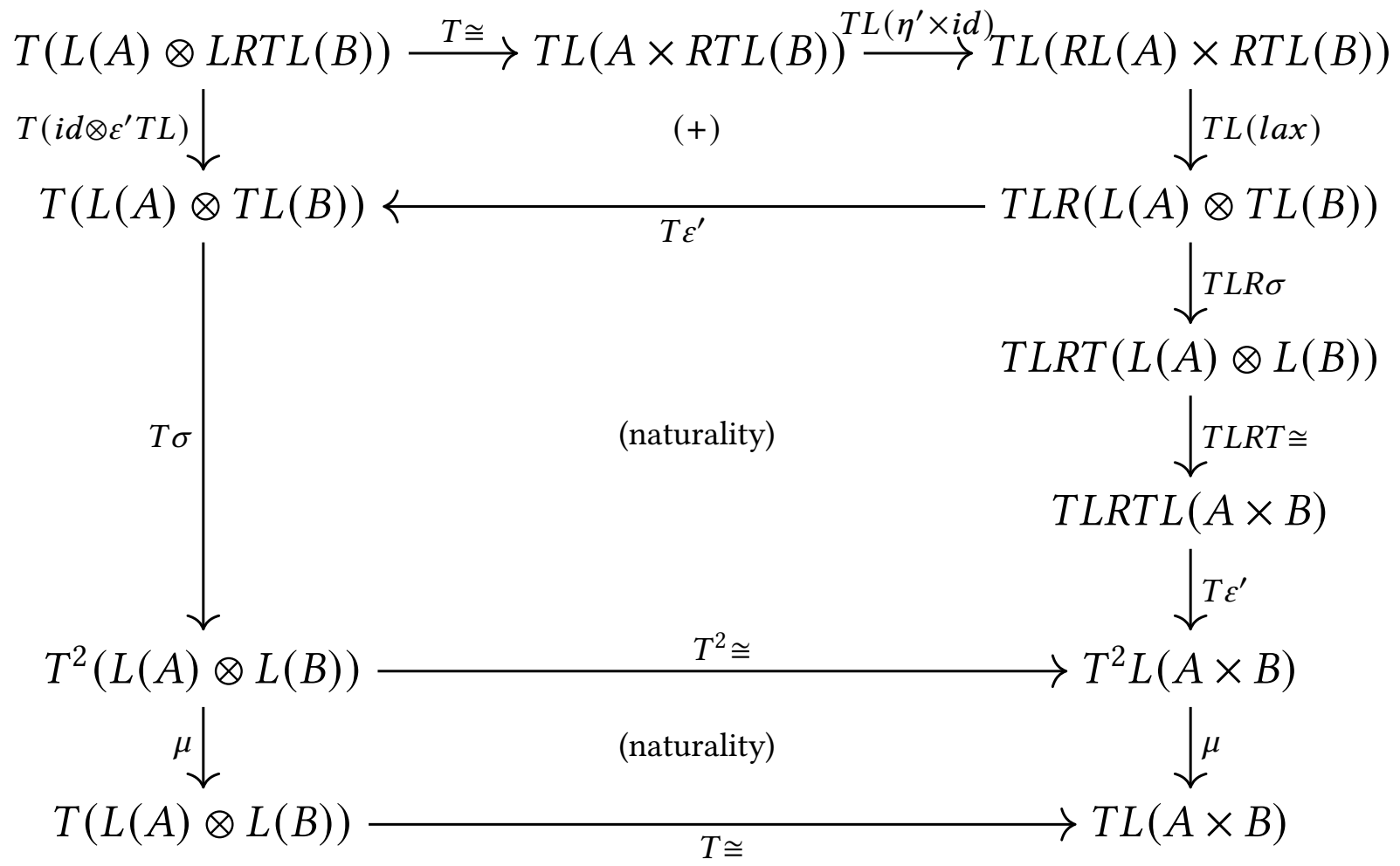
  \[\begin{tikzcd}[cramped]
	    {T(L(A) \otimes LRTL(B))} & {TL(A \times RTL(B))} & {TL(RL(A) \times RTL(B))} \\
	    {T(L(A) \otimes TL(B))} && {TLR(L(A) \otimes TL(B))} \\
	    && {TLRT(L(A) \otimes L(B))} \\
	    && {TLRTL(A\times B)} \\
	    {T^2(L(A) \otimes L(B))} && {T^2L(A \times B)} \\
	    {T(L(A) \otimes L(B))} && {TL(A \times B)}
	    \arrow["{T\cong}", from=1-1, to=1-2]
	    \arrow[""{name=0, anchor=center, inner sep=0}, "{T(id \otimes \varepsilon' TL)}"', from=1-1, to=2-1]
	    \arrow["{TL(\eta' \times id)}", from=1-2, to=1-3]
	    \arrow[""{name=1, anchor=center, inner sep=0}, "{TL(lax)}", from=1-3, to=2-3]
	    \arrow[""{name=2, anchor=center, inner sep=0}, "{T\sigma}"', from=2-1, to=5-1]
	    \arrow["{T\varepsilon'}", from=2-3, to=2-1]
	    \arrow["{TLR \sigma}", from=2-3, to=3-3]
	    \arrow[""{name=3, anchor=center, inner sep=0}, "{TLRT\cong}", from=3-3, to=4-3]
	    \arrow["{T\varepsilon'}", from=4-3, to=5-3]
	    \arrow["{T^2\cong}", from=5-1, to=5-3]
	    \arrow[""{name=4, anchor=center, inner sep=0}, "\mu"', from=5-1, to=6-1]
	    \arrow[""{name=5, anchor=center, inner sep=0}, "\mu", from=5-3, to=6-3]
	    \arrow["{T\cong}"', from=6-1, to=6-3]
	    \arrow["{(+)}"{description}, draw=none, from=0, to=1]
	    \arrow["{\text{(naturality)}}"{description}, draw=none, from=2, to=3]
	    \arrow["{\text{(naturality)}}"{description}, draw=none, from=4, to=5]
  \end{tikzcd}\]
  We conclude by presenting the proof of the diagram $(+)$.
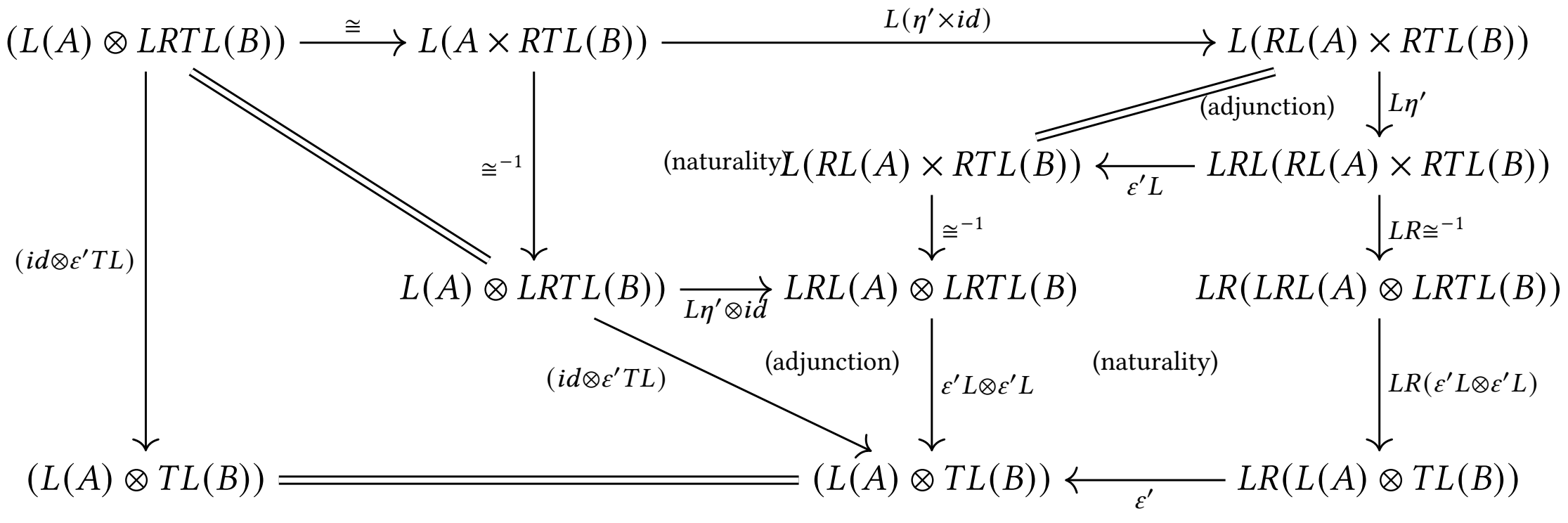
  \[\begin{tikzcd}[cramped]
	    {(L(A) \otimes LRTL(B))} & {L(A \times RTL(B))} && {L(RL(A) \times RTL(B))} \\
	    && {L(RL(A) \times RTL(B))} & {LRL(RL(A) \times RTL(B))} \\
	    & {L(A) \otimes LRTL(B))} & {LRL(A) \otimes LRTL(B)} & {LR(LRL(A) \otimes LRTL(B))} \\
	    \\
	      {(L(A) \otimes TL(B))} && {(L(A) \otimes TL(B))} & {LR(L(A) \otimes TL(B))}
	      \arrow["\cong", from=1-1, to=1-2]
	      \arrow[equals, from=1-1, to=3-2]
	      \arrow["{(id \otimes \varepsilon' TL)}"', from=1-1, to=5-1]
	      \arrow["{L(\eta' \times id)}", from=1-2, to=1-4]
	      \arrow[""{name=0, anchor=center, inner sep=0}, "{\cong^{-1}}"', from=1-2, to=3-2]
	      \arrow[""{name=1, anchor=center, inner sep=0}, equals, from=1-4, to=2-3]
	      \arrow[""{name=2, anchor=center, inner sep=0}, "{L\eta'}", from=1-4, to=2-4]
	      \arrow[""{name=3, anchor=center, inner sep=0}, "{\cong^{-1}}", from=2-3, to=3-3]
	      \arrow["{\varepsilon'L}", from=2-4, to=2-3]
	      \arrow["{LR\cong^{-1}}", from=2-4, to=3-4]
	      \arrow["{L\eta' \otimes id}"', from=3-2, to=3-3]
	      \arrow[""{name=4, anchor=center, inner sep=0}, "{(id \otimes \varepsilon' TL)}"'{pos=0.3}, from=3-2, to=5-3]
	      \arrow[""{name=5, anchor=center, inner sep=0}, "{\varepsilon'L \otimes \varepsilon'L}", from=3-3, to=5-3]
	      \arrow[""{name=6, anchor=center, inner sep=0}, "{LR(\varepsilon'L \otimes\varepsilon'L)}", from=3-4, to=5-4]
	      \arrow[equals, from=5-1, to=5-3]
	      \arrow["{\varepsilon'}", from=5-4, to=5-3]
	      \arrow["{\text{(naturality)}}"{pos=0.3}, draw=none, from=0, to=3]
	      \arrow["{\text{(adjunction)}}"', shift left=2, draw=none, from=1, to=2]
	      \arrow["{\text{(adjunction)}}", draw=none, from=4, to=5]
	      \arrow["{\text{(naturality)}}", draw=none, from=5, to=6]
  \end{tikzcd}\]
\end{proof}

\Cref{th:huber} is now a simple consequence of the results above.

\end{document}